# Variation of protein backbone amide resonance by electrostatic field

John N. Sharley, University of Adelaide. arXiv:1512.05488

john.sharley@pobox.com

## Table of Contents

## 1 Abstract

Amide resonance is found to be sensitive to electrostatic field with component parallel or antiparallel to the amide C-N bond, an effect we refer to here as EVPR-CN. EVPR-CN is linear and without threshold in the biologically plausible electrostatic field range -0.005 to 0.005 au. Variation of amide resonance varies Resonance-Assisted Hydrogen Bonding [1], RAHB, such as occurs in the hydrogen bonded chains of backbone amides of protein secondary structures such as beta sheets [2] and non-polyproline helices [3] such as alpha helices [4], varying the stability of the secondary structure. The electrostatic properties including permittivity of amino acid residue sidegroups influence the electrostatic field component parallel or antiparallel to the C-N bond of each amide, giving a novel relationship between residue sequence and protein structure. Additionally, a backbone-based theory of protein folding [5] which includes this effect is presented in Section 8.4.

The significance of EVPR-CN relative to other factors in protein folding depends on field C-N component at each backbone amide at a given time. Calculation indicates that backbone amides do not occupy an intrinsically electrostatically-protected niche. We propose that EVPR-CN warrants investigation in any study of stable protein structure or protein folding pathway [6]. EVPR-CN is somewhat associated with hydrophobia [7], since hydrophobia creates low permittivity environments. EVPR-CN is more directionally and hence structurally specific than hydrophobia.

Hypotheses concerning the stability of beta sheets and amyloid fibrils [8] and of protein complexation and molecular chaperone function [9] are offered. An analogous effect in nitrogenous base pairing [10] is proposed.

EVPR-CN is energetically significant in biologically plausible electrostatic fields even without considering a hydrogen bonding context, and a hypothesis concerning the stability of polyproline helices types I and II is offered.

- resonance assisted hydrogen bond
- backbone torsion
- base pairing
- stark effect

## 3 Notation

“->” denotes Natural Bond Orbital [11], NBO, resonance-type charge transfer and “|” denotes NBO steric exchange repulsion. “(“ and “)” enclose specification of an orbital type and follow an atom name for single-center NBOs and a pair of atom names separated by “-“ for two-center NBOs.

Examples: N(lp) for the amide nitrogen lone pair NBO, O(lp-p) for the oxygen p-type lone pair NBO, O(lp-s) for the s-rich lone pair NBO, C-O(p)* for the pi carbonyl antibonding orbital NBO and N(lp)->C-O(p)* for the primary amide resonance type charge transfer.

## 4 Overview

The cooperativity of hydrogen-bonded chains of protein backbone amides depends on the resonance at each amide in the chain. This is an example of Resonance-Assisted Hydrogen Bonding, RAHB. We investigate variation of amide resonance and RAHB due to electrostatic field.

Resonance-type charge transfer from the nitrogen lone pair NBO to the carbonyl pi antibonding NBO is primary to amide resonance. An NBO charge transfer variation of 0.001 electrons is nominated as the threshold of chemical significance, having associated energy of ~0.6 kcal/mol (p104-105 of [12]).

A striking change in amide resonance occurs when an electrostatic field is applied parallel or antiparallel to the amide C-N vector, an effect referred to here as EVPR-CN. Using two monovalent ions to create an electrostatic field, we find variation in the primary amide charge transfer of 0.2 electrons, some 200 times the nominated level of chemical significance, and note that this is not the full extent of what is physically possible. Using a uniform electrostatic field, we find that the variation in charge transfer is linear with the field magnitude in the biologically plausible range of -0.005 to 0.005 au. In this range the primary amide charge transfer in N-methylformamide varies by 0.042 electrons. An applied field magnitude of 0.000238 au varies the primary amide charge transfer by 0.001 electrons. We demonstrate changes to RAHB in a chain of N-methylformamides in the presence of an electrostatic field.

EVPR-CN is a previously unrecognized factor in protein folding. The significance of EVPR-CN depends on the C-N vector component of electrostatic field environmental to backbone amides.

The directional sensitivity of the resonance of the amide group to electrostatic field allows amino acid residue specification of protein structure through residue electrostatic properties including permittivity. We propose that EVPR-CN is central to the structure and function of proteins, and that evolution selected a polymer having an amide group in its hydrogen-bonded chains for this effect. The amide group is a remarkably simple molecular mechanism for introducing directional sensitivity to electrostatic field resulting in variation in hydrogen bonding cooperativity, and it may be that no simpler organic molecular mechanism exists for this.

Rose *et al*. [5] propose a backbone-based theory of protein folding in which "the energetics of backbone hydrogen bonding dominate protein folding" and backbone hydrogen bonding is a "universal folding mechanism", but do not propose that this mechanism is directly varied by residue sidechain properties. Electrostatic variation of backbone amide resonance and hence RAHB depends on the electrostatic properties including permittivity of the sidechains in the residue sequence. In addition to EVPR-CN, we propose a backbone-based theory of protein folding in which the universal folding mechanism becomes peptide group (PG) resonance (PR) rather than backbone hydrogen bonding (Section 8.4).

We offer hypotheses concerning the stability of beta sheets, amyloid fibrils, polyproline helices of types I and II and of protein complexation and molecular chaperone action, and of an analogous effect in nitrogenous base pairing.

## 5 Introduction

In some protein secondary structures, such as beta sheets and non-polyproline helices, there exist hydrogen bonded chains of backbone amide groups. These distinctive protein hydrogen bonding patterns were discovered by Pauling *et al.* in 1951 [2, 4]. The hydrogen bonds in these chains are cooperative, with cooperativity mediated by the resonance of the backbone amides. The phenomenon of hydrogen bonding cooperatively depending on resonance is referred to as RAHB.

Any variation of backbone amide resonance is of relevance to hydrogen bonding in backbone amide RAHB chains and hence the stability of RAHB protein secondary structures, the appearance of secondary structures in proteins and the stability of the folded protein structure.

Variation in the primary amide resonance-type charge transfer, N(lp)->C-O(pi)*, is associated with a variation in energy which can be considerable and must be taken into account when considering protein structure stability, regardless of whether this variation is due to EVPR-CN, RAHB or any other source.

Variation of resonance by electrostatic field in general is known, but we are not aware of a report of quantification of this effect in amides or the peptide groups of polypeptide backbones, which is surprising since the resonance of amide or peptide groups is large. The general effect has been described by Shaik and co-workers and Coote and co-workers and investigated by these groups as a mean of catalysis [13-19]. Though electrostatic catalysis is used in enzymes [20-24], it is the least developed form of synthetic chemistry [17]. Karafiloglou observed that an electrostatic field has an important directionally-modulated on delocalization energy and hence the weights of alternative resonance structures even in a non-polar molecule, noting that the same principles must apply to other pi-systems [25]. The general effect appears to be in want of a name and we propose “Field-Dependent Resonance Mixing”. The underlying physics is the very familiar influence of electrostatic field on charge transfer [26]. Electrostatic field variation of amide resonance in particular is mentioned in lecture notes offering an account of the means by which hydrogen bonding to backbone amides changes amide resonance (i.e. the mechanism of RAHB), remarking that “The resonant state could be influenced by an external field in different ways, but the primary cause is hydrogen bonding” [27], though no quantification of the influence of external field on amide resonance is cited. As far as we are aware, no quantification of the variation in backbone amide resonance due to electrostatic field has been reported.

# 6 Software

Methods used in experiments are as implemented by Gaussian 09 D.01 [28], Orca 3.0.3 [29-31] and TeraChem 1.5K [32-35]. Unless otherwise stated, default grids and optimization and SCF convergence limits were used, except that the Orca option VeryTightSCF was used throughout as were cartesian coordinates for geometry optimization with TeraChem.

A development version of NBO [36] was used for its XML [37] output option. The XML was queried with XQuery 3.0 [38] or XSLT 3.0 [39] as implemented by Saxon-PE 9.6.0.4 [40], and the results imported into Excel 2013 [41].

Molecular coordinates are depicted by UCSF Chimera 1.10.2 [42].

# 7 Results and discussion

## 7.1 Constrained to O-C-N plane

We used N-methylethanamide to study amide resonance in the presence of hyperconjugative interactions with the amide carbonyl orbitals, and N-methylformamide otherwise. These molecules have greater amide resonance than formamide and are more representative of protein backbone amides.

With the atoms of N-methylformamide that are in the O-C-N plane in the absence of an electrostatic field constrained to that plane, and one of the methyl hydrogens constrained to this plane in cis configuration with the N-H bond, but all bond lengths and bond angles free to re-optimize, variation in amide resonance due to variation of electrostatic field direction and magnitude is calculated at SCS-MP2 [43] and MP2 [44] with the aug-cc-pVTZ [45] basis set. The term “X-Y vector” refers here to the normalized vector from the X nuclear center to the Y nuclear center. Figure 2 shows that in 360 degree rotation in 10 degree intervals in the amide O-C-N plane, the maximum increase in amide resonance as indicated by C-O(pi)* NBO occupancy occurs when the electrostatic field vector has the same direction as the C-N vector and maximum decrease in the opposite direction. Refining this rotation to 1 degree intervals in a range of -10 to 10 degrees from the C-N vector in the O-C-N plane, the peak is found at 0 or 1 degree (Ap1:Figure 17). Conducting a similar experiment with N-methylethanamide with a hydrogen of the extra methyl group constrained to O-C-N planarity in trans configuration to C-O, the peak is 5 degrees away from the C-N vector, passing through the C-O bond (Ap1:Figure 18). The substituent at the amide carbon makes a difference, but so little elevated is the peak at 5 degrees over the value at 0 degrees, being less than 0.0001 electron which is an order of magnitude lower than the nominated level of chemical significance, that we will take the C-N vector as being a good approximation to the field direction giving maximum effect.

Figure 3 shows that the vectors for maximum resonance increase and decrease in 360 degree rotation in the plane though the O-C-N plane normal and C-N are the same as occur in rotation in the O-C-N plane. Figure 4 shows the results of 360 degree rotation of the electrostatic field normal to the C-N vector about that vector which are that these field directions make little difference to amide resonance.

Available physically determined data for electrostatic variation in proteins [26, 46-50] give that the range of magnitudes without regard to direction is at least -0.005 to 0.005 au.

The variation of C-O(pi)* NBO occupancy with fields of C-N vector and range of magnitudes between minus 0.005 au and 0.005 au is shown in Figure 5, demonstrating a linear variation to C-O(pi)* NBO occupancy with field. DLPNO-CCSD(T) [31] with aug-cc-pVTZ and def2-TZVPP [51] basis sets was used over coordinates optimized at SCS-MP2/aug-cc-pVTZ, again showing linearity. DFT methods B3LYP [52], B97-D3 [35], BLYP [53, 54] and PBE0 [55] in combination with basis sets aug-cc-pVTZ, def2-TZVPP and 6-311++G** [56] all show this same linearity (Ap1:Figure 19), showing that DFT captures this effect well. Ap1:Figure 20 shows the linearity of the variation in C-O(pi)* NBO occupancy with variation of magnitude of N-C vector field in N-methylethanamide at SCS-MP2 and MP2/aug-cc-pVTZ.

The SCS-MP2 and MP2 data shown Figure 5 was obtained with N-C-O, N-C-H, C-N-H and C-N-CA angles free to re-optimize, and there is variation of these angles in the field. At 0.005 au, the N-C-O angle is 124.81 degrees, for the zero field case is 124.25 degrees and for -0.005 au is 123.68. SCS-MP2 and MP2 data was obtained again with these angles fixed to the zero field case, but the data was not discernibly different from that with angles free and is not shown.

We must state to forestall non-confirmatory report that we found an absence of concurrence between the uniform electrostatic field results calculated by Orca [29] and Gaussian [28]. In Gaussian, we used Field=Read and 6 decimal places. Uniform electrostatic field implementation was found by the Orca developers to be in error at Orca 3.0.2, and this was revised for 3.0.3. Because of the discrepancy between Gaussian and Orca, we also conduct computational experiments using ions, and then find that Orca and Gaussian results are in agreement and in accord with Orca's uniform electrostatic field results. We use Orca 3.0.3 and Gaussian 09 D.01 for these experiments, since these have the most robust NBO interfaces of the quantum chemistry packages available to us.

Figure 6, Ap1:Figure 23, Ap1:Figure 24 and Ap1:Figure 25 show the C-O(pi)* NBO occupancy resulting from constraining the ions Li+ or F- to the C-N line, constraining an ion to be either at distance from C further away from N or an ion at distance from N further away from C or both. As can be expected, the largest variation in primary amide charge transfer occurs between the cases of Li- and F+ bracketing C-N versus that with the positions of the ions swapped giving a C-O(pi)* NBO occupancy difference of 0.2 electron (Figure 6), some 200 times the nominated level of chemical significance being 0.001 electron. We have not explored fields of larger and increasingly biologically implausible magnitude and do not give an upper bound on the magnitude of the variation that is physically possible.

Ap1:Figure 21 and Ap1:Figure 22 show the variation in Natural Localized Molecular Orbital [11], NLMO, dipole moment magnitude with field. NLMOs are unitarily equivalent to canonical molecular orbitals and the vector sum of their dipole moments for the molecular system is that also given by canonical molecular orbitals. Significant contributions of the dipoles of lone pairs to the overall dipole of formamide are noted on p150-151 of [12]. The dipole of the oxygen lone pairs of formamide are aligned with the C-O axis, and importantly, the dipole of N(lp) is aligned with the C-N axis. This is surely associated with the N(lp)->C-O(p)* NBO charge transfer. Since this charge transfer is contained within the N(lp) NLMO, the associated dipole is best viewed in terms of NLMOs. The change in this dipole is directly physically related to electrostatic field, whereas the change to N(lp)->C-O(p)* does not by itself reveal the underlying physics.

Ap1:Figure 26 shows the Second-Order Perturbation Theory Energy [11], SOPT energy, associated with the primary amide charge transfer for Li+ without F-, giving variation of up to 16 kcal/mol in the range of distances considered. The equation given in Figure 1 for second-order correction has the square of the Fock NBO off-diagonal matrix element in the numerator and the gap in the energy levels between acceptor and donor orbitals in the denominator. Ap1:Figure 27 to Ap1:Figure 32 show variation in energy level of N(lp), C-O(pi)*, the gap between them, and the square of Fock NBO matrix element with ions placed in the C-N line. Electrostatic variation of resonance-type charge transfer might be dissected as being primarily due to variation in the overlap of the NBOs which then causes variation in the energy levels of the NBOs, but change in the dipole of the resonance-type charge transfer is perhaps more compelling.

$$E_{i\to j}{}^{(2)} = -n_i{}^{(0)} \frac{\langle \phi_i{}^{(0)} | \hat{F} | \phi_{j*}{}^{(0)} \rangle^2}{\epsilon_{j*}{}^{(0)} - \epsilon_i{}^{(0)}}$$

Figure 1. NBO Donor-Acceptor Second-Order Correction (Eq 1.24 p19 [11]) for stabilizing interaction between filled donor orbital $\phi_i^{(0)}$ and an unfilled acceptor orbital $\phi_j^{*(0)}$ with energy levels $\epsilon_i^{(0)}$ and $\epsilon_j^{*(0)}$ and orbital occupancy $n_i$ and Fock operator $\hat{F}$

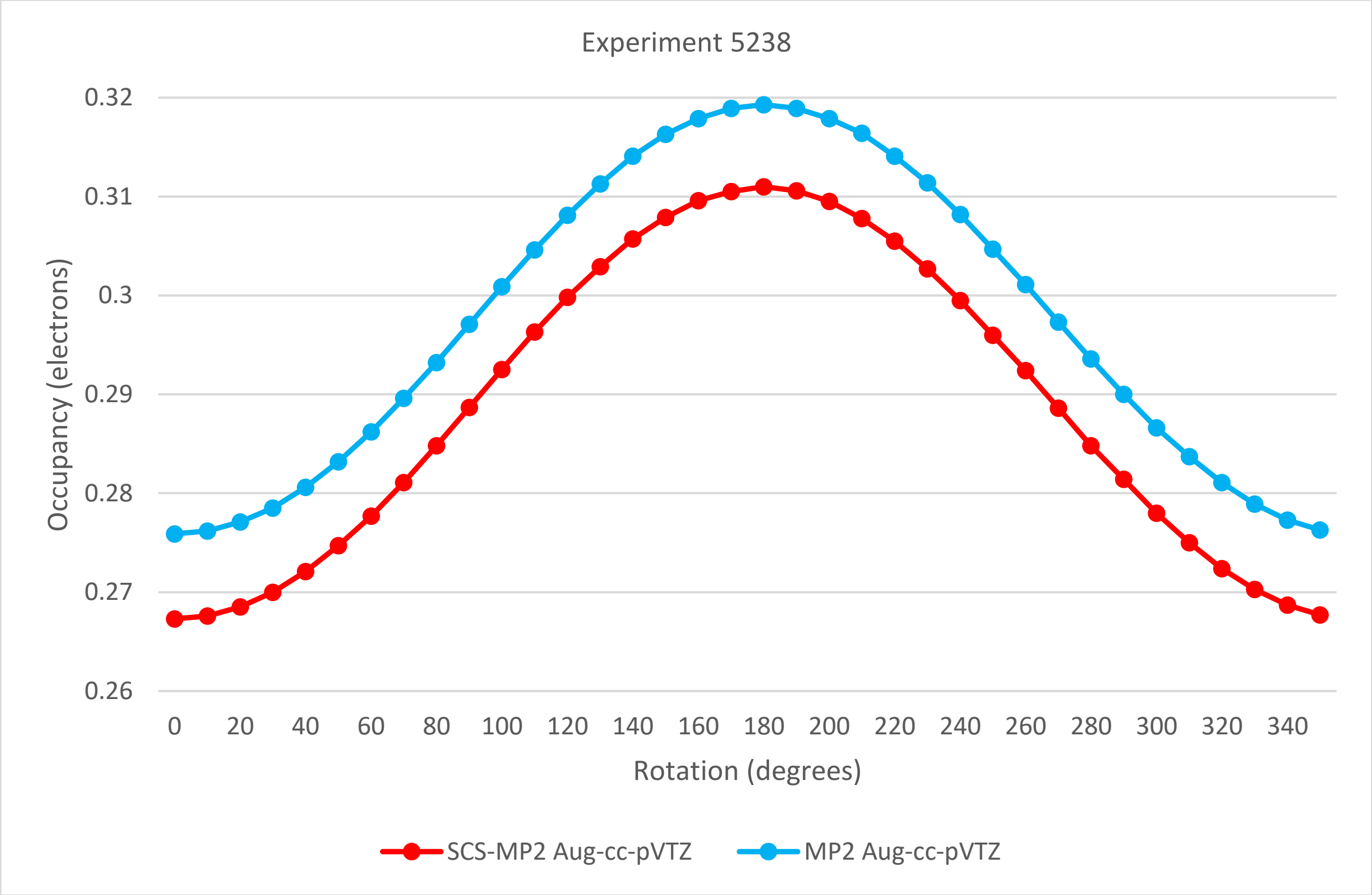

Figure 2. C-O(pi)* NBO Occupancy in N-methylformamide in 0.005 au Uniform Electrostatic Field Rotated in O-C-N Plane from N-C Vector

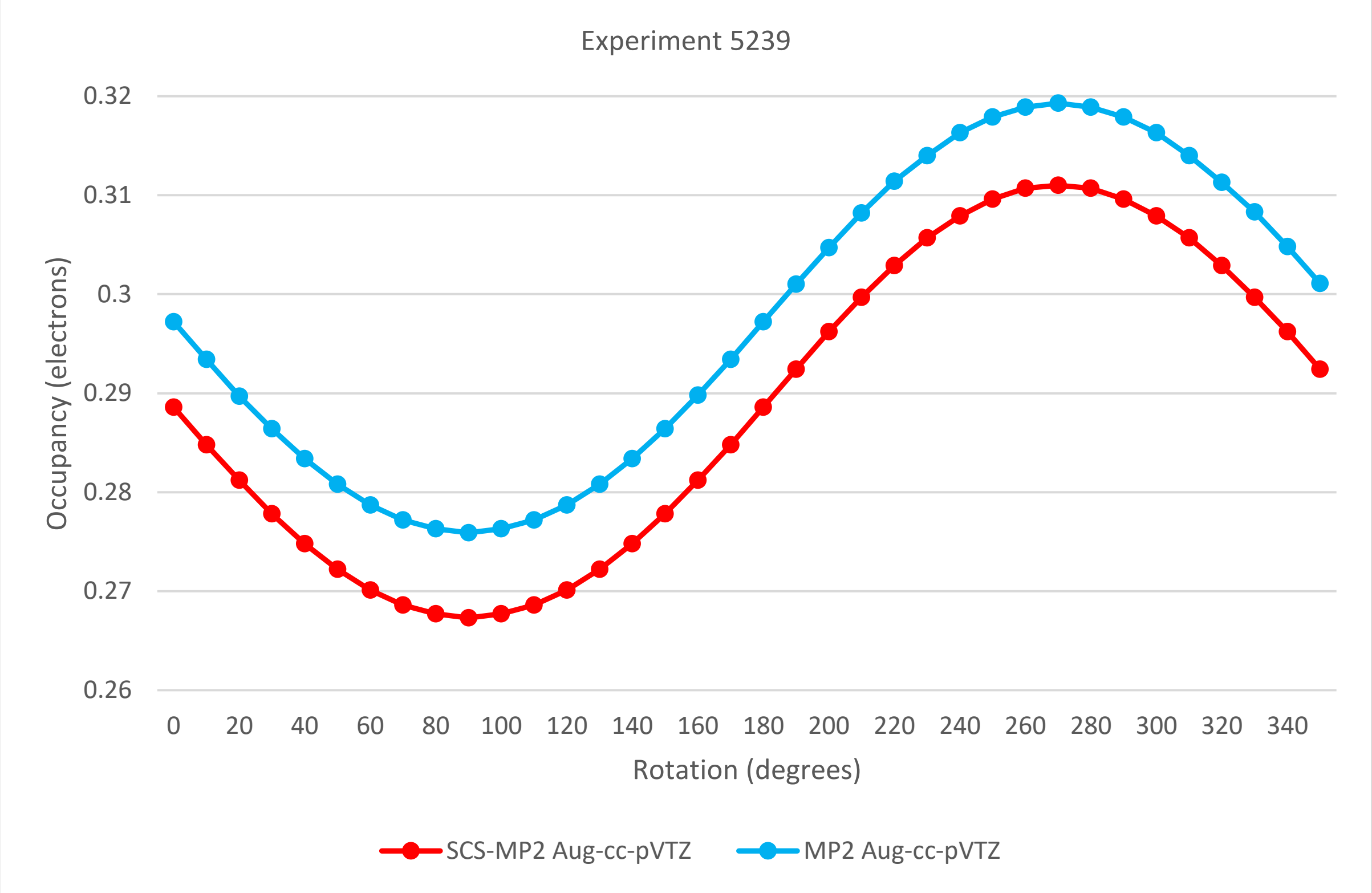

Figure 3. C-O(pi)* NBO Occupancy in N-methylformamide in 0.005 au Uniform Electrostatic Field Rotated in Plane Containing O-C-N Normal and N-C Vector Starting from O-C-N Normal

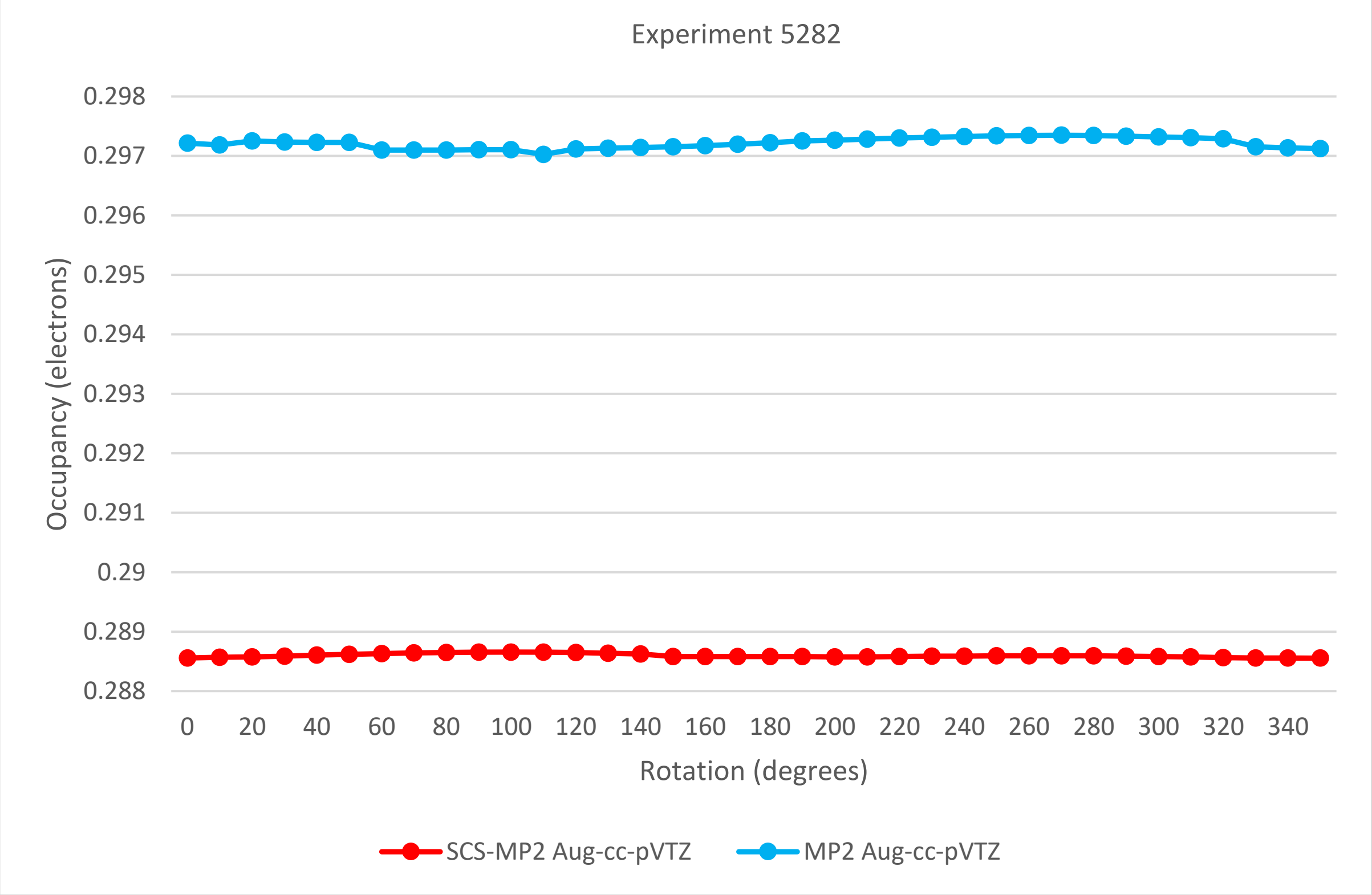

Figure 4. C-O(pi)* NBO Occupancy in N-methylformamide in 0.005 au Uniform Electrostatic Field Orthogonal to C-N Vector and Rotated from O-C-N Plane

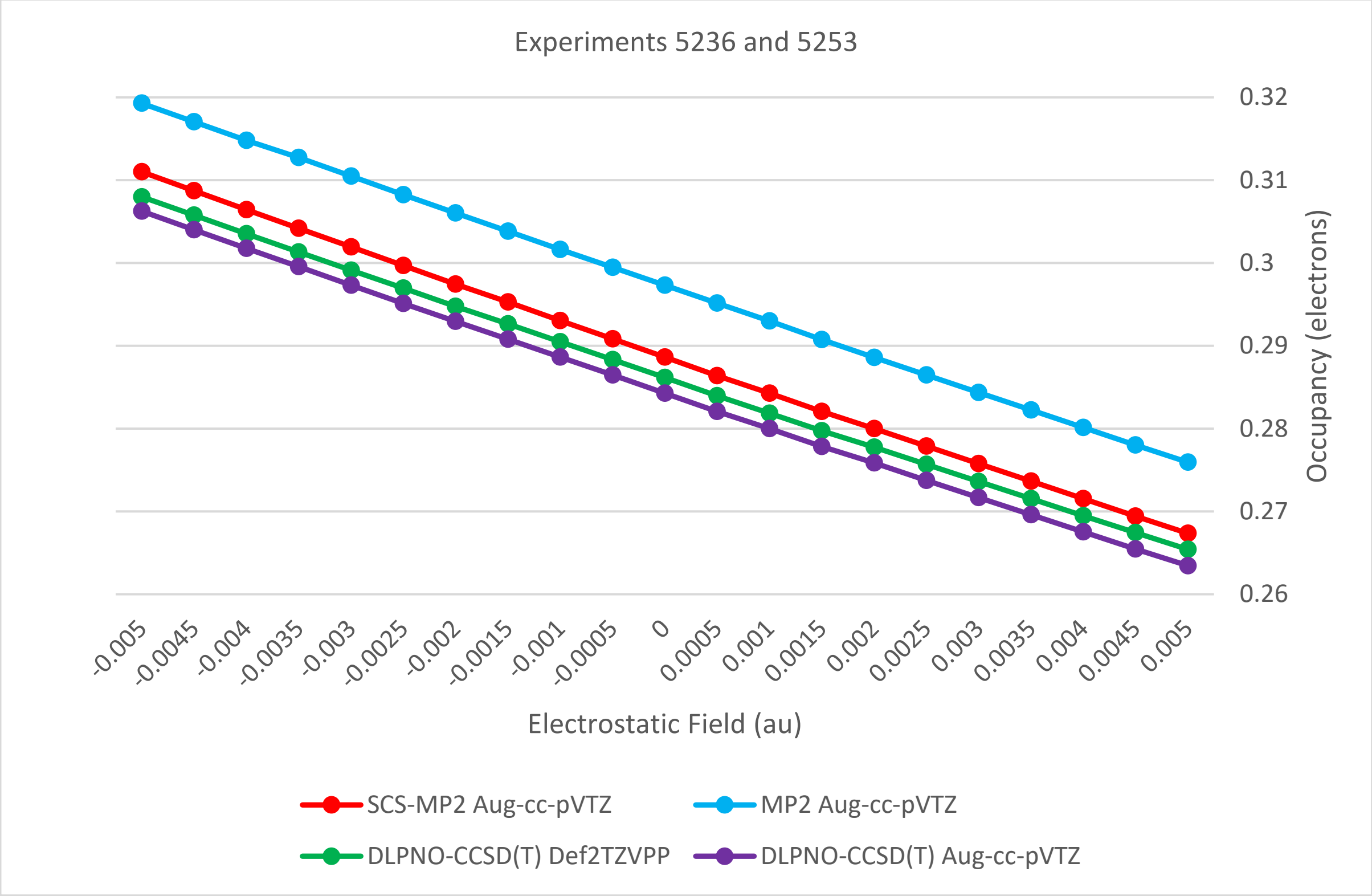

Figure 5. C-O(pi)* NBO Occupancy in N-methylformamide in Uniform Electrostatic Field with N-C Vector

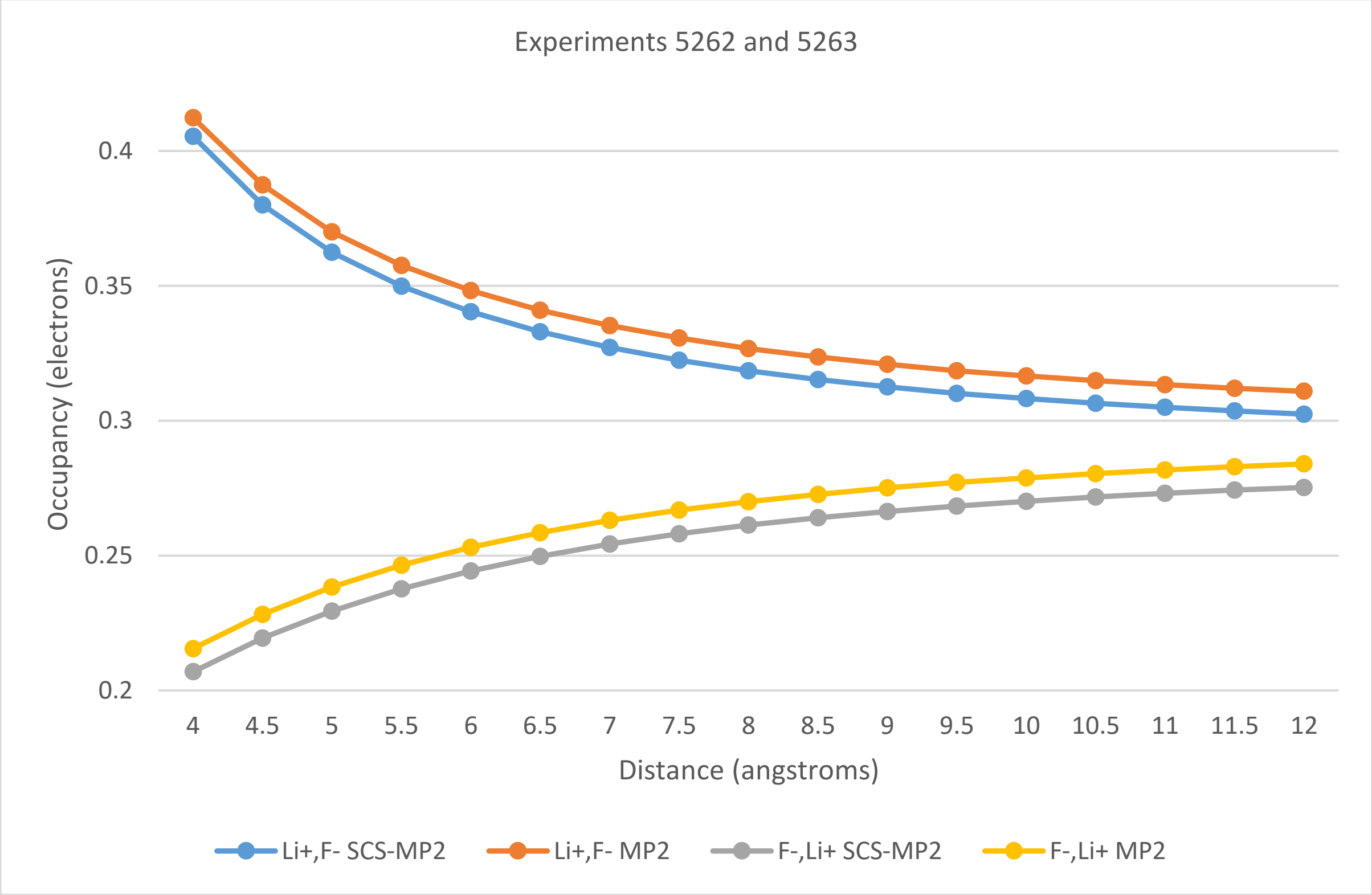

Figure 6. C-O(pi)* NBO Occupancy in N-methylformamide with Ions on C-N Line, right-listed Ion at Distance from N, left-listed Ion at Distance in Other Direction from C with aug-cc-pVTZ

## 7.2 Partial sp3 hybridization at N

Amide O-C-N normal electrostatic field can cause partial sp3 hybridization at N depending on amide resonance, but may also arise from mechanical torsion arising in the molecular context. However the sp3 hybridization arises, amide resonance is reduced since the bond order of C-N is reduced. Another view of this is that overlap of the N(lp) with C-O(pi)* becomes less favourable, since N(lp) takes on some s character in keeping with partial sp3 hybridization at N. Reduction in amide resonance enables sp3 hybridization at N, and increasing amide resonance inhibits sp3 hybridization at N [11]. In the absence of amide resonance, N would completely sp3 hybridize.

With sp3 hybridization at N introduced by rotating O-C-N-H and O-C-N-CA so that H and CA move away from the O-C-N plane, Ap1:Figure 33 shows that there is no change in sp3 hybridization when O-C-N normal electrostatic field of 0.005 au is applied, and Ap1:Figure 34 shows there is no change in primary amide charge transfer. Ap1:Figure 35 shows that with O-C-N-H and O-C-N-CA dihedral angles 10 degrees from planar an O-C-N normal field varying from -0.005 to 0.005 au results in no change in the primary amide charge transfer. It can be concluded that the change in amide resonance in O-C-N normal field is entirely due to the partial sp3 hybridization at N.

Ap1:Figure 36 shows the change O-C-N-H dihedral angle from planarity with unconstrained H, CA and HA as a 0.005 uniform field is rotated in the plane containing the O-C-N normal and C-N vector, and Ap1:Figure 37 shows the primary amide charge transfer. SCS-MP2 is somewhat smoother in calculating the O-C-N-H dihedral, though the primary amide transfer varies smoothly for MP2 as well. In can be concluded that the maximum and minimum primary amide charge transfers in these experiments occurs when the field passes through the O-C-N plane even when the O-C-N-H and O-C-N-CA dihedrals are unconstrained.

Figure 7 shows the primary amide charge transfer variation as a Li+ or F- ion moves in a line parallel to N-C but shifted 5 angstroms above the O-C-N plane. The ion moves from 0.5 angstroms before the intersection with the O-C-N normal from N to 2.0 angstroms after. The intersections of similar methods for different ions gives a position neutral to charge transfer variation. For both methods used, the neutral point is close to 0.7 angstroms from above the point above N, and at neutral point the occupancy values are very close to those for no field (Figure 5). At this neutral point, the C-N bond length is 1.350 angstroms, so the neutral point is a slightly closer C than N.

Ap1:Figure 38 shows that there is no difference in the primary amide charge transfer resulting from having Li+ and F- on the O-C-N normal from N equidistant from N but on different sides of O-C-N plane compared to constraining the O-C-N-H and O-C-N-CA dihedrals to the result of the presence of the ions but removing the ions. Ap1:Figure 39 shows that this is not the case when only one ion is used. It can be concluded that the two ions cause a field that passes through the O-C-N plane normal to that plane but the single ion does not and provides a component parallel to the N-C vector, thus decreasing the amide resonance further than that due to sp3 hybridization at N only. Ap1:Figure 40 shows the similar result with the ammonium ion used in place of Li+, where the distance shown is to the ammonium nitrogen. Charge transfer from amide N(lp) NBO to any ammonium N-H can be taken to be slight at 5.0 angstroms. Ap1:Figure 41 shows the difference in primary amide charge transfer between the ion present and ion removed but dihedrals retained cases for 4 methods and 2 basis sets.

Ap1:Figure 42 shows the variation of N(lp) NBO s character with variation in O-C-N normal uniform field rather than ions with unconstrained O-C-N-H and O-C-N-CA dihedrals, and figure Ap1:Figure 43 shows the difference in primary amide charge transfer.

As given by [11] and our data which is not shown, increase in amide resonance such as by RAHB inhibits sp3 hybridization at N. Once an RAHB chain has formed and resonance increased, sp3 hybridization at N will be prevented. However, if in the presence of an electrostatic field with component normal to the O-C-N plane the RAHB chain is weakened for example by thermal jolting and partial sp3 hybridization becomes possible at any point in the chain, significant loss of RAHB may result. Also, H-N, C-O or the whole amide group may rotate in the electrostatic field, weakening hydrogen bonds. A field normal to the amide plane can be expected to introduce an RAHB chain nucleation barrier.

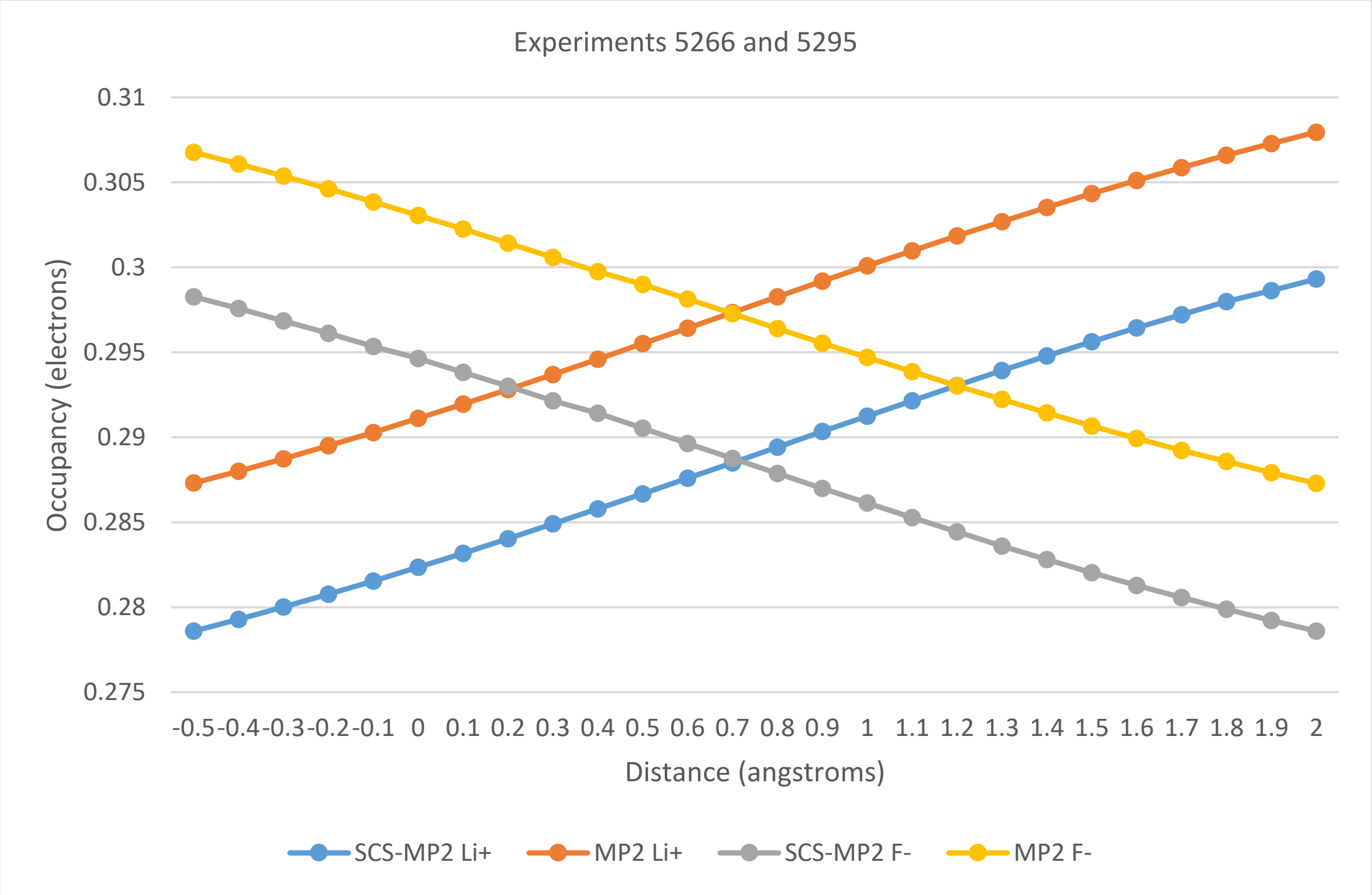

Figure 7. C-O(pi)* NBO Occupancy in N-methylformamide Constrained to Planarity with Li+ on Line Parallel to N-C Vector 5 Angstroms Above O-C-N Plane at Distance from Point Above N

## 7.3 Other geometry variations

Ap1:Figure 44 shows the primary amide charge transfer with rotation of the methyl group in N-methylformamide in orthogonal electrostatic fields. There is very little difference in the primary amide charge transfer between the cases of the field being 0, the field being along the y axis and the field being along the z axis. The case with the field along the x axis shows the expected variation for field aligned with the N-C vector. The N(lp) NBO's busy donation to the methyl group is not varied by the presence of electrostatic field of 0.005 au.

Ap1:Figure 45 shows the primary amide charge transfer with twisting about the C-N bond in N-methylformamide in orthogonal electrostatic fields. There is a slight separation of the curve for z-axis field from that of 0 and y-axis fields, but this is beneath the level of chemical significance.

Ap1:Figure 46 shows the primary amide charge transfer with rotation of the carbonyl substituent methyl in N-methylethanamide in orthogonal electrostatic fields. The separation of the curves for y-axis field from the 0 and z-axis fields is likely due to the optimal field vector being offset by 5 degrees from C-N in N-methylethanamide, with variation in primary amide charge transfer at twice the level of chemical significance. The slight separation of the 0 and z-axis field cases is likely due to variation in hyperconjugation between the methyl group and the carbonyl group.

Ap1:Figure 47 shows the primary amide charge transfer in N-methylethanamide with pyramidalization at the carbonyl carbon formed by rotating the CA-N-C-CA dihedral while maintaining the CA-N-C-O dihedral. The explanation offered is as for the case in the previous paragraph.

## 7.4 Torsional steering

A means by which change in resonance might change the torsional barrier across a single bond connecting to the atom which is central to resonance, which we refer to here as torsional steering, is discussed by [11] p693-702. An example of this is a methyl group bonded to the amide carbon, forming a bond similar to CA-C in proteins. A torsional barrier might be introduced through hyperconjugation between a CA-H bond or antibond and C-O antibonds or bonds. This is entirely plausible, and appreciable charge transfer CA-H->C-O(sigma)* is seen in Ap1:Figure 46, though the torsional steering proposal suggests that greatest stability is found with H-CA in cis with C-O rather than our finding that greatest stability occurs in trans. This difference in findings in undoubtedly due to our use of SCS-MP2/aug-cc-pVTZ rather than B3LYP/6-311++G**. Figure 8 shows the energy associated with N-methylethanamide with variation in O-C-CA-H dihedral from the trans to cis conformations. The trans conformation is ~0.3 kcal/mol lower in energy than cis. This is about an order of magnitude less than the torsional barrier of the torsional steering proposal, and slope has opposite sign. The main point of our experiment is to find what change to torsional barrier results from change to amide resonance. The different electrostatic fields cause markedly different amide resonance, and it can be seen that the curves for the fields do not differ other than being offset on the y-axis, that is, the torsional barrier is the same at each field and hence resonance. At the method and basis set used, these findings are not in accord with the torsional steering proposal.

The difference in molecular energy levels between the field strengths at a given dihedral angle may be taken as the upper bound of the variation in energy associated with the primary amide charge transfer, since purely electrostatic interactions may also be present. The change in energy associated with amide resonance between N-C vector field of -0.005 au and 0.005 au is ~7 kcal/mol per backbone amide. In terms of protein stability, this is considerable and occurs even without any hydrogen bonding to the amide O or H.

In reproducing these data, be aware that while the N-methylformamide experiments used dummy atoms to introduce constraint to planarity constraints and constrain the N-C bond to the x-axis, the N-methylethanamide experiments use two helium atoms at 8 angstroms from N, and the two helium atoms are on the y and z axes respectively.

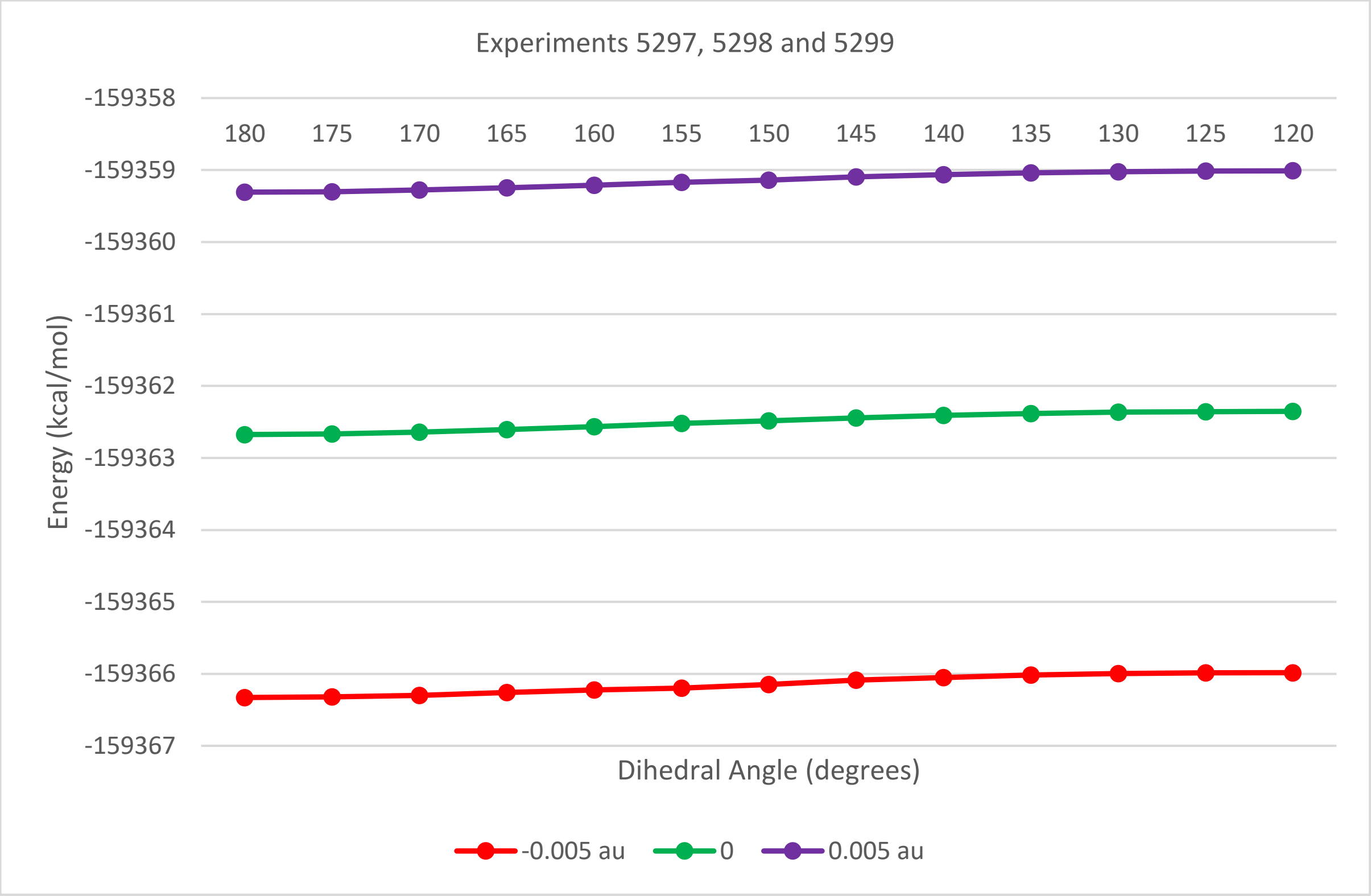

Figure 8. Energy Associated with N-methylethanamide at O-C-CA-H Dihedral Angle in Uniform Electrostatic Field with N-C Vector at SCS-MP2/aug-cc-pVTZ

## 7.5 Backbone amide nitrogen as hydrogen bond acceptor

If the backbone amide nitrogen participated directly in hydrogen bonds, amide resonance would be reduced, since charge transferred from N(lp) for the hydrogen bond would diminish the disposition of N(lp) to participate in the primary amide charge transfer, an instance of the busy donor effect [11]. This manner of variation of backbone amide resonance is proposed by Weinhold and Landis [11] as a means by which their resonance-based torsional effect comes into play during protein folding. In this section, we investigate the existence of this hydrogen bonding in the folded state on the possibility that phenomena occurring during folding might be seen in some aspect of folded state such as random coil.

In a collection of experiments, the central atom of a molecule of water, hydrogen sulfide, methane, ammonia and the ammonium ion was constrained to the O-C-N normal from N, but its distance from N was not constrained. The hydrogen bond length and C-O(pi)* occupancy following geometry optimization is shown in Ap2:Table 2. The hydrogen bond in the case of methane which represents a residue sidechain methyl group is of a length not conducive to charge transfer from N(lp) to methane, and the C-O(pi)* occupancy is largest. Modestly less C-O(pi)* occupancy is seen with ammonia which represents H-N bonds in uncharged groups. Less again C-O(pi)*occupancy is seen with hydrogen sulfide which represents a non-ionized cysteine head group. The C-O(pi)* occupancy with water is less again, and at 0.025 electrons less than the C-O(pi)* occupancy with methane this leads to the expectation that backbone amide N(lp) in RAHB secondary structures will be protected from access by water. Greatest reduction in C-O(pi)* occupancy occurs with the ammonium ion. As demonstrated above, a positively charged ion on the O-C-N normal from N creates an electrostatic field with C-N vector component at the midpoint of the C-N, reducing amide resonance directly and by pyramidalization at N. Also, positive charge in the hydrogen bond donor assists hydrogen bonding generally. We investigate hydrogen bonding with backbone amide N by positively charged residue sidegroups.

Spatial queries over an extract of the Protein Data Bank [57], PDB, to discover the incidence of charged arginine or lysine hydrogen bonding to backbone amide nitrogen were performed. X-ray crystallography [58] structures were not considered since coordinates for the amide proton were required. The extract was for solution Nuclear Magnetic Resonance [59], NMR, PDB files containing protein with no modified residues and no nucleic acid and at the date it was executed (2014-09-08) it returned 8,378 files, and only the first model of each was considered. These models were automatically checked for errors, such as non-contiguous residue sequence number, missing atoms and helices with C-terminal residue sequence number not larger than the N-terminal residue sequence number, eliminating 1,690 models and passing 6,688 models. The following constraints were imposed on identifying hydrogen bonds: N-H..amide N to less than 45 degrees, angle between amide normal and amide N to H vector not in the range 45 to 135 degrees and amide N..H distance to less than or equal 3.0 angstroms. In alternative search, no angular restrictions were imposed and the distance was restricted to 5 angstroms. The 3 polar hydrogens of charged lysine and the 5 of charged arginine were evaluated.

There is no instance of ammonium group of lysine or guanidinium group of arginine hydrogen bonded to backbone amide N according to these criteria in any of the 6,688 PDB NMR models queried. It is likely that more standard amide hydrogen bonding and increased resonance is always more favourable than N lone pair participation in hydrogen bonding which reduces resonance. The proximity of these groups to the backbone amide O-C-N normal from N might occur during protein folding but does not persist even in random coil.

## 7.6 RAHB chains in electrostatic field

We demonstrate the significant consequences of electrostatic field for RAHB in hydrogen bonded chains of amides. With co-planar N-methylformamide units hydrogen bonded C-O...H-N in the manner of RAHB protein secondary structures with the hydrogen bond angle set to 5.209 degrees so that all amide C-N vectors are parallel, the effect of electrostatic field on RAHB in multi-amide systems of molecules was investigated. In this multi-amide experiment, all atoms except 2 methyl hydrogens in each molecule constrained to the common planarity of all the molecules, and the amide N-C-O and C-N-H angles were constrained to that optimal for the molecules in isolation and the hydrogen bond angles constrained to the initial setting. No bond length was constrained, and the only non-dummy atom constrained to coordinates was that of the nitrogen of the first unit. This multiple N-methylformamide chain is shown in Figure 9.

Ap1:Figure 48 supports the use of RI-SCS-MP2/aug-cc-pVDZ with Coulomb and correlation auxiliary basis sets for experiments in which a uniform electrostatic field varies amide resonance. This pairing of method and basis set is used for its computational efficiency at larger atom counts.

Variation in C-O(pi)* NBO occupancy at each amide in this N-methylformamide chain geometry optimized for three different field directions, with constraints described above so that the chain remains linear are shown in Figure 10 to Ap1:Figure 55. The first field (Figure 10) has N-C vector, and the second (Ap1:Figure 49) is orthogonal to this. The second field has no N-C component, but both have component along the hydrogen bonds, so a third field (Ap1:Figure 55) is used which is orthogonal to the H-N vectors so that change is due to variation of amide resonance rather than hydrogen bond resonance. The data shown in Figure 10 and Ap1:Figure 55 is in accord with expectation, being that amide resonance peaks in the middle of the hydrogen bonded chain, and is varied as for the monomer case by electrostatic field. Also, RAHB in the chain is varied with hydrogen bond resonance alone (Ap1:Figure 49).

Hydrogen bond lengths for the three experiments described in the previous paragraph are shown in Figure 11, Ap1:Figure 50 and Ap1:Figure 56. It is apparent that in this unidirectional RAHB chain that the hydrogen bond length becomes less related to the C-O(pi)* occupancy along the chain and more related to O(lp) occupancy buildup which occurs because the final unit has no further acceptor to which to donate charge. It is clear that electrostatic field can eliminate the energetic penalty associated with this charge buildup, and this is in accord with a positive charge cap at the end of the chain being stabilizing [60]. A negative cap at the beginning of the chain would also assist in so much that it contributed to field throughout the chain. Of course, an extra charge donating unit, rather than only electrostatics, at the beginning of the chain would increase the resonance of the chain. Note that hydrogen bond length variation is somewhat muted because the inter-amide charge transfer and steric interactions give a net energetic result close to zero and the binding energy is primarily associated with the N(lp)->C-O(pi)* of the amides on each side of the hydrogen bond [61].

The remainder of the figures in this section relate to the third experiment in which the field was orthogonal to the H-N bonds to eliminate variation to hydrogen bond resonance. Ap1:Figure 51 shows the oxygen Natural Atomic charge along the chain, and Ap1:Figure 52 shows the amide proton NAO charge along the chain. The Natural Atomic charge of first amide proton in the chain is not varied by this electrostatic field, hence it is not varied by change in amide resonance.

Ap1:Figure 53 and Ap1:Figure 54 show the change in occupancy in O(sigma-lp) and O(pi-lp) NBOs in the H-N vector orthogonal field. The O(sigma-lp) NBO of the terminal units are not varied by this field, and in the non-terminal units there is slight variation. In the final unit, the occupancy is higher than in other units, in keeping with the unavailability of an acceptor. In contrast, the terminal O(pi-lp) NBO occupancy is markedly less than that of the other units which is in keeping with decreased C-O(pi)* NBO occupancy, but also declines at the terminal unit with increasing C-O(pi)* NBO occupancy due to field. This divergence of the occupancy of the 2 oxygen lone pairs could be explored further, but we do not do so here. Confirmation that electrostatic field orthogonal to the H-N vector does not differentially vary the O(sigma-lp)->H-N* and O(pi-lp)->H-N* resonance-type charge transfers might also be sought.

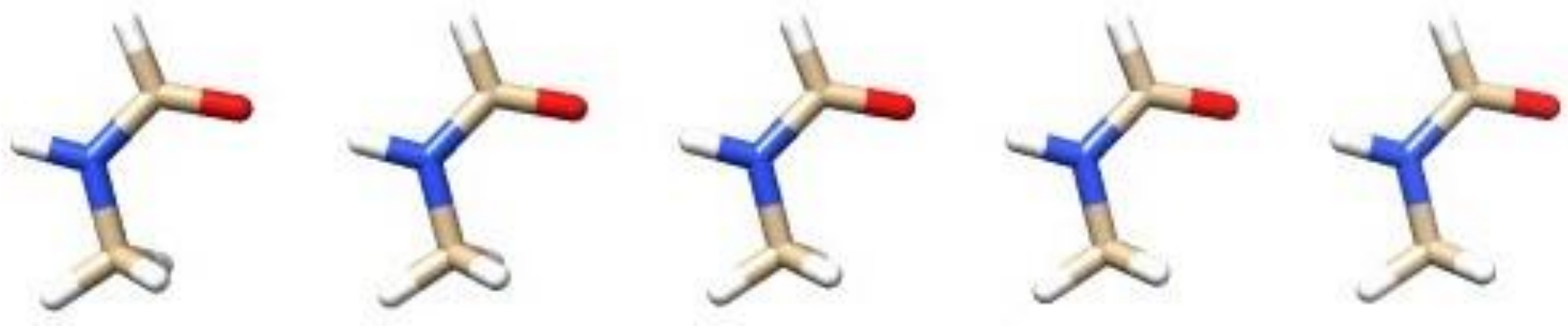

Figure 9. N-methylformamide Hydrogen Bonded Chain

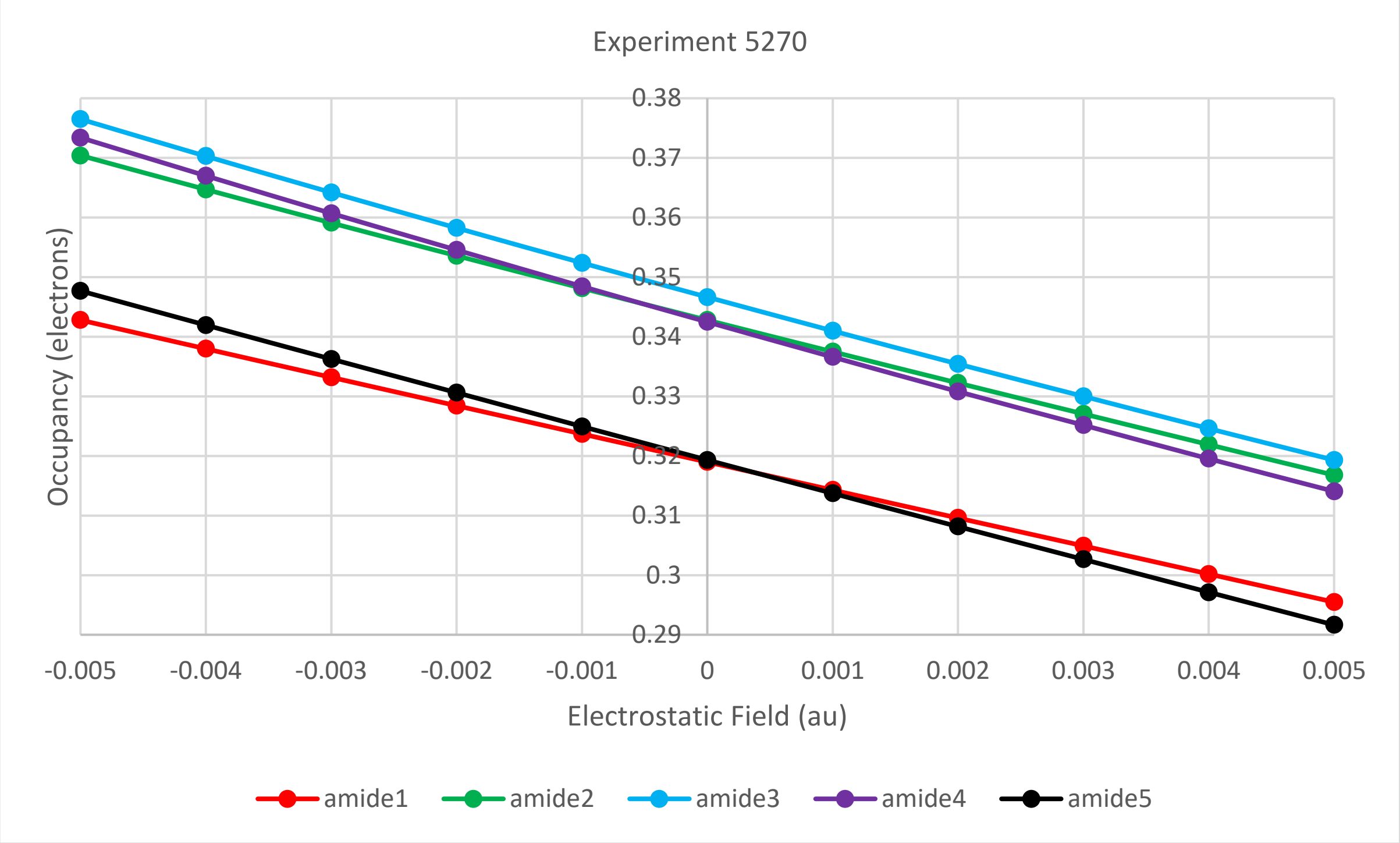

Figure 10. C-O(pi)* NBO Occupancy in Hydrogen Bonded N-methylformamides with Common N-C Vectors and O-C-N-H Planes in Uniform Electrostatic Field with N-C Vector at RI-SCS-MP2/aug-cc-pVDZ with Coulomb and Correlation Auxiliary Basis Sets

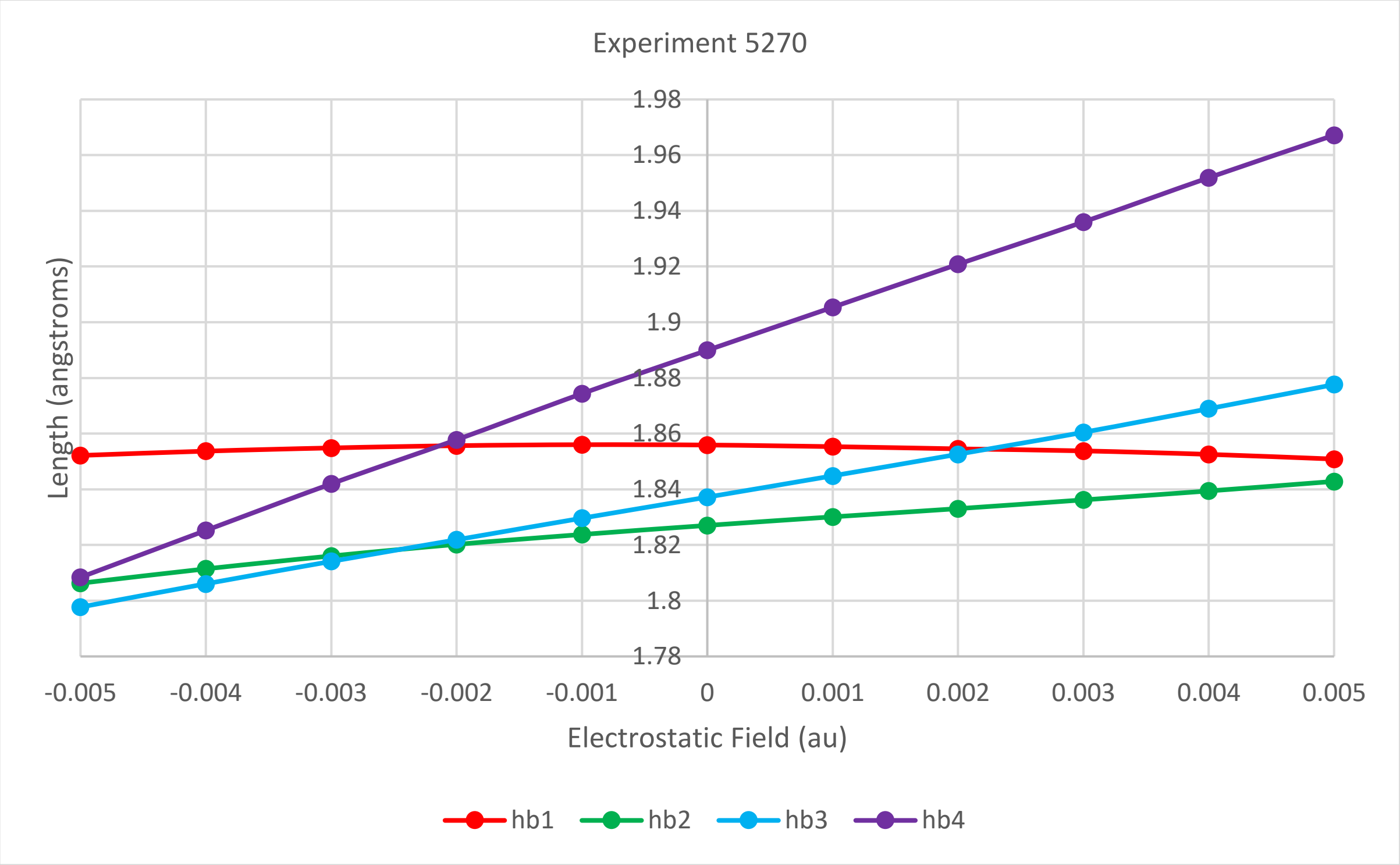

Figure 11. Hydrogen Bond length in Chain of N-methylformamides with Common N-C Vectors and O-C-N-H Planes in Uniform Electrostatic with N-C Vector at RI-SCS-MP2/aug-cc-pVDZ and Coulomb and Correlation Auxiliary Basis Sets

## 7.7 Electrostatic field vectors at backbone amides

The foregoing demonstrates or otherwise implies the relevance of electrostatic field to RAHB and hence to RAHB protein secondary structures. This leads to an interest in observation of the electrostatic field vector at backbone amides in proteins, and that the backbone amide is not protected from variation in electrostatic field with C-N vector component.

Emphasizing that Newtonian or classical simulators have an uncertain account of resonance, but wishing to obtain some estimate of electrostatic field vectors at backbone amides for proteins, we used the Tinker 7.1.2's [62] AMOEBA Protein 2013 [63] force field and Generalized Kirkwood continuous solvent [64] to estimate the electrostatic field vectors at backbone amide nitrogen (Figure 12) and carbon (Figure 57) in accordance with the calculation described in [65] for 10 minimized protein structures having initial coordinates derived from the PDB, entries 1D27, 1H1J, 1IMP, 1UUA, 2JOF, 2LHD, 2LJI, 2LT8, 2LX9 and 2LXR.

The variation in C-N component of electrostatic field is larger than the general intra-protein fields observed by [26, 46-50]. These calculations offer no support for a notion that the C-N component of electrostatic field is constant or protected at protein backbone amides. An appreciably smaller field than calculated by AMOEBA with Generalized Kirkwood continuous solvent can be expected to result in significant influence on protein structure.

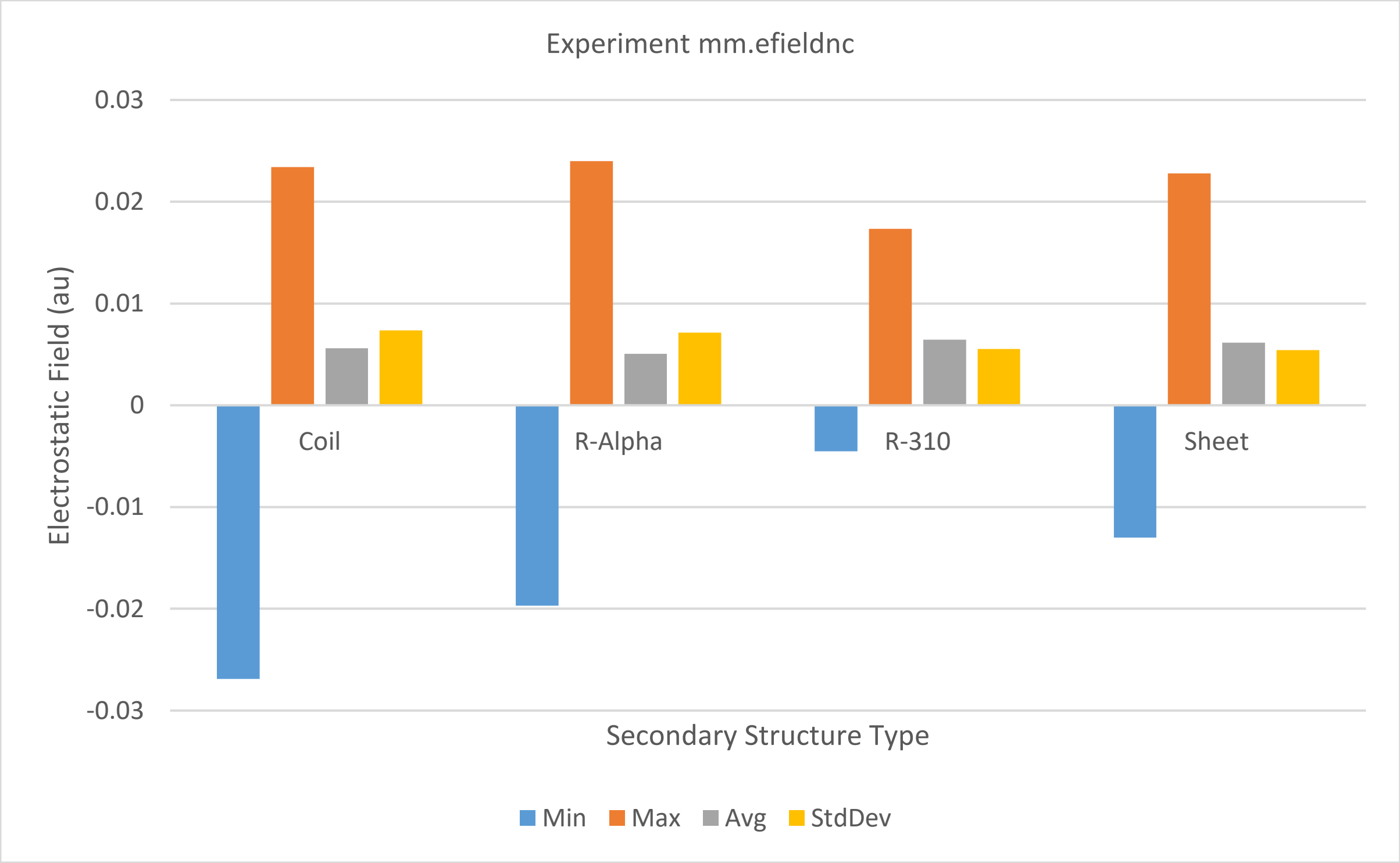

Figure 12. C-N Vector Component of Electrostatic Field at N in Backbone Amides of 10 Small Proteins as Calculated with AMOEBAPRO-2013 with Generalized Kirkwood Continuous Solvent

## 7.8 Protein beta sheet

The multi-amide structure described above (Figure 9) differs from a chain of backbone amides in a parallel beta sheet only in that the units of the multi-amide structure are N-methylformamide rather than backbone amides. The C-N vectors within a RAHB chain of backbone amides of a parallel beta sheet are all aligned. As for the multi-amide experiments above, the resonance of every backbone amide in a chain in a parallel beta sheet will be varied in a similar manner by a uniform electrostatic field, the varied resonances cooperatively determining the extent of hydrogen bonding in the chain. The results described for the multi-amide structure may be taken as indicative of the behaviour of a backbone amide chain in a parallel beta sheet in an electrostatic field.

The backbone amide C-N vectors of a parallel beta sheet (Figure 13) HB chains an even number of chains away from a nominated chain are all aligned. The backbone amide C-N vectors of chains an odd number of chains away from a nominated chain are also all aligned, but these vectors have a component antiparallel to those of those an even number of chains away. A uniform electrostatic field may be applied that increases RAHB in one set of chains, even or odd, but to different extent reduces it in the other set. This leads to a hypothesis given in Section 8.1 (Hypotheses/Beta sheet).

An antiparallel beta sheet differs from a parallel beta sheet with respect to patterns of backbone amide C-N vectors (Figure 14). In a single antiparallel beta sheet RAHB chain, the C-N vectors are alternately substantially orthogonal. A uniform electrostatic field may be applied such that the resonance of alternate amides is increased or decreased with little to no direct change to resonance of the other amides in the chain. Chains in the sheet an even number of chains away from a nominated chain have a similar pattern of amide C-N vectors, so the RAHB of these two chains may be similarly modulated by a field. C-N vectors of amides on chains adjacent a nominated chain but on the same beta strand are largely orthogonal but on adjacent beta strands they are largely opposed, so RAHB in the chains an even number of chains away from the nominated chain may be varied in opposition to chains an odd number of chains away.

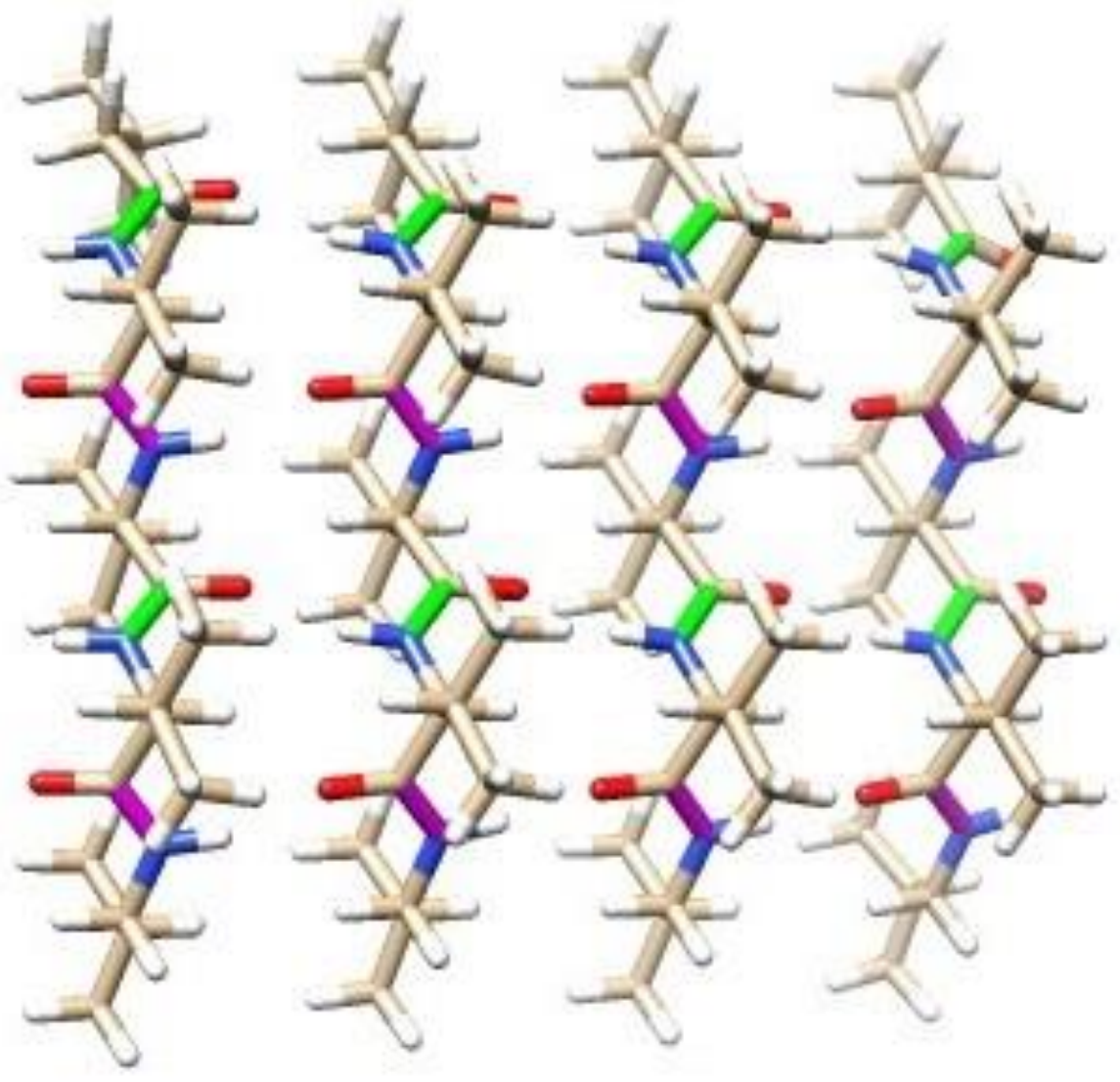

Figure 13. 2 sets (magenta, chartreuse) of parallel C-N Bonds in Parallel Beta Sheet

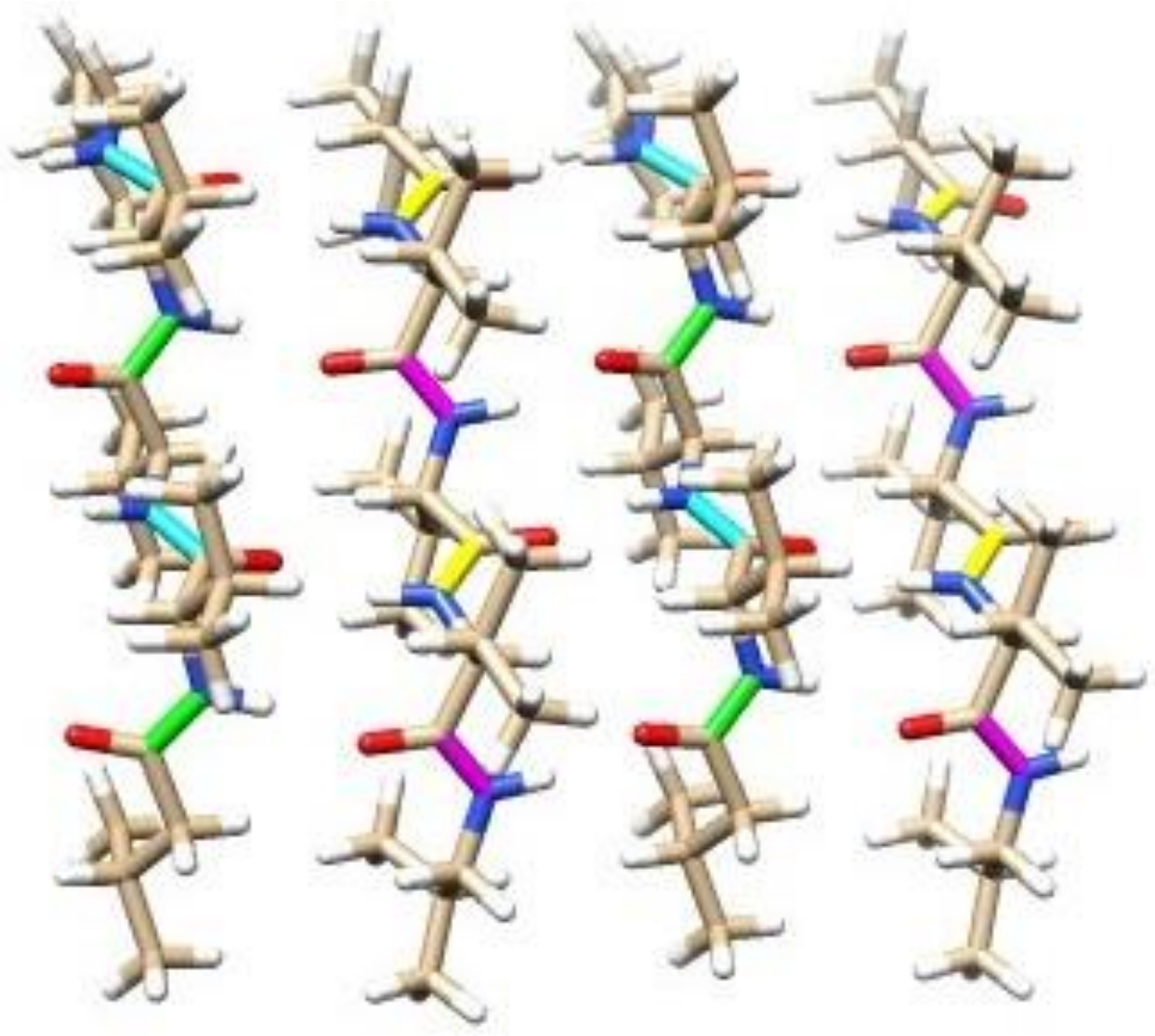

Figure 14. 4 sets (yellow, magenta, light blue, bright green) of parallel C-N Bonds in Antiparallel Beta Sheet

## 7.9 RAHB protein helices

In a RAHB protein helix, the helix macro dipole is collinear with the helix axis which is not altogether orthogonal to the backbone amide C-N vectors. The electrostatic field associated with the helix macro dipole [66] will reduce the resonance of the backbone amides and hence cooperatively reduce RAHB in the RAHB chains or spines as they are sometimes referred to in the context of helices. Charge caps, negative charge at the N-terminus of the helix and positive charge at the C-terminus, can compensate for the field associated with the helix macro dipole and hence assist RAHB. Where the charge caps are also involved in charge transfer, they tend to balance the RAHB terminal charge deficits which arise in unidirectional RAHB because charge transfer into and out from these sites does not cancel [11].

The average angle between the helix axis and the C-N vectors differs between the helix types 3-10, alpha and pi and the field associated with the helix macro dipole also differs between these types because of a difference in divergence of average C-O and H-N vectors from the helix axis and differences in hydrogen bonding arising from these geometry differences, and the variation in backbone amide resonance caused by the helix's intrinsic electrostatic field differs between the helix types.

RAHB helices may be less susceptible to destabilization by uniform electrostatic field not collinear with a helix axis than a beta sheet is to field with any direction due to the lower proportion of C-N vectors in a helix backbone chain being aligned. In an alpha helix for example, the backbone amide C-N vectors are approximately aligned only when the helix turn of ~3.6 backbone amides approximates an integer. Uniform electrostatic field not collinear with the helix axis may be more destabilizing for some helix types than others due to their different turns. The different RAHB protein helix types, 3-10, alpha and pi have different numbers of spines. The 3-10 helix has only 2 spines, which may make it more susceptible to RAHB irregularity caused by electrostatic field.

Charge positioned away from the helix axis, such as on an amino acid residue sidechain, varies each backbone amide resonance depending on the vector of the non-uniform electrostatic field at the amide. Reduction of RAHB in a spine requires reduction of resonance in only one amide, and an equal enhancement of resonance in another amide of the spine does not necessarily compensate, depending on the position of the two amides in the spine. A decrease in resonance at one unit can be expected to make RAHB behave more as two separate RAHB chains either side of that unit.

The electrostatic field created by charged sidechains may be modulated by being surrounded by high permittivity medium such as occurs by immersion in water, and by salt-bridging. An absence of these modulations of electrostatic field caused by charged sidechains invokes the energetic cost of reducing helix RAHB.

## 7.10 Proline

Where the C-terminal-side residue in a peptide bond is the imino acid proline, a N(lp)->C-O(p)* interaction still exists. A field parallel to the C-N bond will tend to reduce variation in peptide bond twist away from that optimal for that donor-acceptor interaction and increase the energetic barrier of a cis-trans or trans-cis transition, and a field antiparallel to the C-N bond will tend to increase variation in peptide bond twist and decrease energetic barrier of a cis-trans or trans-cis transition, which is certainly relevant to protein folding and may be exploited by prolyl isomerases [67].

## 7.11 Permittivity

The hydrophobic interior of a protein has a low permittivity and well supports an electrostatic field. The whole of the sidechain of hydrophobic residues has a low permittivity, and the hydrophobic stalk of every residue type other than glycine and proline has the low permittivity of the alkanes. This is much smaller than the relative permittivity of water which at 80 degrees F is ~80 times that of vacuum [68]. Formamide, which is the functional group of sidechain amides, has a yet higher permittivity and so is more capable of damping a field than water.

Where protein-protein binding interfaces are largely hydrophobic, the hydrophobic interiors of the proteins may be effectively joined, with the electrostatic field of one interior extending into the other.

A charged residue without neutralizing salt-bridge partner in an otherwise hydrophobic protein, or similarly unpaired charged residue at a hydrophobic binding interface that excluded solvent on binding to partner protein would exert long-range electrostatic effects. A change to internal protein electrostatic environment may cause change to secondary structure and conformational change.

Conformational change due to unpaired charged lysine was explored in [69].

## 7.12 Protein folding, conformational and allosteric change

The change in protein structure due to change in electrostatic field may be quite extensive and long-ranged, for a change in field may change secondary structure, which may in turn change the position of functional groups of sidechains which then changes the field, changing yet more secondary structure and so on. Electrostatic field variation of backbone amide resonance can be expected to be important in conformational change in proteins. Similar remarks apply to allosteric change [70].

The electric flux local to amide group will vary during protein folding, with the amide groups preferring to orient to the local flux so as to maximize amide resonance and the energy associated with resonance charge transfer and hydrogen bonding. Variation in the total energy associated with backbone amide resonance during folding may prove a revealing metric, as might the sum for all backbone amides of the C-N component of electrostatic field.

In summary, it is being proposed that the amide local electrostatic environment which is substantially determined by residue sidegroup charge, polarity and permittivity is an important factor in protein folding, structure and function. How important it is relative to other factors in protein folding is yet to be determined, but there is an elegance to this mechanism, offering explanation as it does of the fundamental features of nature’s primary structural polymer.

## 7.13 Unstructured regions

Electrostatic field in a region may prevent formation of secondary structure in a length of backbone due to diminishment resonance of backbone amides in secondary structure conformations, and less structured conformations allow larger total resonance.

## 7.14 Sources of electrostatic field variation in proteins

There are many sources of variation of electrostatic field in proteins. The functional groups of amino acid residue sidechains may be charged or polar. Post-translational modification may change charge or polarity. Phosphorylation and de-phosphorylation figure frequently in signal transduction [71]. Forming or breaking of salt bridges, such as between lysine and glutamine, changes the field. Solvation or de-solvation of charged or polar functional groups of sidechains changes electrostatic field. Binding by desolvated ions, such by the divalent calcium cation, $Ca^{2+}$, to Calmodulin [72] brings large changes to a field. Also, a single solvation shell is unlikely to fully dampen the field created by a divalent cation.

## 7.15 Absence of charged residues

In experiments by Kurnik *et al*. detailed in [73], S6 was deprived of all charged sidechains and neutrally C-terminally capped, and was found to fold normally. The absence of formal charge except N-terminally does not imply there is no electrostatic field environmental to backbone amides, for dipolar residues are present and ions from solution may associate with the protein. There is no threshold for the effect we describe and change of resonance in one backbone amide in an RAHB chain changes resonance and hydrogen bonding throughout the chain.

Charged residues which are salt-bridged or solvated may have modest bearing on electrostatic field at backbone amides, and their removal may make no difference to fold. However, this finding warrants including S6 in any physical survey of the C-N vector electrostatic field variation at C, N and the midpoint of C and N. If a protein's internal electrostatic field at each peptide group is negligible, folding would be possible due to intrinsic amino acid preferences with other causes such as steric interaction with the backbone [74] and steric blocking of backbone hydration [75] and hydrophobic [76] patterning of the amino acid sequence.

## 7.16 Development of methods

Successes in understanding protein folding can be had by modelling pure electrostatics and hydrophobia alone [77, 78]. Classical calculation is oblivious to peptide resonance, EVPR-CN and RAHB, and these omissions will limit what accuracy can be had.

Programs entered into the Critical Assessment of protein Structure Prediction [79] contests might be improved as a result of considering how well these programs capture EVPR-CN and RAHB. Similar remarks apply to Computational Protein Design [80], including de novo protein design.

We await the availability of quantum methods suitable for the quantification of EVPR-CN and its consequences in whole explicitly solvated proteins and protein complexes, noting the large errors of established DFT methods in calculating resonance in beta sheets [81].

On the advent of accurate wavefunction methods with analytic gradients that scale to RAHB protein secondary structures and preferably whole proteins with medium and large non-Pople basis sets, observations may be made of the response of RAHB protein secondary structures and amyloid fibril to an electrostatic field. DLPNO-MP2 [82] and the most recent DLPNO-CCSD(T) implementation [83] do not have analytic gradients and need further constant-factor reduction in runtime but are promising. Improvement in non-wavefunction methods due to Density Perturbation Theory [84] seems promising. Accurate, linear scaling and low prefactor methods are highly desirable so that extensive explicit solvent may be used to immerse protein complexes for molecular dynamic calculations. The need for benchmark of the accuracy of all methods applied to proteins in calculating variation of backbone amide resonance is indicated. For this purpose and otherwise, development of physical methods suitable for accurate observation of backbone amide resonance and of the electrostatic field at the mid-point of the backbone amide C-N bond is highly desirable.

# 8 Hypotheses

## 8.1 Beta sheet

We predict that it is the similarity of RAHB of the backbone amide chains of a beta sheet which primarily determines the stability of that sheet and hence what residue sequences form a beta sheet. A variation in the resonance of amides of a backbone strand will cause irregularity in structure and variation in hydrogen bonding between a pair of strands. A hydrogen bond of greater binding energy than immediately adjacent hydrogen bonds in binding two backbone strands will tend to determine local geometry in its favour rather than in that of its adjacent hydrogen bonds.

In an antiparallel beta sheet, there is steric clash between the HA atoms of adjacent strands which is partially resolved by twisting of the strands and offsetting the strands in favour of the HA hydrogen bonding with O on the adjacent strand in competition with the H. If one backbone hydrogen bond has more favourable geometry than those immediately adjacent between the same pair of strands, the cooperative RAHB in its chain is advantaged leading to further geometry improvement in the favour of that chain. The adjacent chains are disadvantaged since their geometry is necessarily de-optimized. With sufficient RAHB mismatch, the beta sheet is unstable.

In a parallel beta sheet, the steric clash is not between HA atoms and is confined to that between HA and the H and O of the adjacent strand. There is an additional consideration being that even in the absence of steric considerations, only every second hydrogen bond between adjacent strands can be optimized due to the alternating distance between amide oxygen and subsequent amide proton of one strand facing the same neighbouring strand.

Greater balance in the hydrogen bonding between two strands will give more regular structure. The hydrogen bonded chains may traverse more than two strands with the hydrogen bonds between each pair of strands balanced, but balance between different pairs of strands is not required.

The RAHB chains of a possible beta sheet might be seen as independently pursuing RAHB maximization where geometry optimization cannot be had by all the chains, and should there be difference in the field at the backbone amides of the chains, the RAHB in some chains will exceed that of others. The RAHB chains are in competition for optimal geometry.

The hypothesis of the present section is irrespective of the means by which amide resonance and RAHB is varied. Amino acid residue sidechains vary the field and permittivity to backbone amides, and in this manner sidechains influence what sequences form beta sheets. A non-uniform electrostatic field could be crafted to particularly challenge or encourage the structural integrity of a beta sheet. Non-electrostatic influences on the formation of beta sheets are known, such as steric conflict with the backbone in other conformations [74] and blocking of backbone hydration [75].

In previous work [81], we concluded that established non-double hybrid DFT methods significantly underestimated resonance in amides when the carbonyl bonds engage in hyperconjugative interactions such as occur in beta sheets and geometry optimization using these DFT methods should be taken as significantly underestimating RAHB.

## 8.2 Amyloid fibril

Amyloidogenicity is determined by the balance in hydrogen bonding between two beta strands as in a beta sheet, but requires a yet finer degree of balance. The electrostatic environment provided by a strand's sidechains is repeated for each strand since the strands have the same residue sequence, so any variation in backbone amide resonance is cooperatively amplified by RAHB. A single backbone amide resonance variation in a beta sheet RAHB chain leads to changes in resonance throughout the chain, and the same inducement to variation at every backbone amide resonance as in the case of amyloid fibrils leads to a more considerable variation in RAHB throughout the chain. Also, the similarity of residue sequence of each strand does not permit any partial normalization of RAHB by the sidechains of adjacent strands.

Layering of the beta sheets of an amyloid filament so as to provide an even electrostatic environment for RAHB in the chains in each sheet will be important in finely balancing the RAHB. There is a tendency for amyloidogenic sequences to be hydrophobic which create little field but are more permissive of electrostatic field than non-hydrophobic residues and so they moderate electrostatic field mostly by increasing the distance between the backbone amides and causes of field.

Also, the chains run the length of the fibril and may exhibit significant RAHB. Elevated RAHB may increase the absolute differences between the RAHB of the chains.

It is noted that there are others factors which may contribute to stability of amyloid fibrils. Hydrogen bonding and dispersion between sheets may exclude water [85]. Where water is excluded, there are prospects for ions also being excluded.

## 8.3 Polyproline helices types I and II

Though there is no peptide group to peptide group hydrogen bonding in polyproline helices, the variation to the primary peptide resonance-type charge transfer in an electrostatic field will be stabilizing or destabilizing, since there is an energy change associated with the change in the primary peptide charge transfer itself. This energy change has almost always been neglected in considering the energetics of protein structure, and hydrogen bonding has been the focus instead. An electrostatic field parallel or antiparallel to the helix axis has a component in common with all C-N vectors in both type I and II polyproline helices, PPI and PPII, and is thus uniformly stabilizing or destabilizing at each peptide bond.

In PPII, the peptide bond dipole is substantially orthogonal to the helix axis. The C-O vector is very close to 90 degrees from the helix axis, with a variation of 120 degrees about the helix axis between subsequent peptide bond dipoles. This means that an electrostatic field parallel or antiparallel to the helix axis is substantially energetically neutral for the carbonyl bond dipoles, unlike alpha helices. The C-N bonds are ~50 degrees from the axis, the cosine being ~0.64 giving substantial component with C-N vector. The variation in peptide bond resonance has an intrinsic associated energy and also varies the propensity of the peptide group to participate in hydrogen bonding with residue side groups or water. Where the helix is solvated, an electrostatic field attenuates quickly with distance and such a helix will be short.

In the much less biologically significant, proline-requiring, right-handed PPI, there is an alignment of carbonyl bonds which resembles alpha helices, and the carbonyl bond dipoles and peptide resonance are both favoured or disfavoured by components of a helix-axis field.

Which type of polyproline helix is preferred may vary with the field strength, since the relation of C-N vectors to the helix axis and psi torsions differ between the two polyproline helix types.

The amide oxygen lone pairs are inequivalent [61]. In polyproline helices the amide oxygen p-type lone pair is available for hydrogen bonding rather than being protected as in beta sheets. Amide resonance and the binding energy associated with N(lp)->C-O(pi)* is maximised by C-O..H-N hydrogen bonding at C-O..N geometry 75 degrees from linear in the amide plane due to increased charge transfer from a lobe of the p-type lone pair [61]. Both lobes of the amide oxygen p-type lone pair are available for hydrogen bonding, and both lobes participating in hydrogen bonds will result in notably greater amide resonance than only one lobe. With an electrostatic field component aligned with the helix axis, a hydrogen bond involving one lobe will be favoured and the other disfavoured, since the charge transfer for the hydrogen bonds is varied by electrostatic field.

In summary, we predict that the structure of polyproline helices, lacking backbone RAHB chains, is determined by helix-axis electrostatic field causing increase of energy variation directly associated with the increase in primary peptide charge transfer and by increased RAHB with water or residue sidegroups. Also, greater access to the backbone amide oxygen p-type lone pair for HB by water allows increased amide resonance and hence binding energy associated with N(lp)->C-O(pi)*.

## 8.4 Protein folding

### 8.4.1 Overview

In a variation and extension of the backbone-based theory of protein folding of Rose *et al*. [5], it is proposed that the universal folding mechanism is peptide group resonance, PR, rather than inter-peptide group hydrogen bonding, IPHB. PR is integral to the backbone whereas IPHB is a backbone-backbone interaction, and so this proposal directly associates the universal folding mechanism with the backbone and widens the scope of the backbone-based folding mechanism to include peptide groups not participating in IPHB. The energy internal to a peptide group associated with its PR is variable with PR and is posited to drive protein folding. The causes of variation in PR are not limited to RAHB and EVPR-CN, and include C-N torsion, pyramidalization at N, the busy donor effect at the N lone pair and interactions with the carbonyl orbitals other than N(lp)->C-O(p)*. Since the binding energy of IPHB primarily resides in PR [61], the energetics of PR subsume those of IPTB. PR is primary in determining the HB binding energy of a peptide group to other peptide groups, water or sidechains.

Each peptide group seeks to increase its PR. As peptide groups are distributed along a polypeptide chain and PR influences RAHB and orientation to electrostatic field, PR is a driver of backbone conformational change, in competition with hydrophobic interactions [76]. The difference in energy associated with variation in resonance of a single peptide group under biologically plausible conditions (the present work and [61]) is comparable to the energy of stabilization of proteins in native conformation (Section 8.4.2).

The amino acid sequence specifies fold, and the sequence is evaluated by the universal folding mechanism that is PR, whereas the universal mechanism of the Rose *et al*. proposal [5] does not directly heed sequence. In keeping with their proposal, there is no encoding of how to fold in amino acid sequences. Separation of the specification of fold from the procedure for how to fold is expected to confer great evolutionary advantage.

We refer to the present theory as the Resonance Theory of Protein Folding, RTPF. It includes variation of PR due to any cause, including EVPR-CN, RAHB, C-N torsion, pyramidalization at N, the busy donor effect at the N lone pair and interactions with the carbonyl orbitals other than N(lp)->C-O(p)*.

### 8.4.2 Peptide group resonance drives protein folding

In the present proposal, the amino acid sequence is evaluated by each peptide group in terms of its PR. This evaluation is not restricted to amino acids local in the sequence. Each peptide group seeks to increase its PR, though increase in one PR may come at the expense of another PR. This search for increased PR may appear to be cooperative or competitive depending on context, but can be understood as independent search by a peptide group of its changing environment. Conformational change due to hydrophobic interactions [76] or sidechain-sidechain interactions may also prevent any given PR or total PR from being monotonically increasing during folding. Patterning of residues according to hydrophobicity and sidechain-sidechain interactions can change the conformational energy landscape and introduce local free energy minima, thus frustrating folding [86, 87]. Folding frustration would tend to be minimized by evolution so that the driving mechanism of folding is not impeded, but this minimization is not itself the driving mechanism.

The variation in one PR due to biologically plausible electrostatic field is ~7 kcal/mol and variation due to RAHB is similar [61], which overshadows the difference between IPHB and peptide group-water hydrogen bonding being ~1 kcal/mol [88, 89] and is comparable to the stabilization of a protein in its native conformation being 5-15 kcal/mol [90].

PR is posited to be primary in determining backbone conformation which it does by controlling RAHB and hence binding energy of IPHB or other HB involving that peptide group. Also, a peptide group will tend to orient its C->N to align with the local electrostatic field so as to its maximise PR. Peptide groups are distributed along a polypeptide chain and are a factor in determining backbone conformation regardless of whether the PR is part of IPHB secondary structure or not.

Chains of IPHB are resonance-assisted, and RAHB chains are cooperative. In the case of IPHB, the resonance is PR. A change in HB or resonance anywhere in an RAHB chain is accompanied by changes throughout the chain. In this sense, an RAHB chain evaluates changes to one of its HBs or resonant units in terms of all the HB and resonant units of the chain. Since the cooperativity of some IPHB chains is greater than others due to this variation of RAHB, IPHB chains of extensible secondary structures have different affinity for extension. Transient secondary structures with IPHB chains of higher RAHB are more likely to extend and those with lower RAHB are more likely to contract in the early stages of folding. Secondary structures are in competition for peptide groups during early folding. This does not mean that a peptide group has moved immediately from one secondary structure to another, rather that extension of one secondary structure can drag the polypeptide chain so that peptide groups are lost from an IPHB chain of another secondary structure.

When the steric exchange energy of an IPHB is deducted from that associated with the charge transfer of that IPHB, the result is very small [61]. The binding energy of an IPHB resides primarily in the resonances of the two peptide groups [61]. In this sense, PR subsumes IPHB.

The amino acid secondary structure preferences as categorized on the basis of chemical structure by Malkov *et al*. [91] influence PR in IPHB chains in specific backbone conformations in various ways, varying the resonance of the peptide groups throughout the IPHB chain. Strand preferring residues have unfavourable steric interactions with the backbone when in the alpha helical conformation [74], disrupting IPHB. Strand preference correlates with steric blocking by sidechain of backbone hydration in that preferred conformation [75], so alpha helical preferring residues in a beta sheet will weaken IPHB unless the surrounding residues are complementary in blocking backbone hydration. These effects are unfavourable to IPHB and hence RAHB and PR. A study of all the ways amino acid secondary structure preferences influence PR is needed.

### 8.4.3 Specification of fold separated from how to fold

In a purely declarative language [92], what is to be accomplished is described in a manner which says nothing about the means by which it is to be accomplished. This absence of procedure for arrival at solution from statements in the language is greatly simplifying of the language. The existence of a procedure for solution is assumed by the language and is universal to all expressions in the language. Rose *et al*. [5] refer to this separate procedure as a universal folding mechanism. This universal folding mechanism is the means by which a procedure for how to fold is not given in sequences of amino acid residues.

Each amino acid specifies variation in PR via modulation of RAHB of IPHB chains in secondary structure conformations and electrostatic properties including permittivity. All of these have a bearing on PR. The evaluation of a statement in this language, a sequence of amino acids, proceeds primarily by each peptide group searching for maximum resonance, and yields the polypeptide backbone fold.

### 8.4.4 Comparison with earlier backbone-based theory

The present theory, RTPF, can be stated as three principles:

(1) There is a binding energy associated with PR itself. This binding energy is primarily that of the amide/peptide donor-acceptor interaction N(lp)->C-O(p)*. In the absence of factors varying PR and hence the energy associated with PR, different DFT methods calculation of this energy range from about 50 to 100 kcal/mol. This is a large donor-acceptor interaction, and there are a number of known sources of variation of this donor-acceptor interaction: electrostatic field with C-N component, hydrogen bonding, any degree of pyramidalization at N, any C-N torsion (amide non-planarity), the busy donor effect at N and interactions with the carbonyl orbitals other than N(lp)->C-O(p)*. All of these vary the same quantity - PR. Further, all of these interact through PR. Even though the primary amide donor-acceptor interaction is internal to the peptide group, this energy variation is as significant as energy variation of same magnitude elsewhere.

(2) The NBO interactions directly between the peptide groups engaged in IPHB (direct donor-acceptor interactions minus direct steric interactions) sum to very close to zero, with the IPHB binding energy residing in the (internal) peptide group resonance. This means that PR gives a good account of IPHB, and that most of the IPHB stabilization is subject to all other sources of variation of peptide group resonance.

(3) Sidechains influence all of the sources of variation of PR. The peptide group is distributed uniformly along polypeptide chains and each peptide group integrates the various sources of variation of PR as experienced by that group. The variation in PR energy at different backbone conformations is significant, and a driving role in backbone conformational change is anticipated. Residue sequence specifies PR value in a structure-dependent manner, then PR drives folding from there. How to fold does not need to be evolved for each new sequence.

A feature of the earlier theory is that the number of possible backbone conformations is constrained by the restriction on the number of alpha helices and beta strands with short links between them possible in any given length of amino acid residue sequence, a constraint referred to here as ABSL. ABSL would limit the search for protein fold to far fewer possible backbone conformations than is given by the Levinthal estimate. One of the subset of backbone conformations allowed by ABSL is selected as native structure by means involving sidechains. Assuming the length of alpha helices and beta strands is fixed at the observed average and that residue sequences linking them are short, there are about a thousand possible backbone folds for a residue sequence of length 100 [93]. A sequence length of 100 was used by Levinthal to estimate the number of conformations that need to be searched if independence of every residue’s phi and psi is assumed. Unlike the number estimated by Levinthal as demonstration of absurdity, the previous backbone theory gives that the possible backbone conformations may be exhaustively searched in plausible time.

Principles (1) and (2) provide inter alia a mechanism for secondary structures formation subject to backbone constraints such as ABSL, so no conflict between principles (1) and (2) and ABSL arises, though these principles do not assume a tendency to uniform length secondary structure elements with short links. However, ABSL tends to constrain the possible conformations to which principle (3) may drive the backbone, and the extent to which ABSL does this is likely to be decided on a sequence-by-sequence basis. Note that formation of secondary structure elements follows from RTPF, so ABSL may be viewed as an intermediate-level and approximate result in terms of RTPF. The length of links between alpha helices and beta strands is highly variable from native structure to native structure, and perhaps half the population of links is greater than 5 residues each (Figure 1A of [94]). The lengths of alpha helices and beta strands varies. It cannot be assumed for all sequences that alpha helices and beta strands are stable once formed and thus be fixed constraints during the remainder of folding. The length of alpha helices and beta strands may vary after initial formation during folding, including being reduced to zero so that the secondary structure element is transient. Search through different sets of alpha helices and beta strands may occur during folding and ABSL does not offer an account of this search whereas principle (3) does.

### 8.4.5 Observation

Support for this theory could be gathered by observing PR throughout folding and correlating it with other factors. Ideally, each PR would be observed throughout the folding of single molecules by physical methods. Increasing the demands on physical methods, these observations need to be compared with energy changes due to changes in hydrophobia, sidechain-sidechain interactions and entropy. Meeting these demands requires observing the structure of a single molecule throughout folding, a long sought development.

Classical calculations have no account of PR variation, and Quantum Mechanical Dynamics, QMD, calculations, have simulated times that are far short of the time to fold even the fastest folding protein, but progress in linear-scaling methods [95] is not abating. PR might usefully be observed during binding or limited conformational change by QMD. Attention is drawn to the large errors of established DFT methods in calculating the electron density associated with PR [81]. These errors are larger than the natural variation in resonance, and these methods are unsuitable for this purpose. Benchmarks are proposed for the development of more accurate DFT methods [81].

When observation of the PR and C-N component of electrostatic field of each peptide group throughout folding becomes possible, either by physical experiment or by accurate QMD simulation of folding with quantum mechanical handling of every water molecule, these quantities might be shown on each peptide group in a 3D visualization of atoms and bonds. More immediately, such visualization is needed for QMD simulation of a few picoseconds e.g. a few thousand frames. In this visualization, a peptide group could be selected for a report of the sources of electrostatic field with C-N component at that peptide ordered by descending significance. The total peptide resonance and the Gibbs free energy would be shown per frame. The total of peptide resonance for each secondary structure type including unassigned would also be shown. Where the QMD method used can give resonance in terms of energy, kcal/mol might be used as units.

### 8.4.6 Summary

RTPF proposes that the universal folding mechanism is peptide groups seeking to increase their resonance. This search may have the appearance of cooperative or competitive behaviour depending on context, but can be understood as an independent search by peptide groups in their changing environment. This search does not yield monotonically increasing PR, for increase in some PRs may be associated with diminishment of other PRs, and hydrophobic and sidechain-sidechain interactions may alter the conformational landscape to diminish individual or total PR. Evolution of sequences to minimize frustration of folding via local minima in the folding free energy surface would still occur, but is not primary to folding. The driving mechanism of folding, energetic favour of increase in PR and hence RAHB, has complexity and kinetics unrelated to the Levinthal estimate [96].

The present variation and extension of the backbone-based theory of protein folding [5] moves the universal mechanism of folding from IPHB to the backbone per se and allows the common folding mechanism of IPHB secondary structure, polyproline helices and non-secondary structure peptides in folding to be seen. The energetics of PR subsume those of IPHB, since IPHB binding energy primarily resides in PR [61]. A peptide group's PR varies as its binding partners change. Resonance of sidechain groups does not drive folding as these groups are not integral to the backbone.

The amino acid sequence is a language which is interpreted by all peptide groups in terms of PR and backbone conformation. This language separates specification of fold from procedure for how to fold which is given by the universal folding mechanism, PR. This separation would confer great evolutionary advantage, since how to fold does not need to be evolved for each new sequence as how to fold is not encoded in the sequence.

Means of observing the resonance of every peptide group in a single molecule throughout folding is needed. QMD simulations could simulate enough time to capture changes to the resonance of peptide groups during binding or limited conformational change. Established DFT methods are unsuitable for this purpose [81].

## 8.5 Molecular chaperones and protein complexes

When one protein binds to another at a largely hydrophobic interface, the hydrophobic interiors of the two proteins may be substantially joined, allowing the electrostatic field of one interior to extend into the interior of the other. Charged residues at the binding interface of either of the proteins will be desolvated on protein-protein binding, and if these are not paired to neutrality after protein-protein binding, will cause electrostatic field in the interiors of both proteins. We predict that hydrophobic protein-protein binding by a protein chaperone allows the chaperone to introduce a defined electrostatic field into the client protein thus inducing reorganization of the client. This may be a multi-step process in which in the first stage the chaperone provides an electrostatic field of sufficient magnitude in the hydrophobic environment that the client is forced to reorganize, a second step in which the chaperone no longer extends a field into the interior of the client thus allowing the client to reorganise according to its own electrostatic field and a third stage in which the chaperone detaches. Both pure electrostatics and electrostatic field variation of backbone amide resonance are effectual in this process.

## 8.6 Nitrogenous base pairing

We note features of nucleic acid bases that are likely also subject to electrostatic field variation of N(lp)->C-X(pi)*, where X is either O or N, charge transfer. In a pair of nitrogenous bases there are a number of bonds that may play the role of amide C-N in electrostatic variation of resonance. In these base pairings, wherever there is a nitrogen lone pair out of the plane of the rings and the atom bearing the lone pair is bonded to a carbon participating in another bond which has double-bond character, significant resonance-type charge transfer can be expected to occur and sensitivity to C-N component electrostatic field is to be anticipated. All of these interactions will modify the properties of the rings. Table 1 shows the energetics of the subset of these interactions which are closely involved with the base pairing from NBO's default vantage point of best or nominated Lewis picture.

Bonds 1 and 2 (both of Figure 15 and Figure 16) in both the guanine/cytosine, GC, pairing and the thymine/adenine, TA, pairing are antiparallel and part of cyclic hydrogen bonding between the bases. Cyclic RAHB will be limited by the least member of the cycle, so an overall weakening of the base-pairing hydrogen bonds could be expected in the presence of electrostatic field with component antiparallel to the C-N of one of these bonds. Also, the N(lp)->C-O(pi)* associated with bond 1 will have different sensitivity to the field than the N(lp)->C-N(pi)* associated with bond 2. The vectors of these bonds with respect to the nucleic acid helix differs between the cases of TA versus AT and similarly for GC versus CG. A molecule which moves along the helix and which has an associated electrostatic field to which TA versus AT and GC versus CG bond 1 and bond 2 vectors are not symmetric will vary inter-base hydrogen bonding differently in accord this asymmetry. Also, electrostatic field aligned with the major groove is reversed by a field is aligned with the minor groove.

The protonated nitrogen on the thymine ring directly involved in inter-base hydrogen bonding has two N(lp)->C-O(pi)* interactions, making it a busy donor. Only one of those interactions is part of cyclic RAHB involving the two base-pairing hydrogen bonds, so an electrostatic field with the C-N vector of the other interaction, bond 3 (Figure 16), could be expected to diminish cyclic RAHB, again with the consequence of facilitating base-pair opening.

The nitrogen of cytosine that would connect to the helix backbone has a N(lp)->C-O(pi)* interaction, across bond 4 (Figure 15), is involved in non-cyclic inter-base hydrogen bonding and can be expected to vary the ring properties of a paired base.

Variation in electrostatic field in nucleic acid polymers occurs when the negatively charged phosphates of the nucleic acid backbone are permitted to attract positive charge, or a charged molecule binds to the major or minor groove of the nucleic acid helix.

Variation of these resonances by electrostatic field may vary the binding energy of base pairing, and hence the energetic barrier of base pair opening for nucleic acid strand separation.

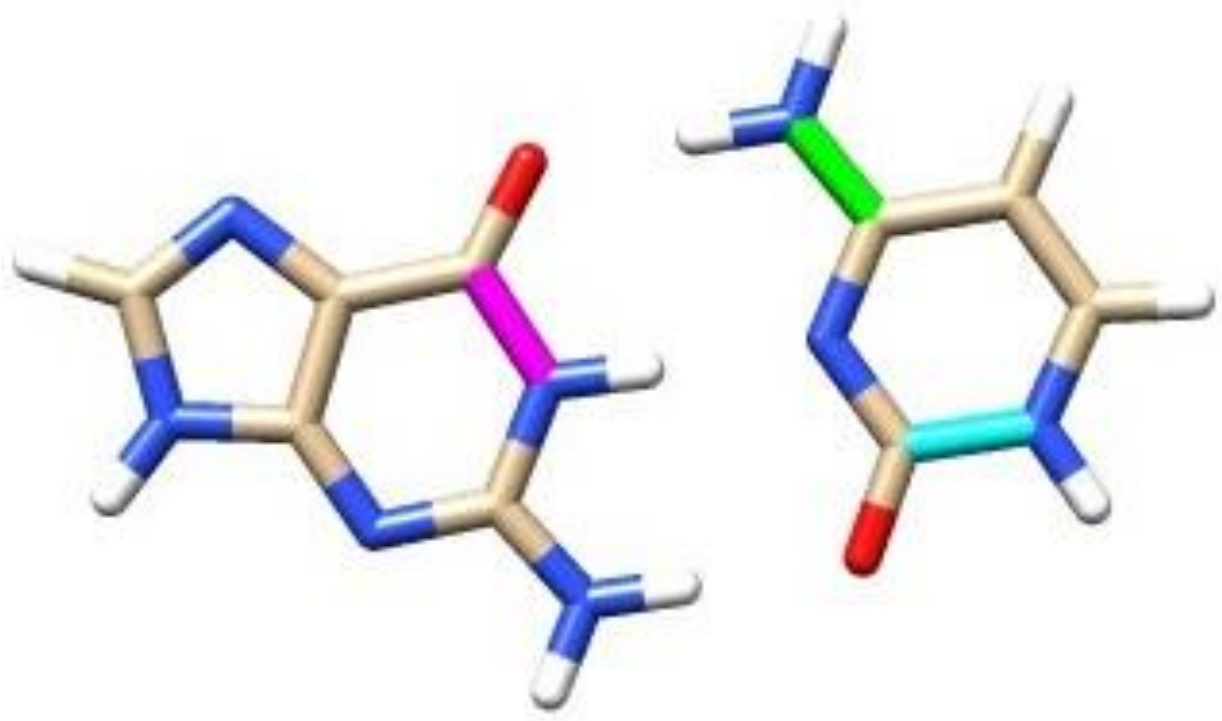

Figure 15. Guanine/Cytosine Pairing. Bond 1 magenta, Bond 2 chartreuse, Bond 4 Light Blue. Major Groove Top, Minor Groove Bottom.

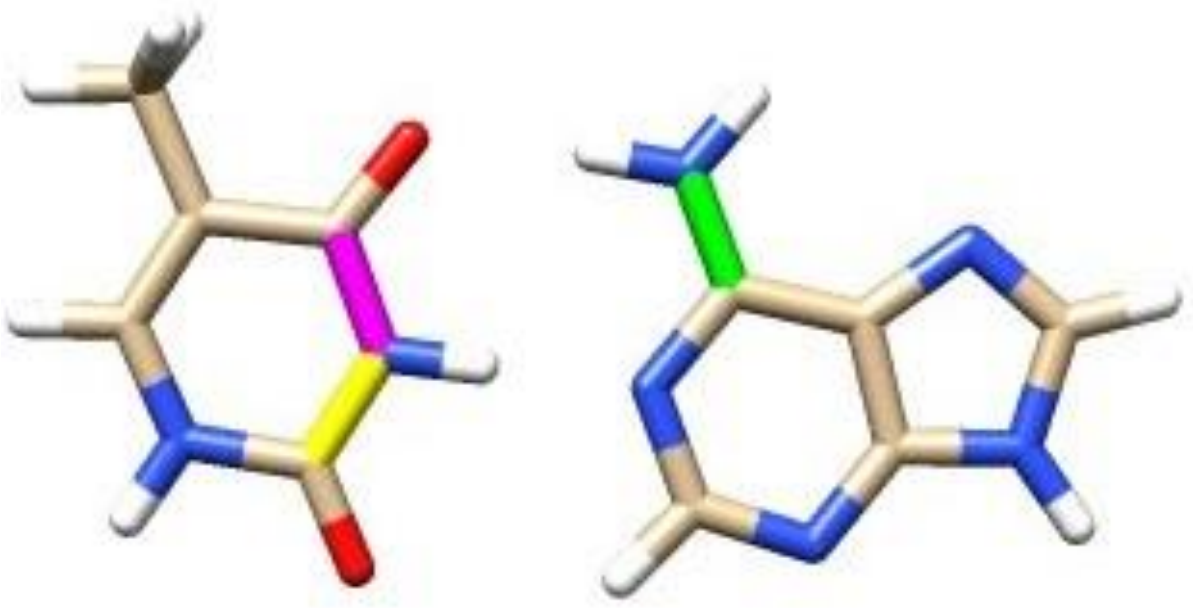

Figure 16. Thymine/Adenine Pairing. Bond 1 magenta, Bond 2 chartreuse, Bond 3 yellow. Major Groove Top, Minor Groove Bottom.

Table 1. Selected N(lp)->C-X(p)* Interactions in GC and AT Base Pairs at LC-wPBE(w=0.4)/6-311++G**

| Base Pair | N-C Bond ID | Type of X | N(lp)->C-X(p)* kcal/mol |
|---|---|---|---|
| GC | 1 | O | 78.38 |
| AT | 1 | O | 87.42 |
| GC | 2 | N | 126.65 |
| AT | 2 | N | 113.26 |
| AT | 3 | O | 88.07 |
| GC | 4 | O | 70.21 |

## 9 Conclusion

Amide resonance is sensitive to electrostatic field with component parallel or antiparallel to the C-N vector. A field of 0.000238 au is sufficient to vary the primary amide resonance-type charge transfer by 0.001 electron, a level we nominate as the threshold of chemical significance. The charge transfer as calculated by both DFT and wavefunction methods increases linearly with field magnitude in the range -0.005 to 0.005 au without threshold. A variation in field between -0.005 and 0.005 au gives a 0.042 electron variation in primary amide charge transfer in N-methylformamide and a 0.2 electron variation in charge transfer is calculated between the cases when the two monovalent ions, Li+ and F-, are placed on the N-C line with one 4 angstroms from C further from N and the other 4 angstroms from N further from C as compared with the ions being swapped in those positions to reverse the field. This arrangement is not offered as biologically plausible but shows that a 0.2 electron variation in the primary amide charge transfer is physically possible though is not the full extent of what is physically possible. Electronic structure of the amide group is not particularly multireference [81] and a field of magnitude 0.005 au causes negligible change to this (data not shown).

If an amide is participating in a protein secondary structure RAHB chain, change in the amide's resonance causes variation in hydrogen bonding in the chain. Electrostatic field with direction that increases the resonance of an amide in the chain without directly diminishing hydrogen bonding stabilizes the RAHB chain. This hydrogen bonding also involves charge transfer from lone pairs, and its charge transfer is varied less in absolute terms at a given field magnitude than that primary to amide resonance. Also, with the electrostatic field orthogonal to H-N so the hydrogen bonding resonance-type charge transfer is not varied with O..H-N angle constrained, the field has a substantial component parallel or antiparallel to the C-N vector.

A consequence of EVPR-CN is the change in energy associated with change in the primary amide charge transfer itself which is irrespective of the amide bonding context, which may be considerable and must then be taken into account when studying the energetics of protein structure.

We discussed stability of beta sheets and non-polyproline helices due to electrostatic field variation of backbone amide resonance, mentioned a connection to conformational and allosteric change and implications for unstructured regions, and offered hypothesis concerning stability of beta sheets, amyloid fibrils and both types of polyproline helices, and of the function of protein chaperones. We predict analogous stability considerations for nitrogenous base pairing in an electrostatic field, and offer a hypothesis in this regard.

It may be that other factors such as hydrophobia, entropy and pure electrostatics are sufficient to determine a fold of some proteins. However, it would be remarkable if there were any protein that provided an electrostatic environment for all its backbone amides such that there was negligible field component parallel or antiparallel to the C-N vector environmental to the amide group. Wherever such a field exists, it necessarily changes amide resonance. The significance of the effect and its consequences for protein folding and conformational change need to be placed in relation to that of other effects influencing protein folding and conformational change. However, this variation of backbone amide resonance is directionally sensitive to the electrostatic field created and permitted by amino-acid residue sidechains, and offers a novel mechanism for the relationship of amino-acid residue sequence and protein fold [97]. This mechanism is a more precise specification of structure than less directionally-sensitive hydrophobia, but is facilitated by hydrophobia's creation of a low permittivity protein interior.

Rose et al. [5] proposed that protein folding is backbone-based and that backbone hydrogen bonding is a universal folding mechanism, but do not propose direct variation of this mechanism by residue sidechains. Variation of backbone amide resonance by electrostatic field is a backbone-based mechanism, with electrostatic properties of residue sidechains directly varying backbone amide resonance, hence varying backbone hydrogen bonding. We proposed the Resonance Theory of Protein Folding (Section 8.4) in which protein folding is driven by PR, where PR is varied by any effect including EVPR-CN, RAHB, C-N torsion, pyramidalization at N, the busy donor effect at the N lone pair and interactions with the carbonyl orbitals other than N(lp)->C-O(p)*.

## 10 Acknowledgements

Prof. John A. Carver is acknowledged for reading this manuscript and offering editing suggestions.

eResearch South Australia is acknowledged for hosting and administering machines provided under Australian Government Linkage, Infrastructure, Equipment and Facilities grants for Supercomputing in South Australia, directing funds to the acquisition of Nvidia Tesla GPU nodes and allocating 64 CPU cores and 256 GB RAM of the NeCTAR Research Cloud (a collaborative Australian research platform supported by the National Collaborative Research Infrastructure Strategy) to the present work.

## 12 Appendix 1

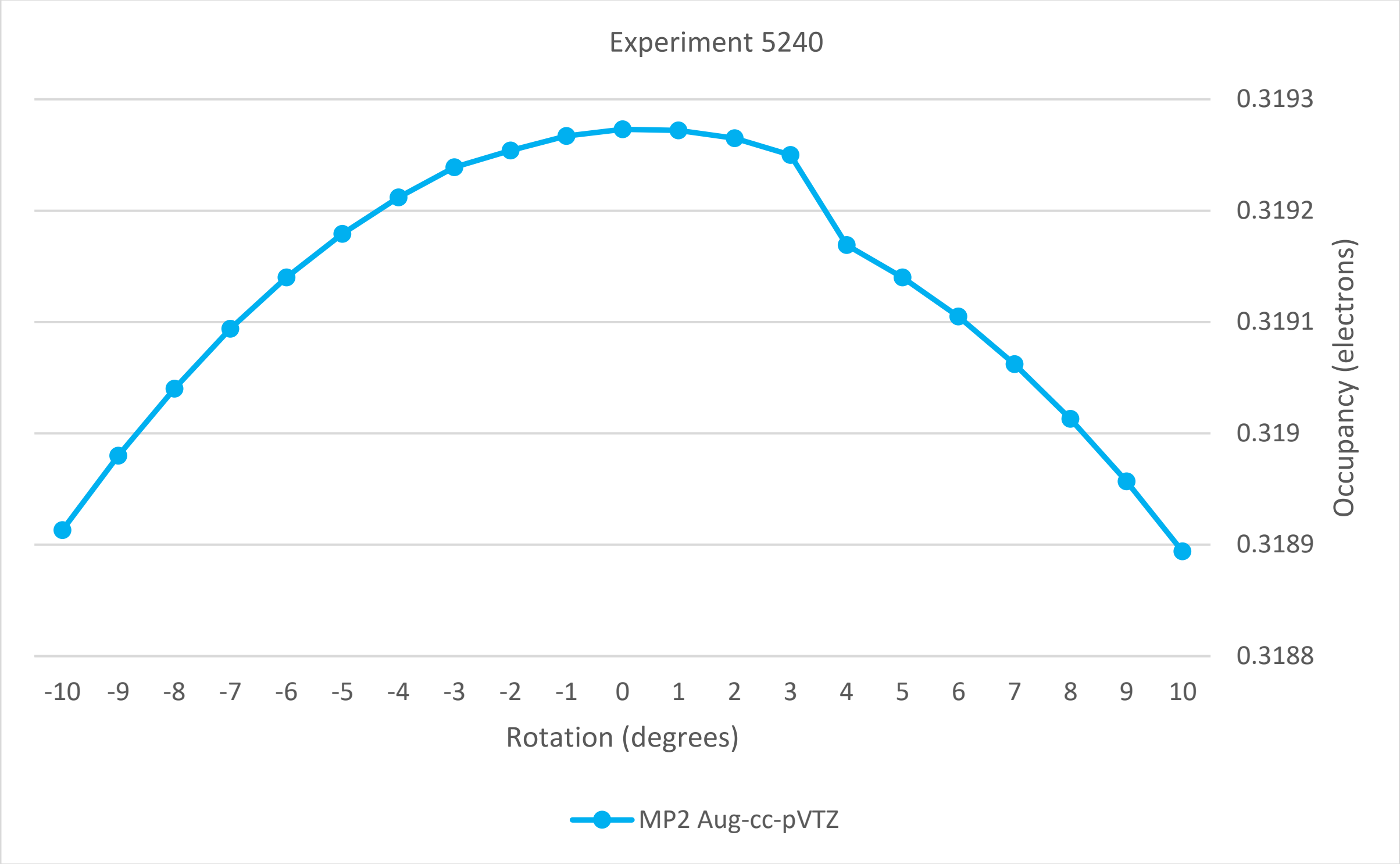

Figure 17. C-O(pi)* NBO Occupancy in N-methylformamide in 0.005 au Uniform Electrostatic Field Rotated in O-C-N Plane Starting from C-N Vector

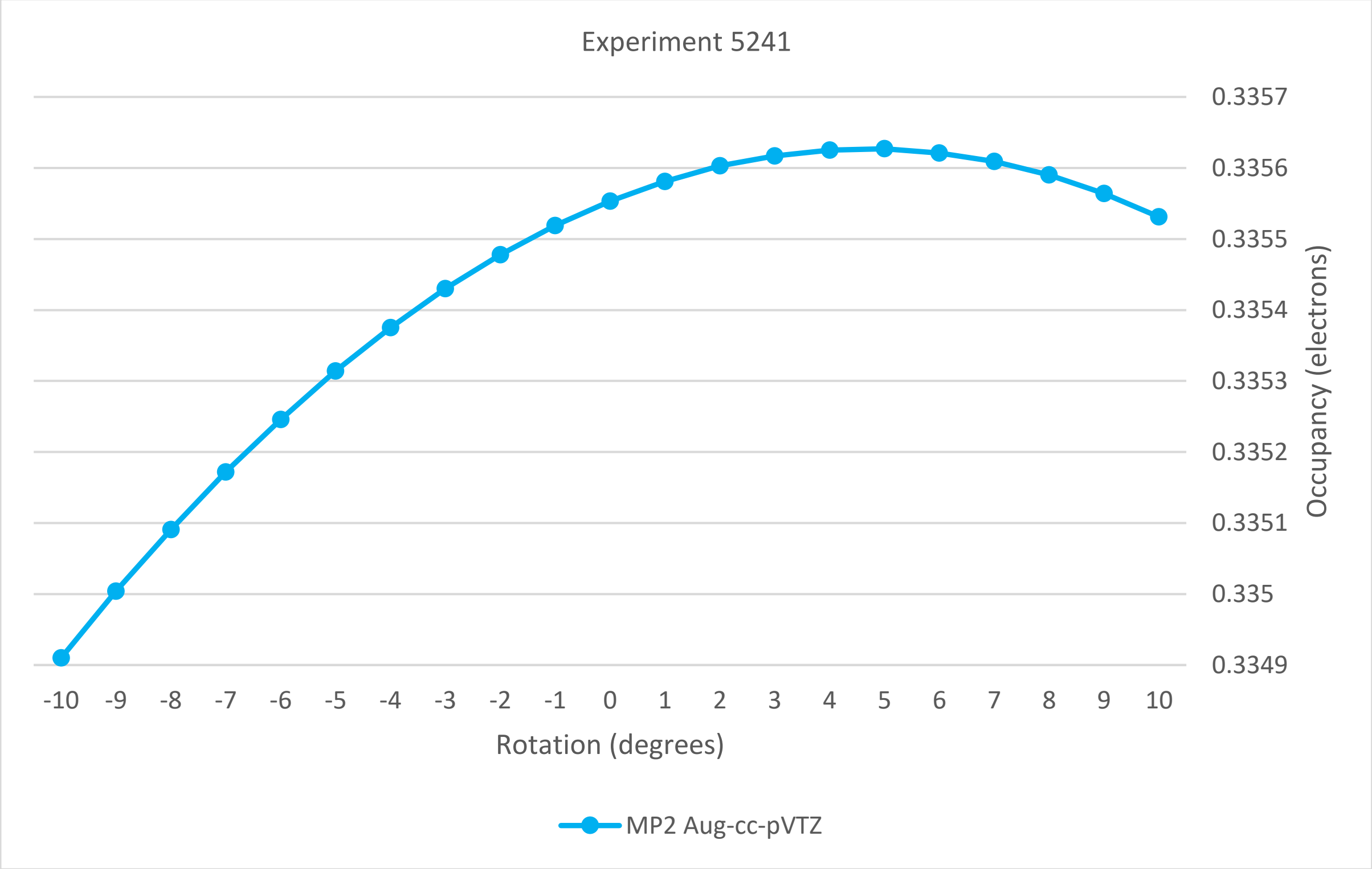

Figure 18. C-O(pi)* NBO Occupancy in N-methylethanamide in 0.005 au Uniform Electrostatic Field Rotated in O-C-N Plane Starting from C-N Vector

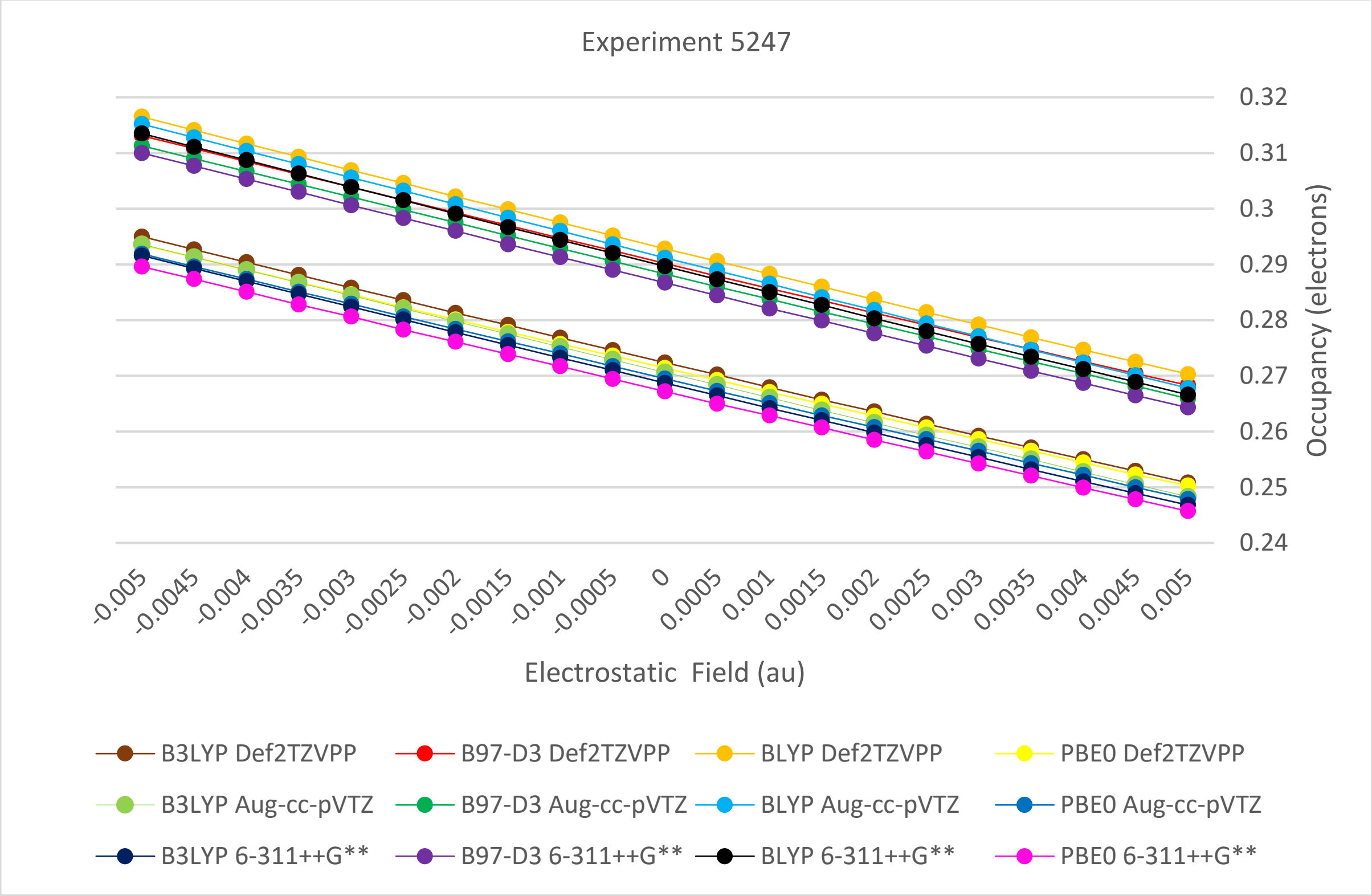

Figure 19. C-O(pi)* NBO Occupancy in N-methylformamide in Uniform Electrostatic Field with N-C Vector

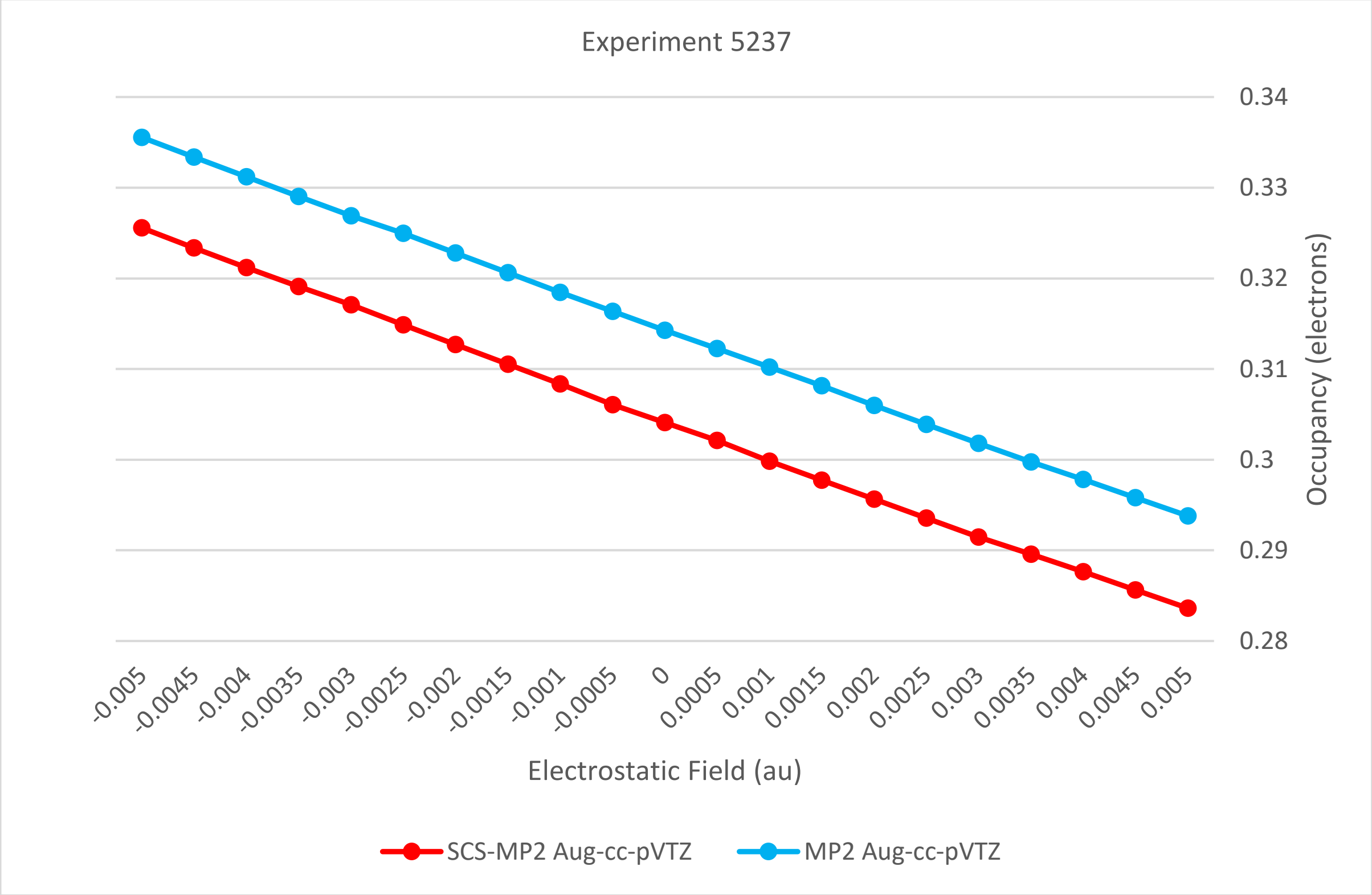

Figure 20. C-O(pi)* NBO Occupancy in N-methylethanamide in Uniform Electrostatic Field with N-C Vector

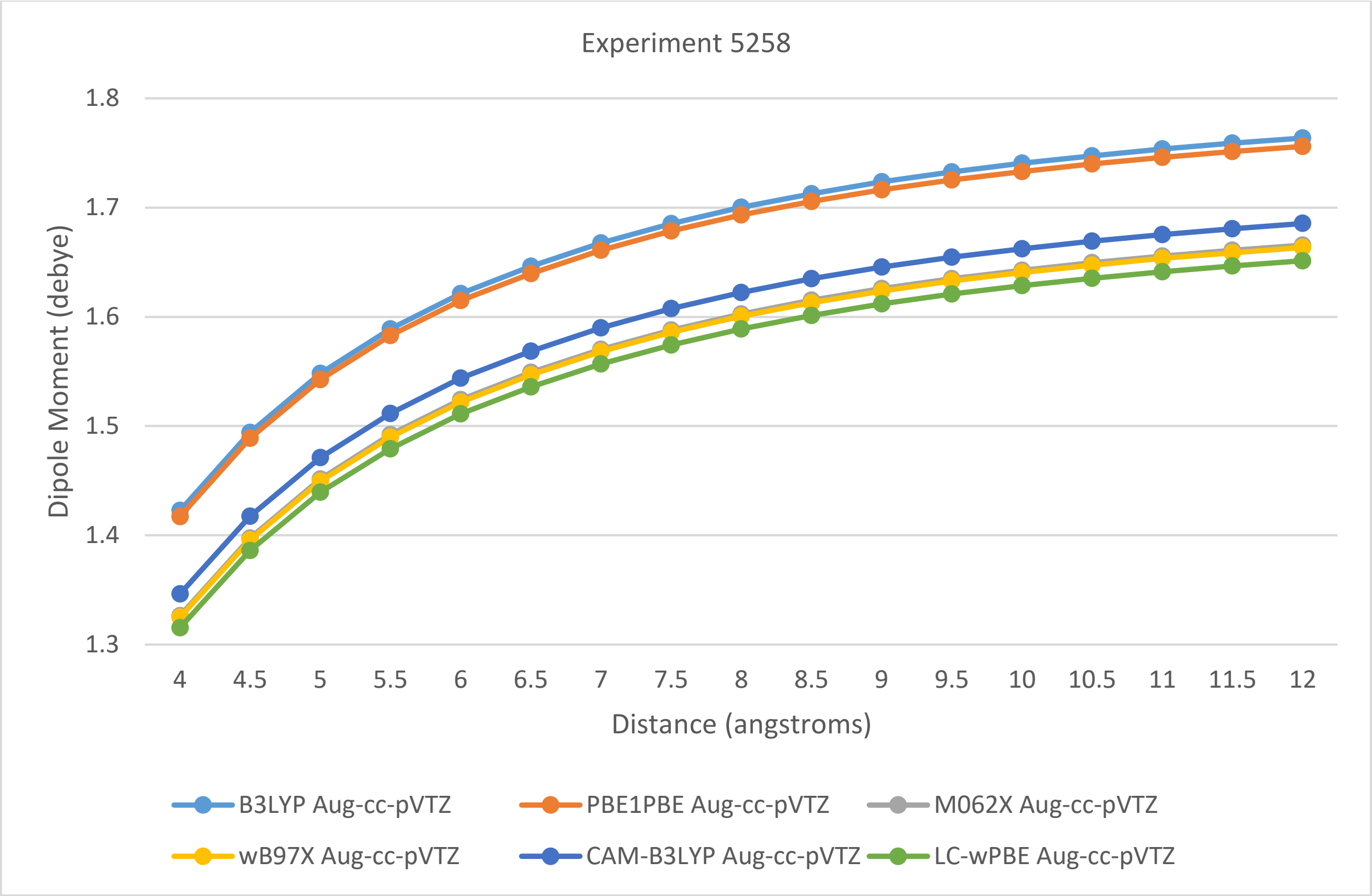

Figure 21. Dipole Moment of N(lp) NLMO in N-methylformamide with Li+ on C-N Line at Distance from N

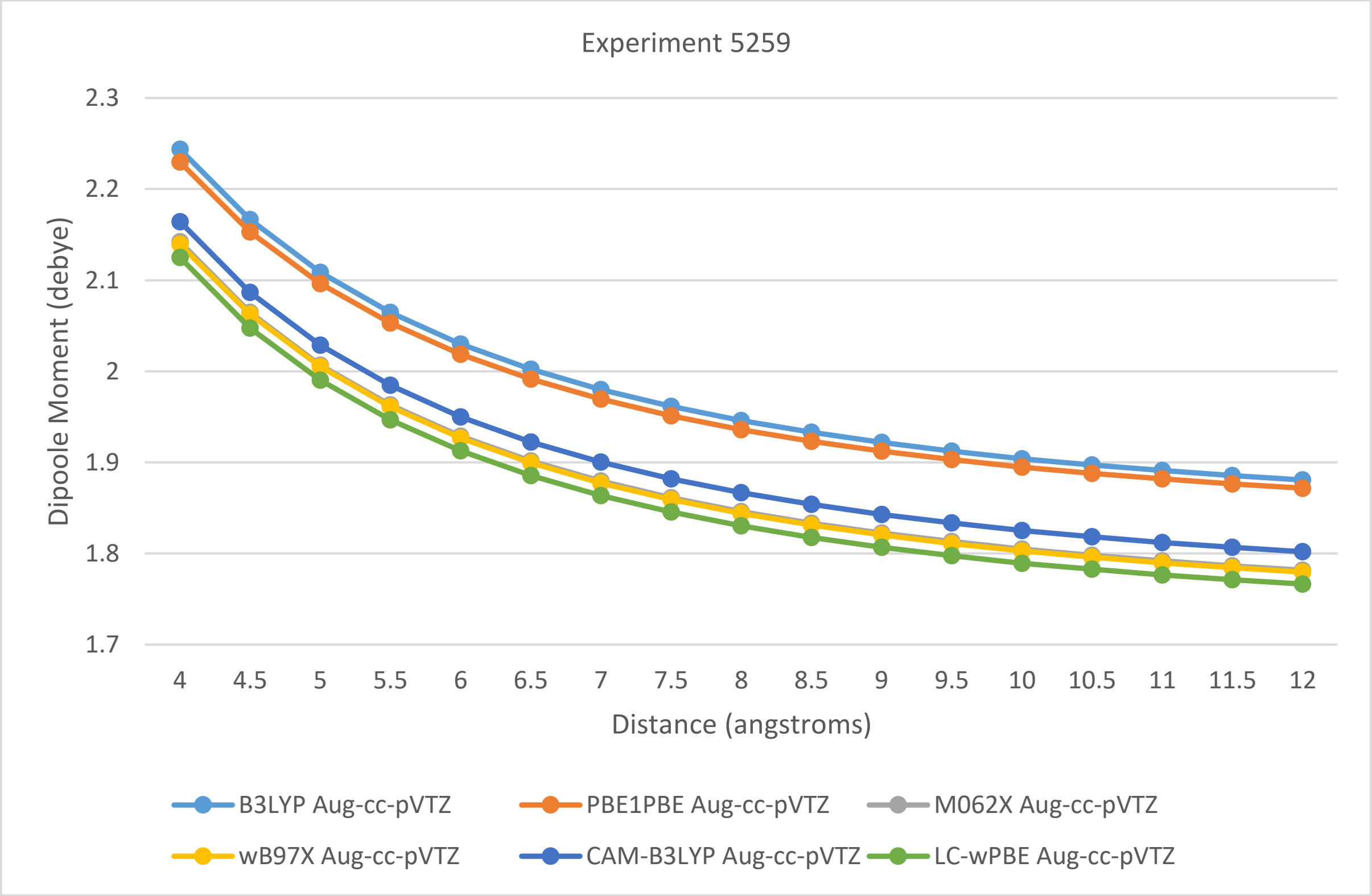

Figure 22. Dipole Moment of N(lp) NLMO in N-methylformamide with F- on C-N Line at Distance from N

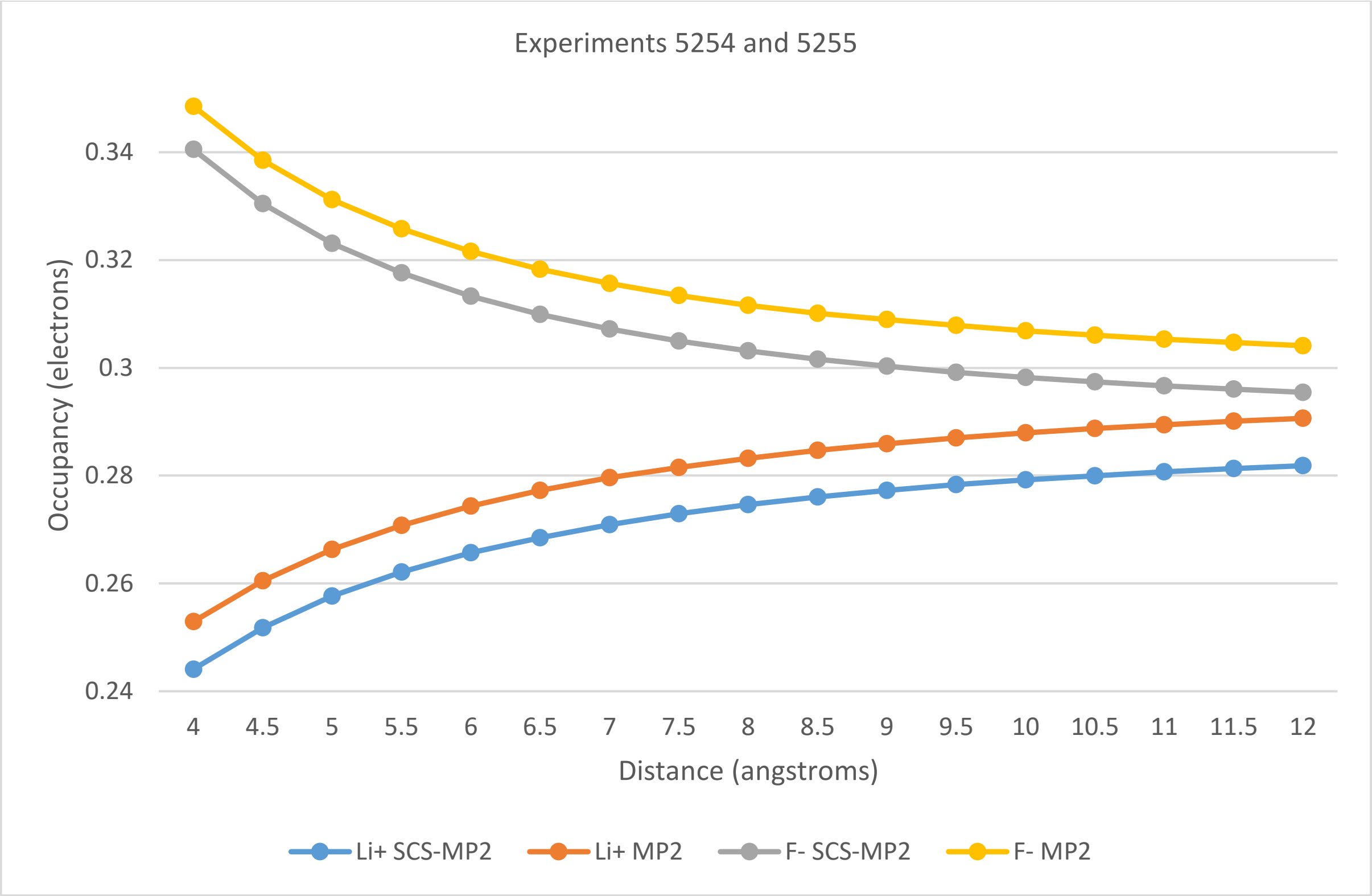

Figure 23. C-O(pi)* NBO Occupancy in N-methylformamide with Ion on C-N Line at Distance from N with aug-cc-pVTZ

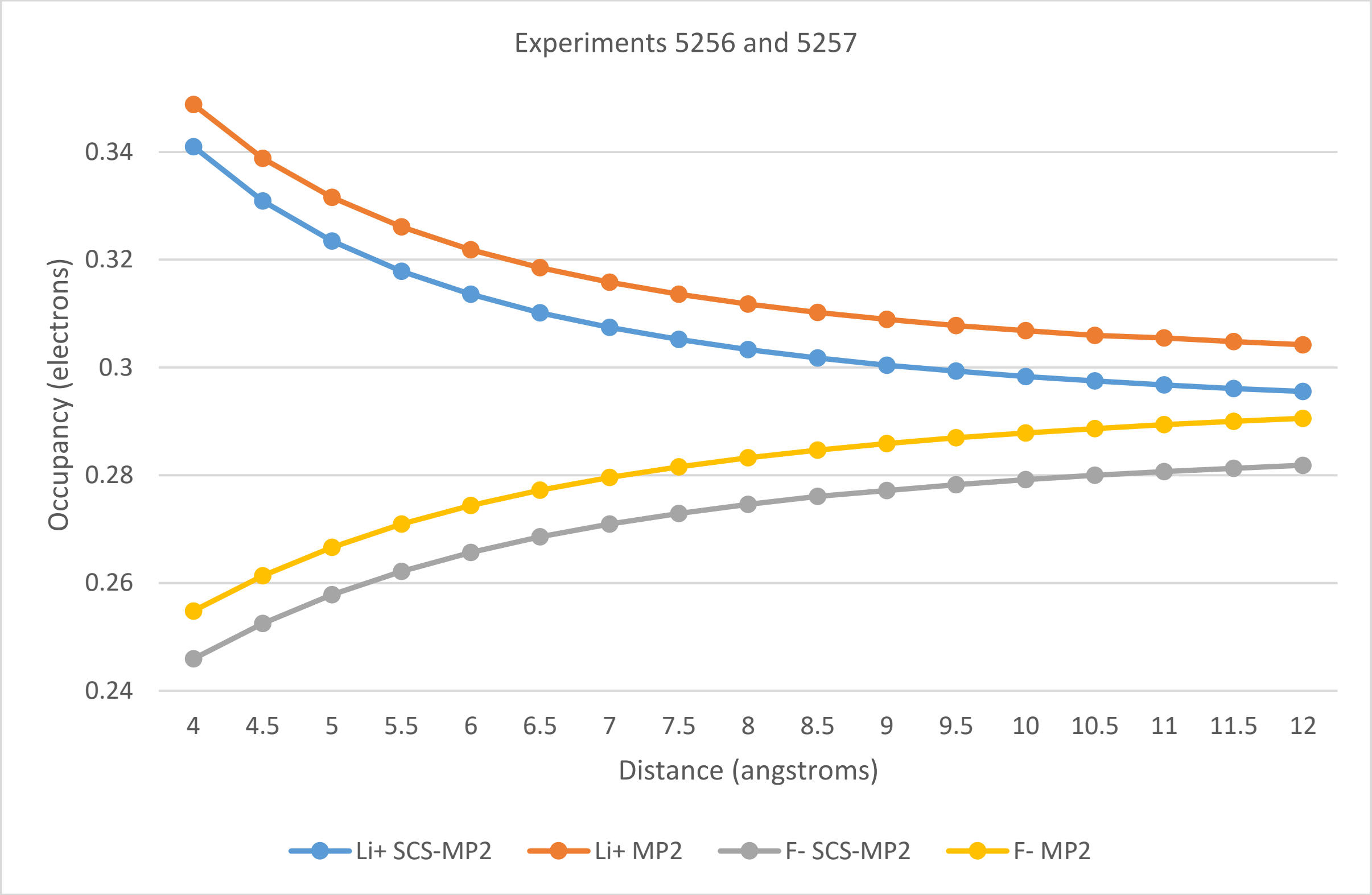

Figure 24. C-O(pi)* NBO Occupancy in N-methylformamide with Ion on N-C Line at Distance from C with aug-cc-pVTZ

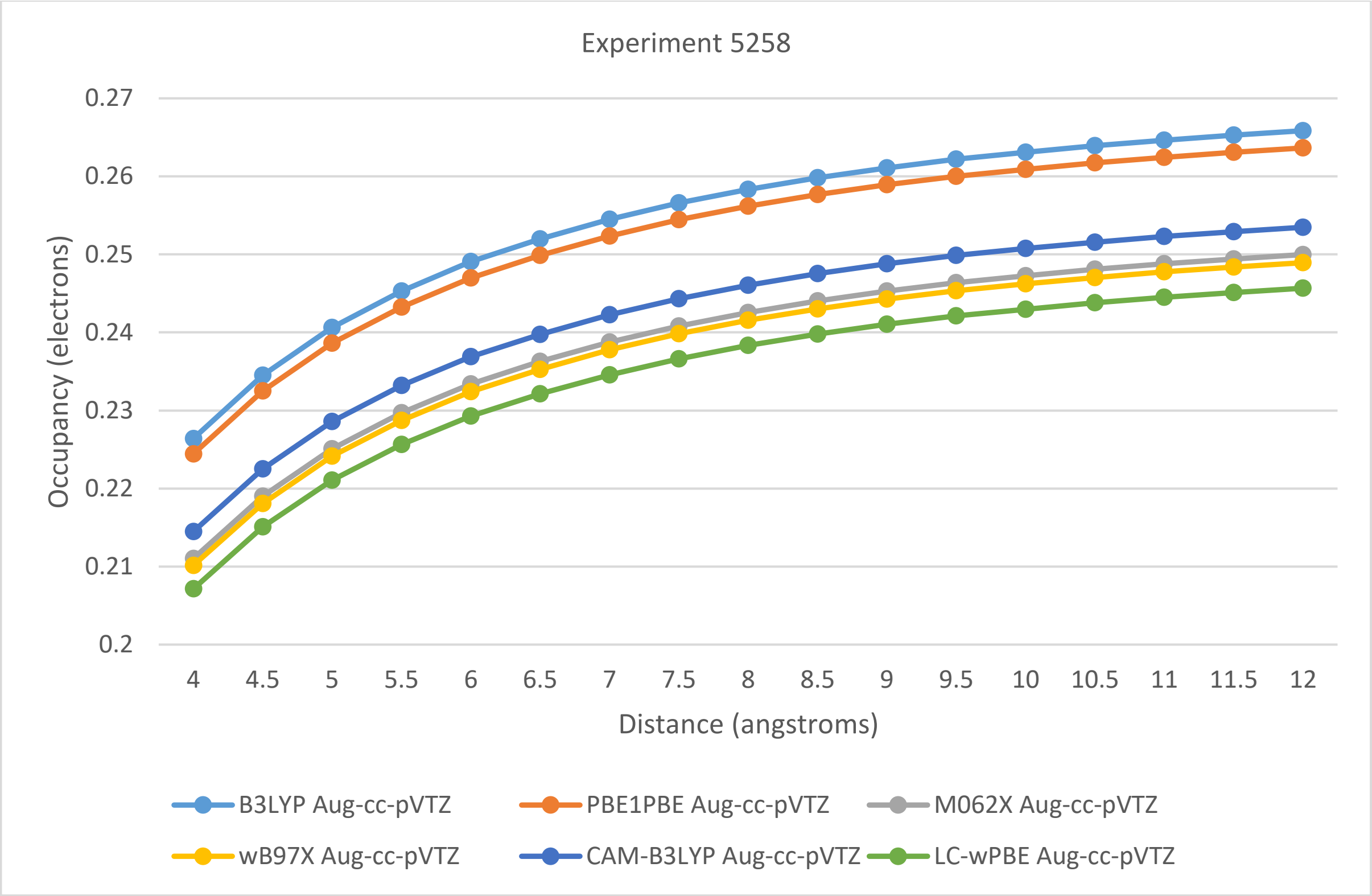

Figure 25. C-O(pi)* NBO Occupancy in N-methylformamide with Li+ on C-N Line at Distance from N using Gaussian SP over Coordinates Optimized by Orca at SCS-MP2/aug-cc-pVTZ

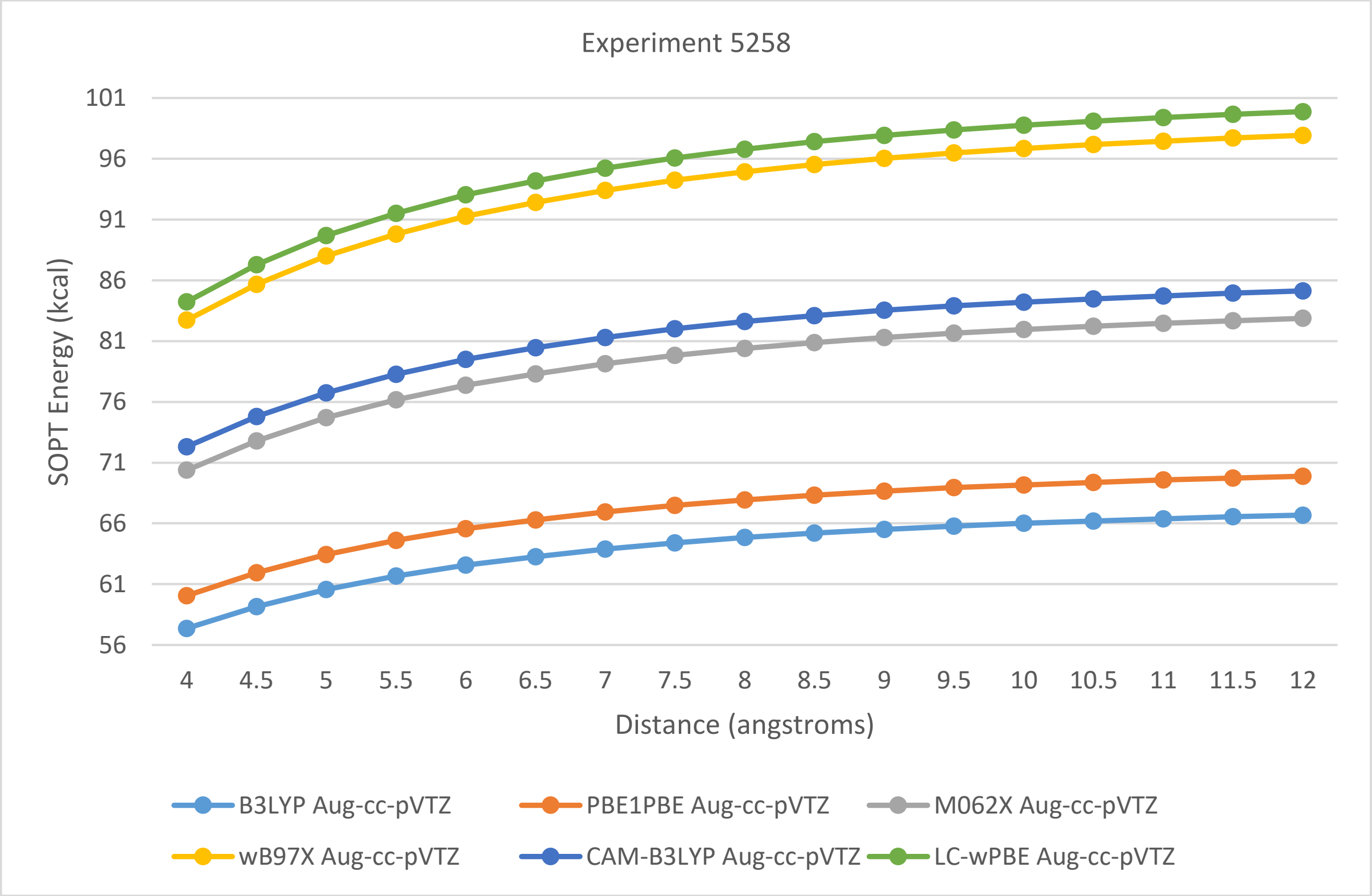

Figure 26. N(lp)->C-O(pi)* SOPT Energy in N-methylformamide with Li+ on C-N Line at Distance from N using Gaussian SP over Coordinates Optimized by Orca at SCS-MP2/aug-cc-pVTZ

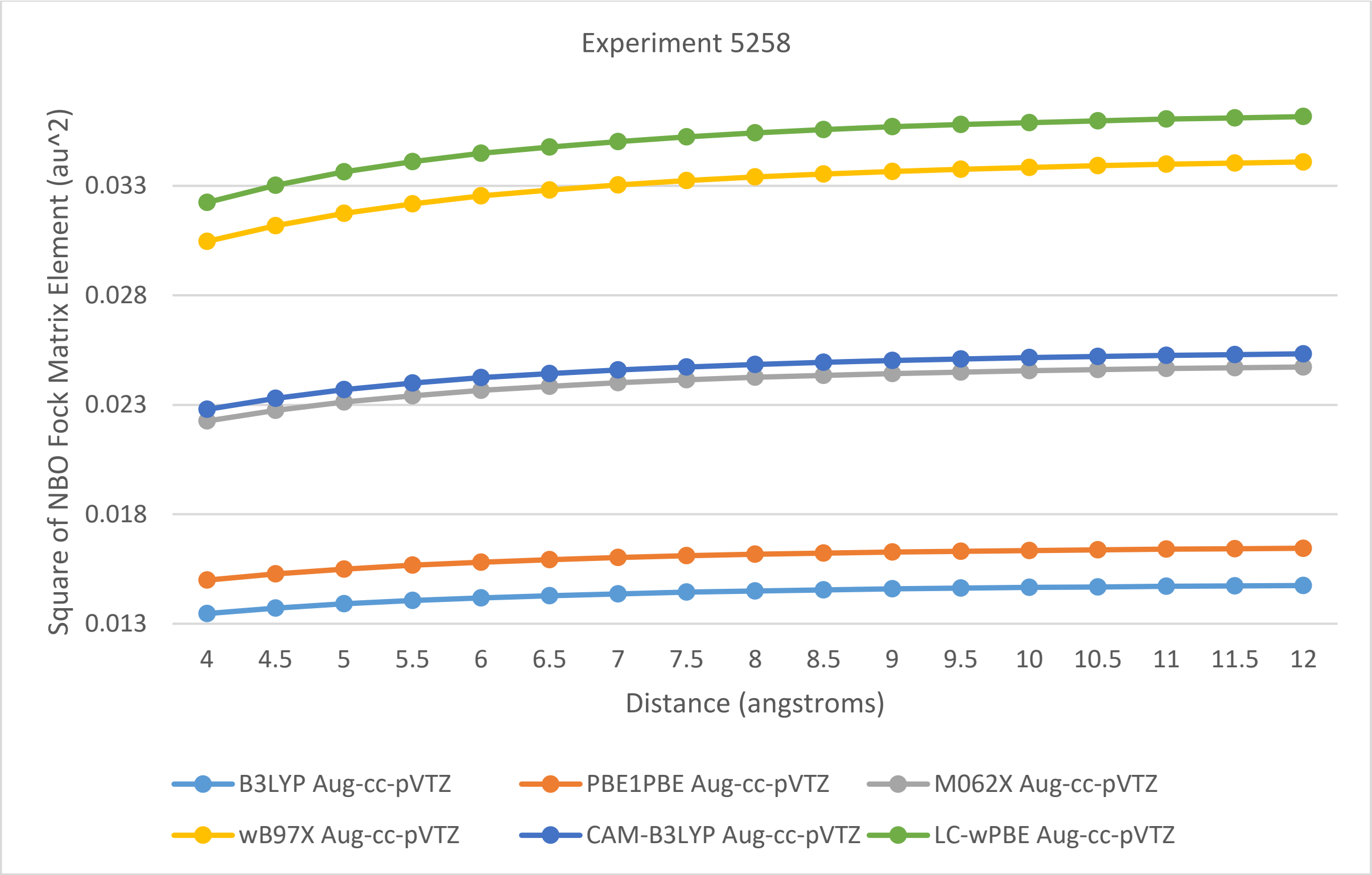

Figure 27. Square of NBO Fock Matrix Element for N(lp)->C-O(pi)* in N-methylformamide with Li+ on C-N Line at Distance from N using Gaussian SP over Coordinates Optimized by Orca at SCS-MP2/aug-cc-pVTZ

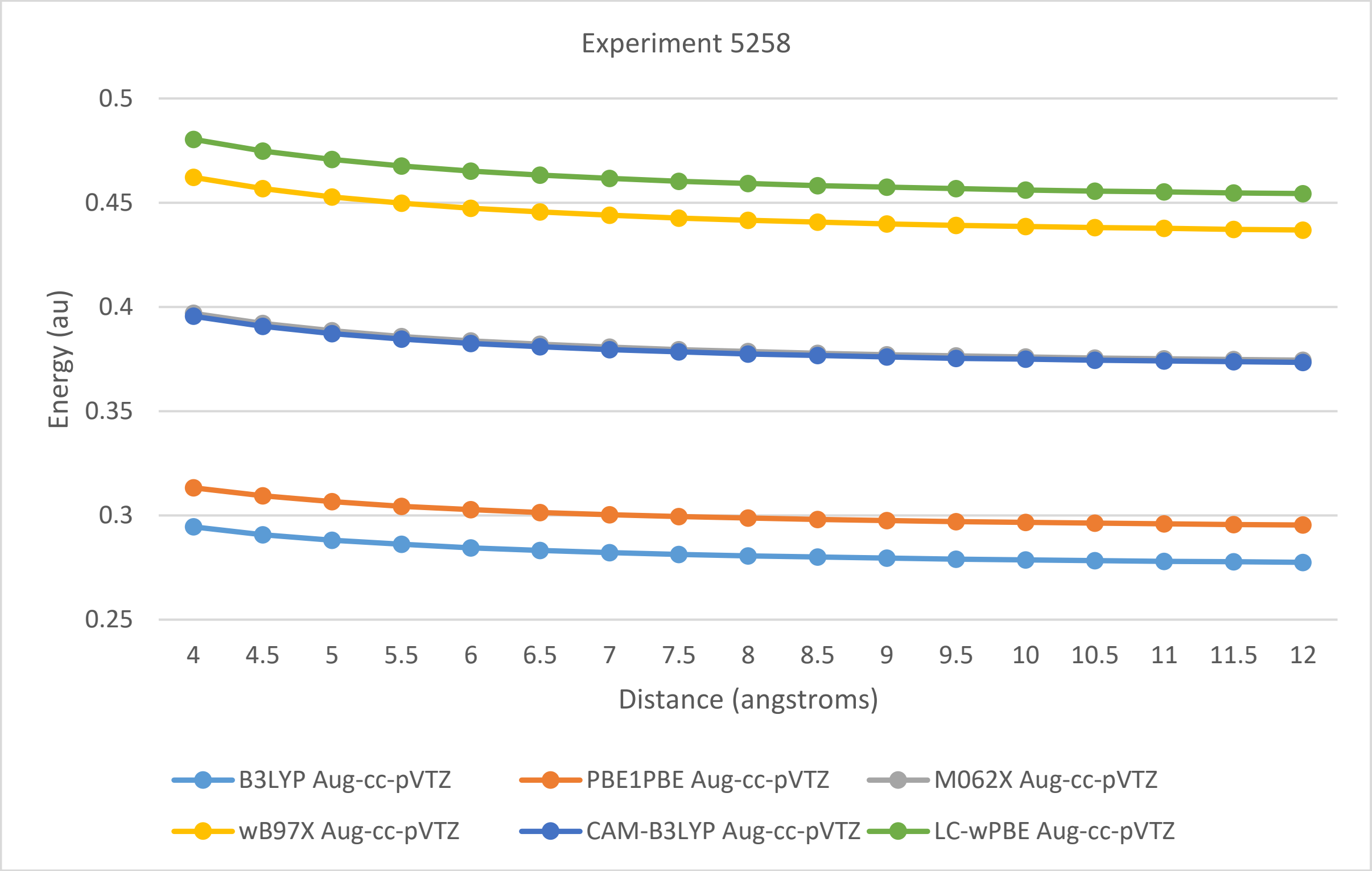

Figure 28. Energy Level of C-O(pi)* Minus Energy Level of N(lp) in N-methylformamide with Li+ on C-N Line at Distance from N using Gaussian SP over Coordinates Optimized by Orca at SCS-MP2/aug-cc-pVTZ

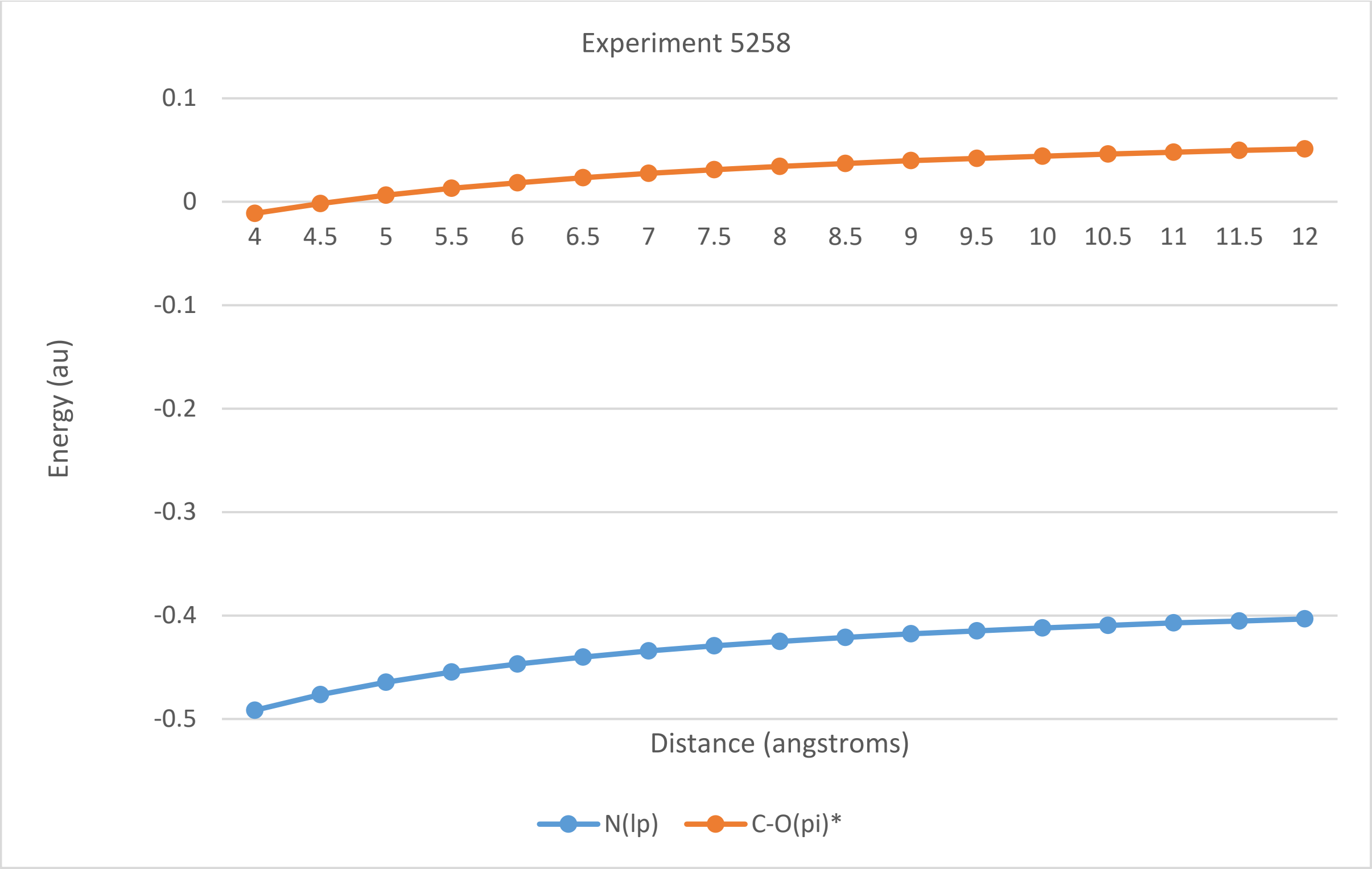

Figure 29. Energy Level of NBOs in N-methylformamide with Li+ on N-C Line at Distance from C using Gaussian SP over Coordinates Optimized with Orca at SCS-MP2/aug-cc-pVTZ

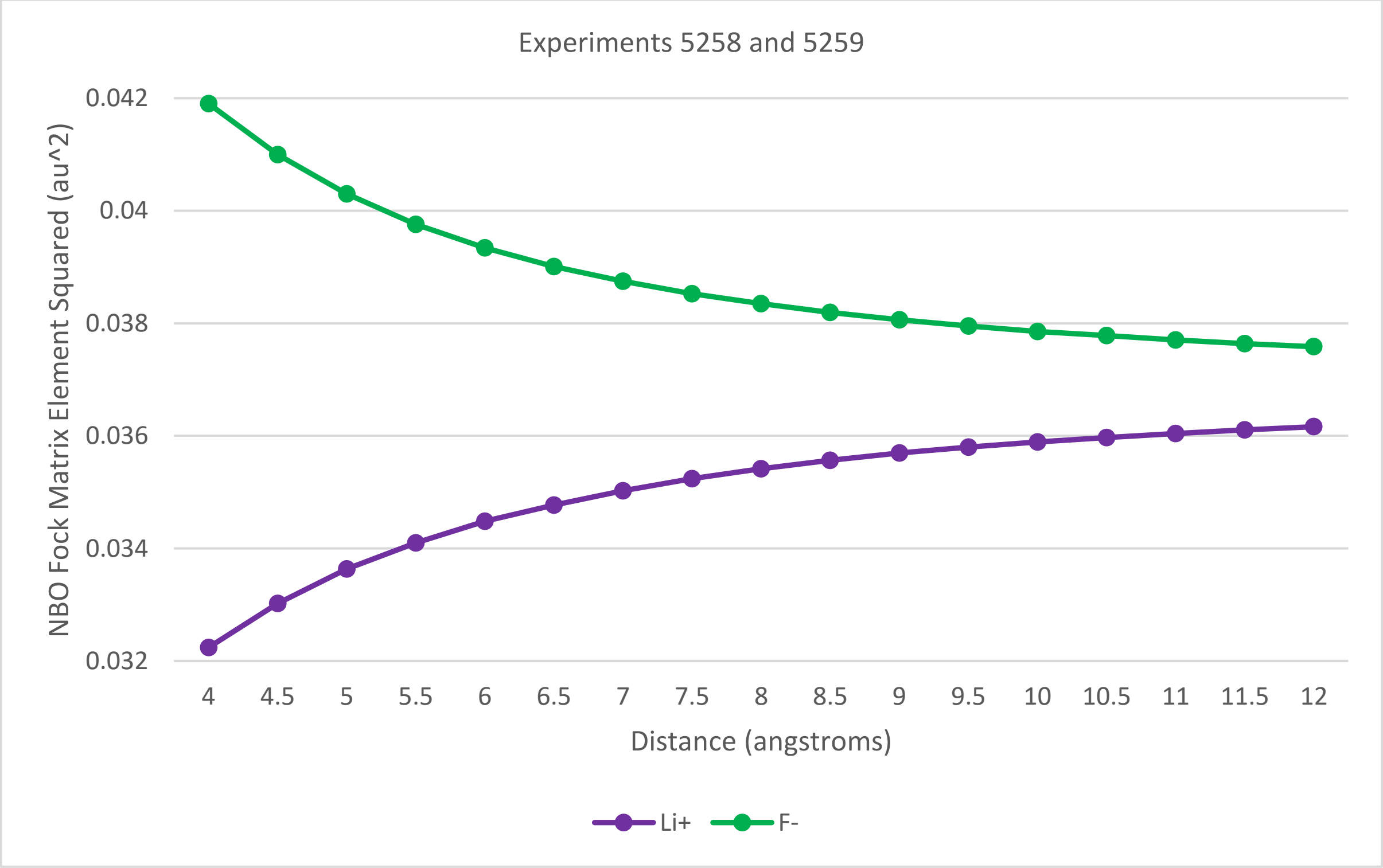

Figure 30. Square of NBO Fock Matrix Element for N(lp)->C-O(pi)* in N-methylformamide with Ion on C-N Line at Distance from N with Gaussian SP LC-wPBE/aug-cc-pVTZ over Coordinates Optimized by Orca at SCS-MP2/aug-cc-pVTZ

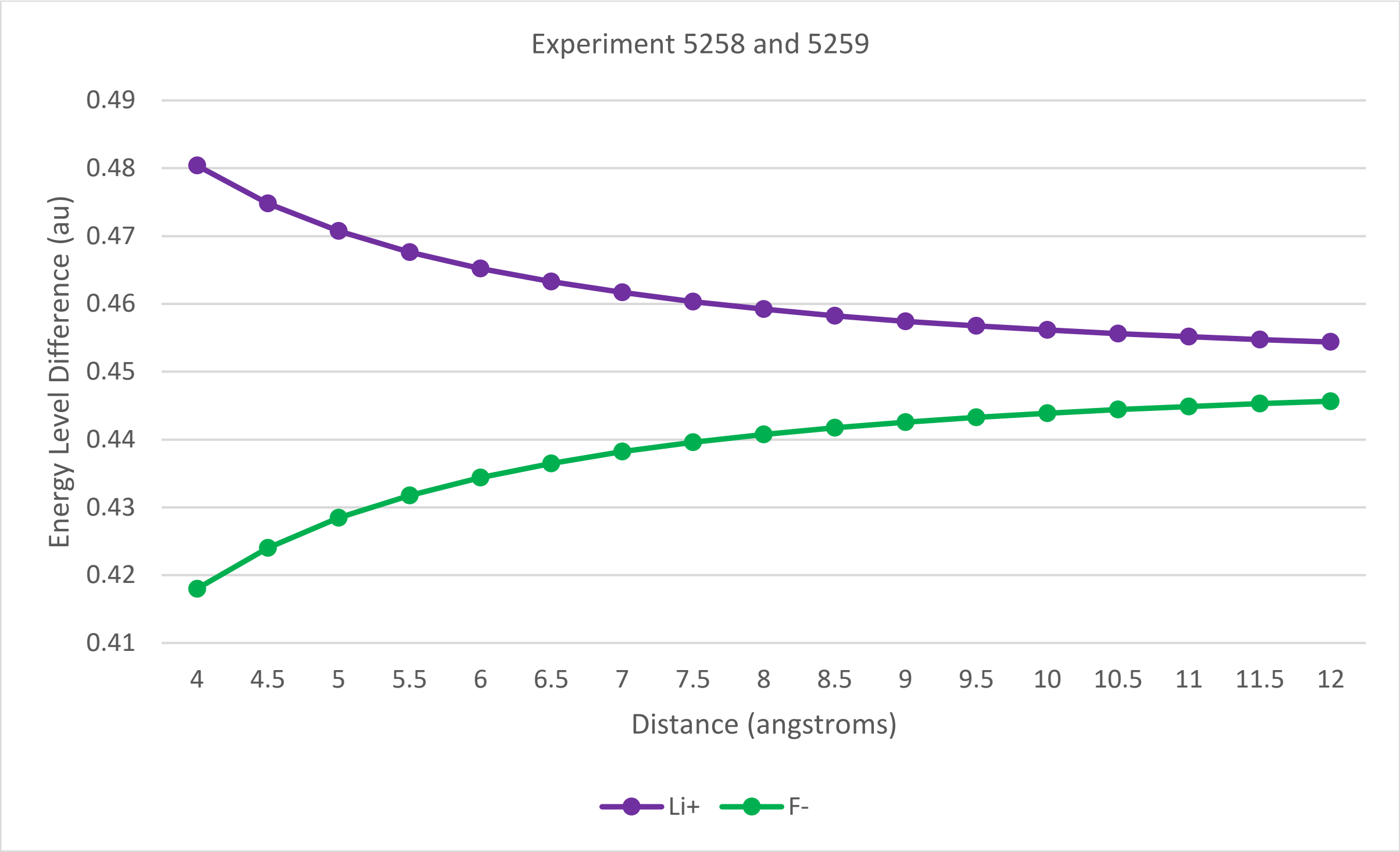

Figure 31. Energy Level of C-O(pi)* Minus Energy Level of N(lp) in N-methylformamide with Ion on C-N Line at Distance from N using Gaussian SP LC-wPBE/aug-cc-pVTZ over Coordiantes Optimized by Orca at SCS-MP2/aug-cc-pVTZ

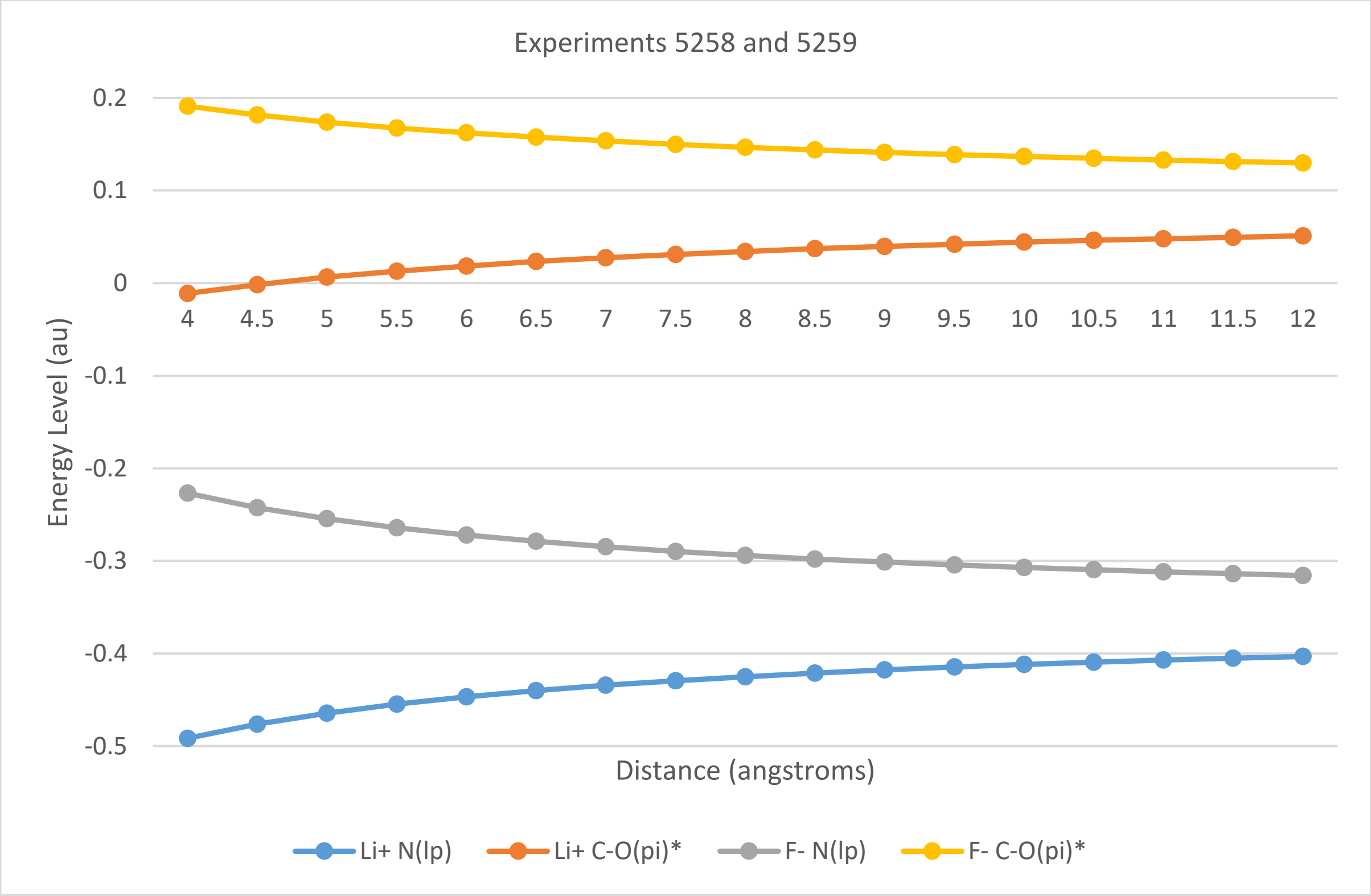

Figure 32. Energy Level of NBOs in N-methylformamide with Ion on C-N Line at Distance from N with Gaussian SP LC-wPBE/aug-cc-pVTZ over Coordinates Optimized by Orca at SCS-MP2/aug-cc-pVTZ

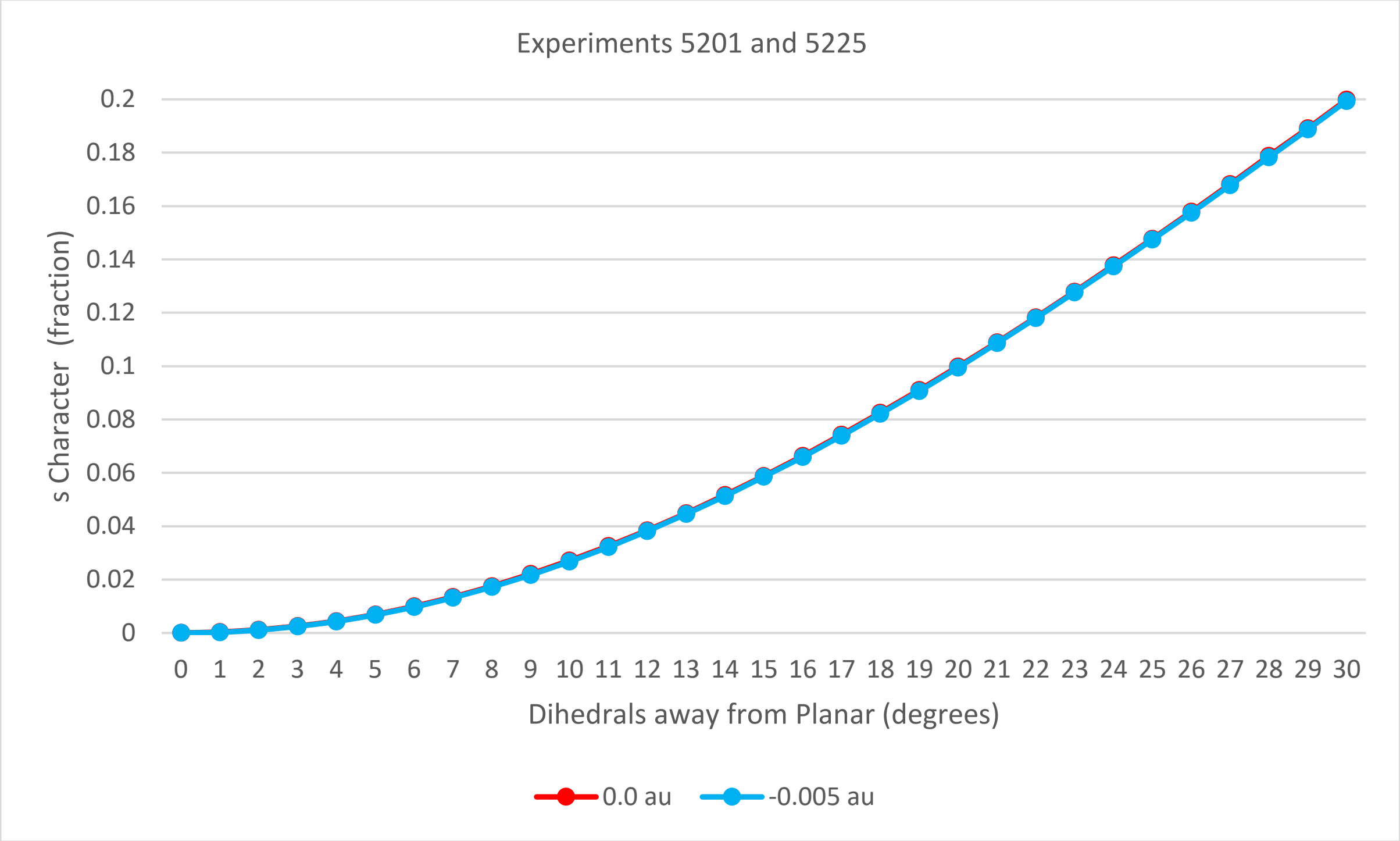

Figure 33. N(lp) NBO s Character in N-methylformamide with O-C-N-CA and O-C-N-H Dihedrals Constrained to Angles Away from Planarity toward sp3 Geometry at N in O-C-N Normal Uniform Electrostatic Field 0.0 and -0.005 au at SCS-MP2/aug-cc-pVTZ

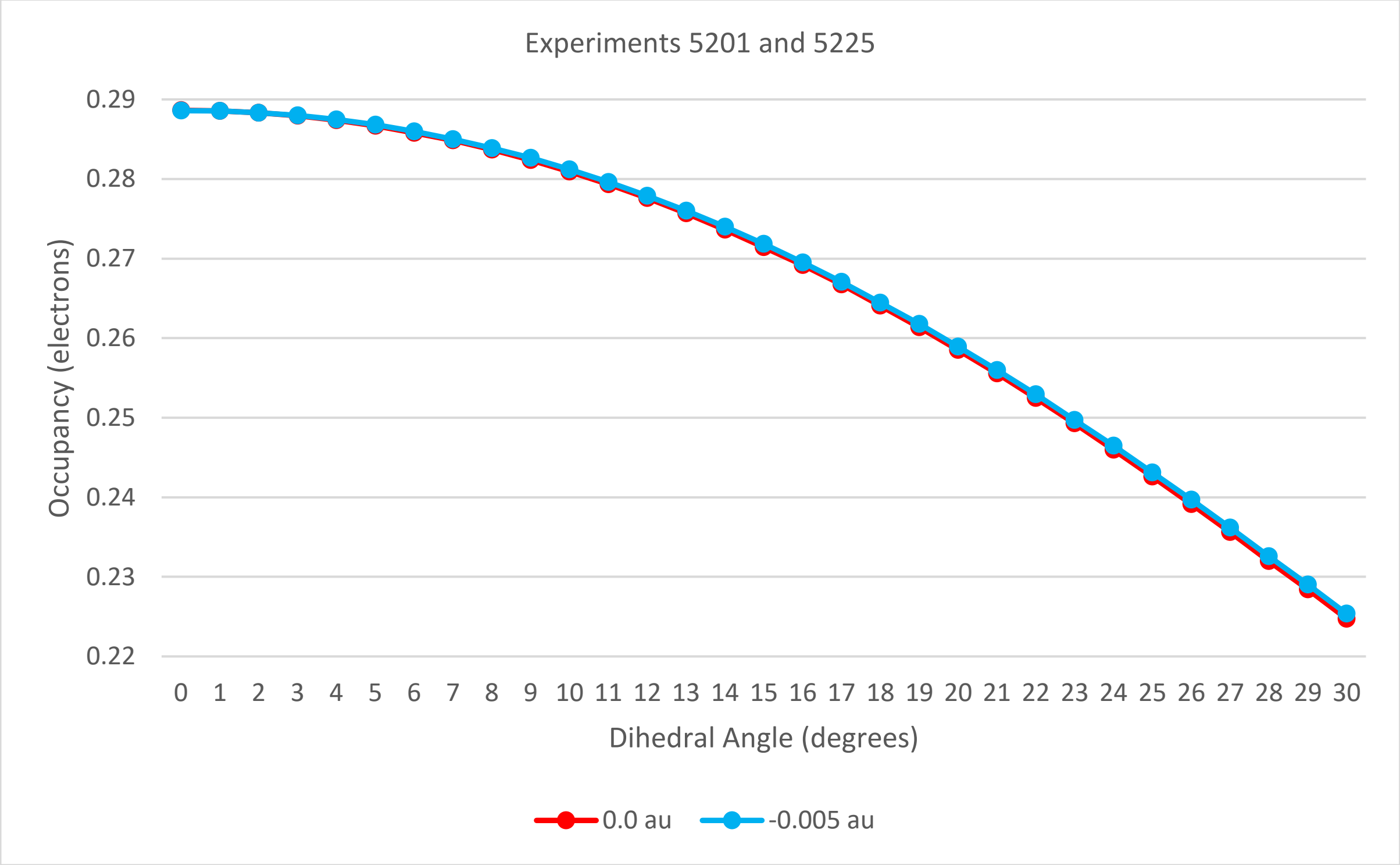

Figure 34. C-O(pi)* NBO Occupancy in N-methylformamide with O-C-N-CA and O-C-N-H Dihedrals Constrained Away from Planar Toward sp3 Geometry at N in O-C-N Normal Uniform Electrostatic Field of 0.0 and -0.005 au at SCS-MP2/aug-cc-pVTZ

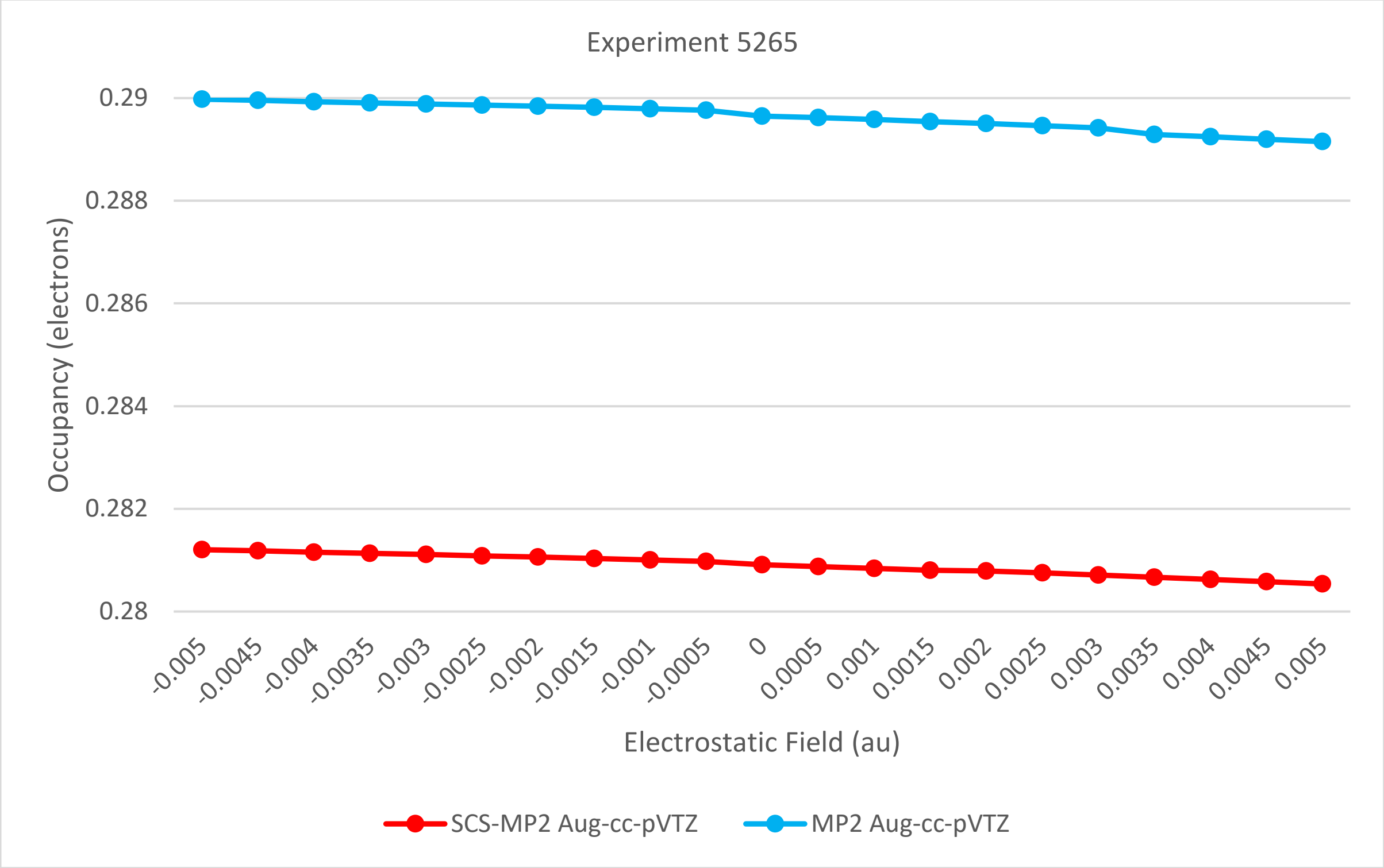

Figure 35. C-O(pi)* NBO Occupancy in N-methylformamide with O-C-N-CA and O-C-N-H Dihedrals 10 Degrees from Planar Toward sp3 geometry at N in O-C-N Normal Uniform Electrostatic Field of 0.0 and -0.005 au at SCS-MP2/aug-cc-pVTZ

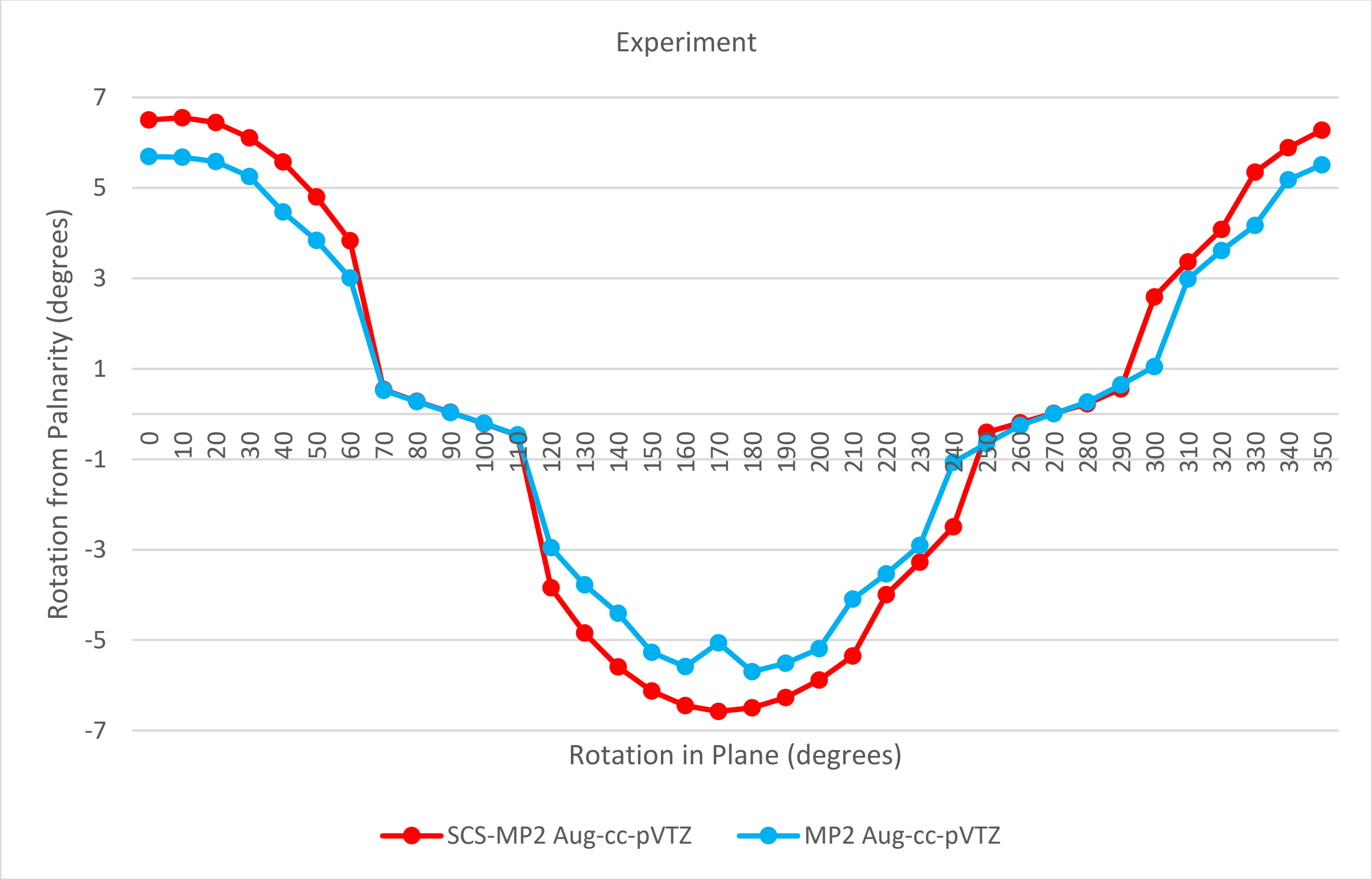

Figure 36. O-C-N-H Dihedral from Planarity in N-methylformamide in 0.005 au Uniform Electrostatic Field Rotated in Plane Containing N-C and O-C-N Normal Starting from O-C-N Normal

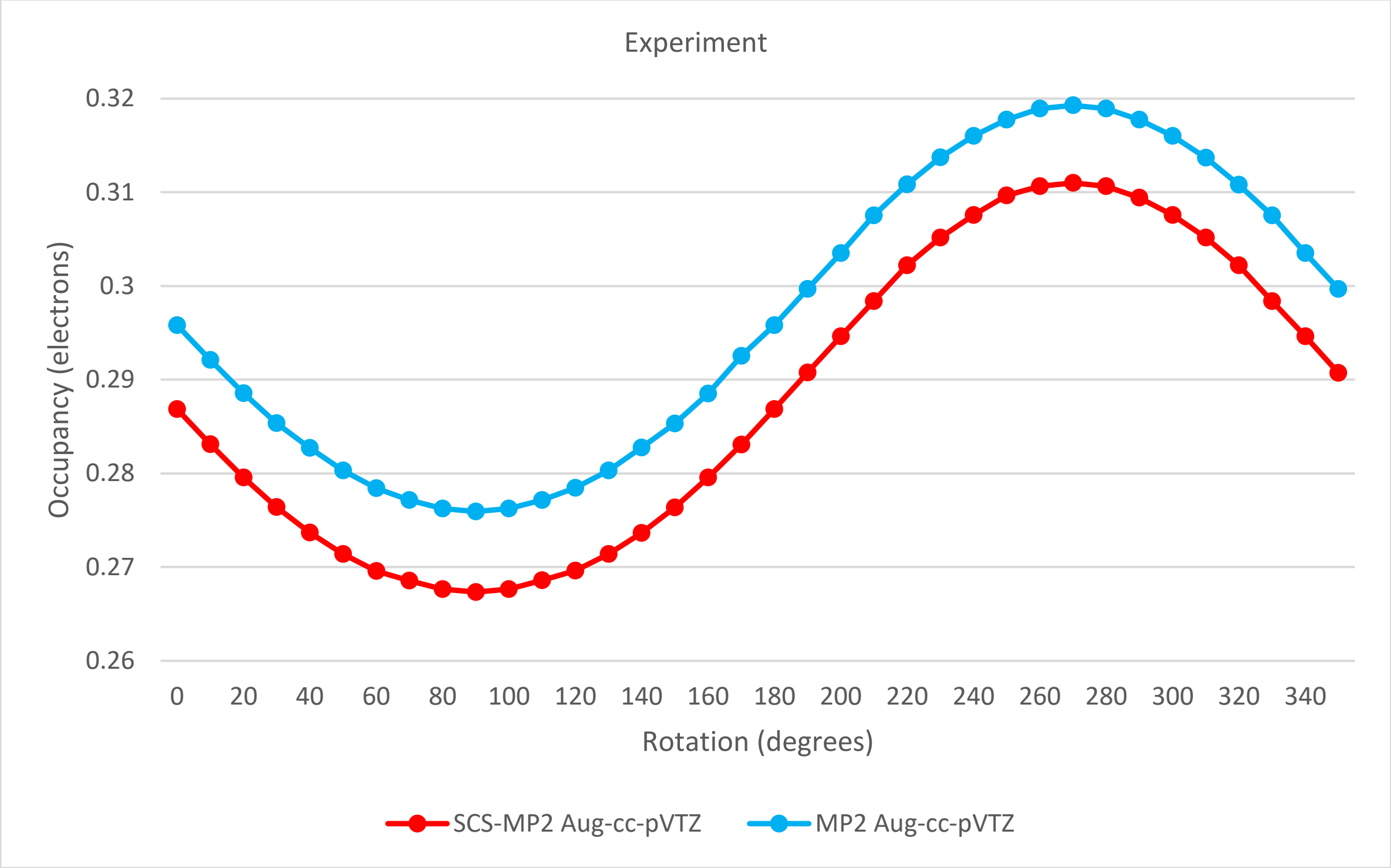

Figure 37. C-O(pi)* NBO Occupancy in N-methylformamide with Unconstrained H, CA and HA in a 0.005 au Uniform Electrostatic Field Rotated in Plane Containing N-C and O-C-N Normal Starting from O-C-N Normal

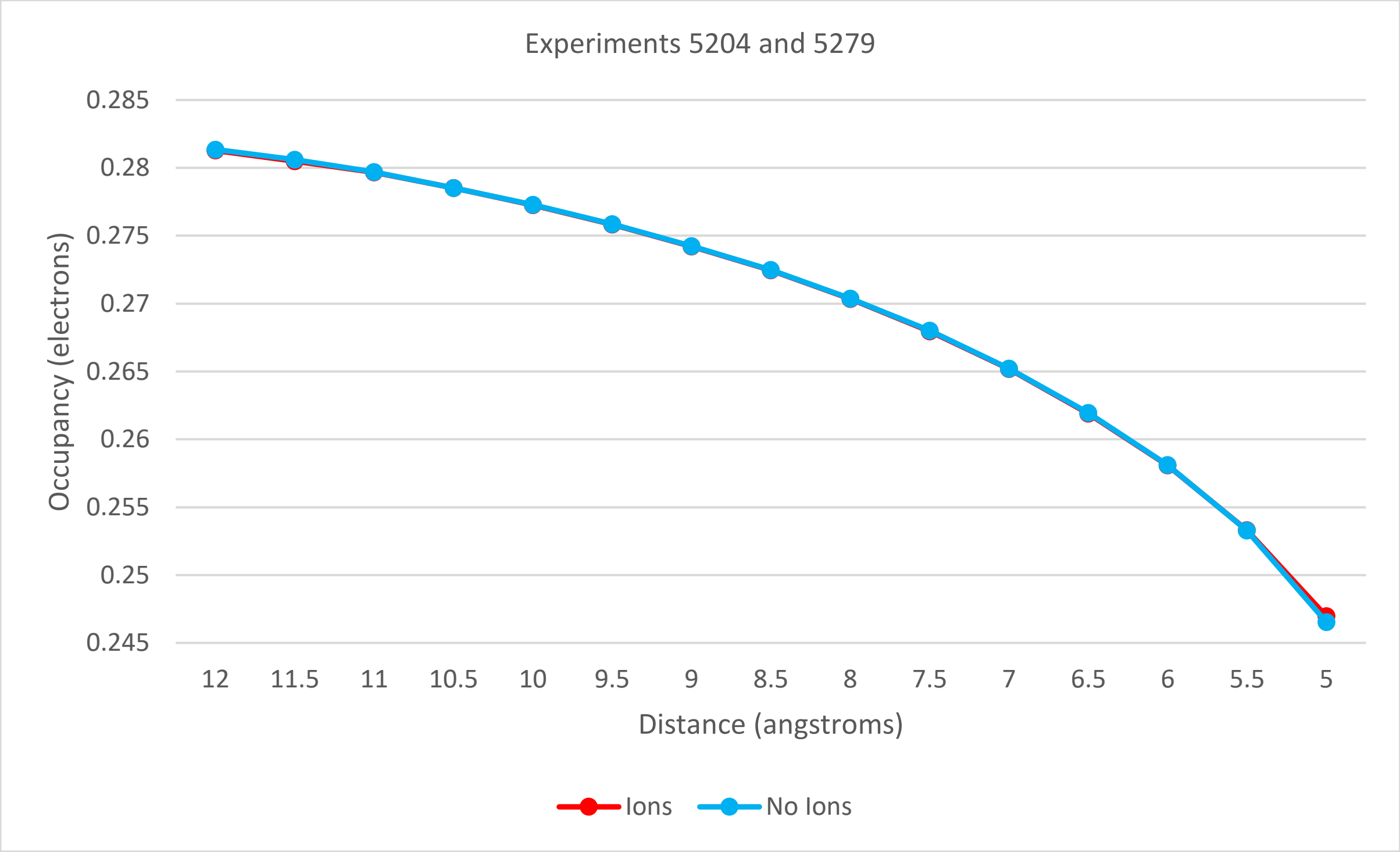

Figure 38. C-O(pi)* NBO Occupancy in N-methylformamide at O-C-N-CA and O-C-N-H Dihedrals Resulting from Li+ and F- On Opposite Sides Equidistant from N on O-C-N Normal with Ions Present and Ions Removed at SCS-MP2/aug-cc-pVTZ

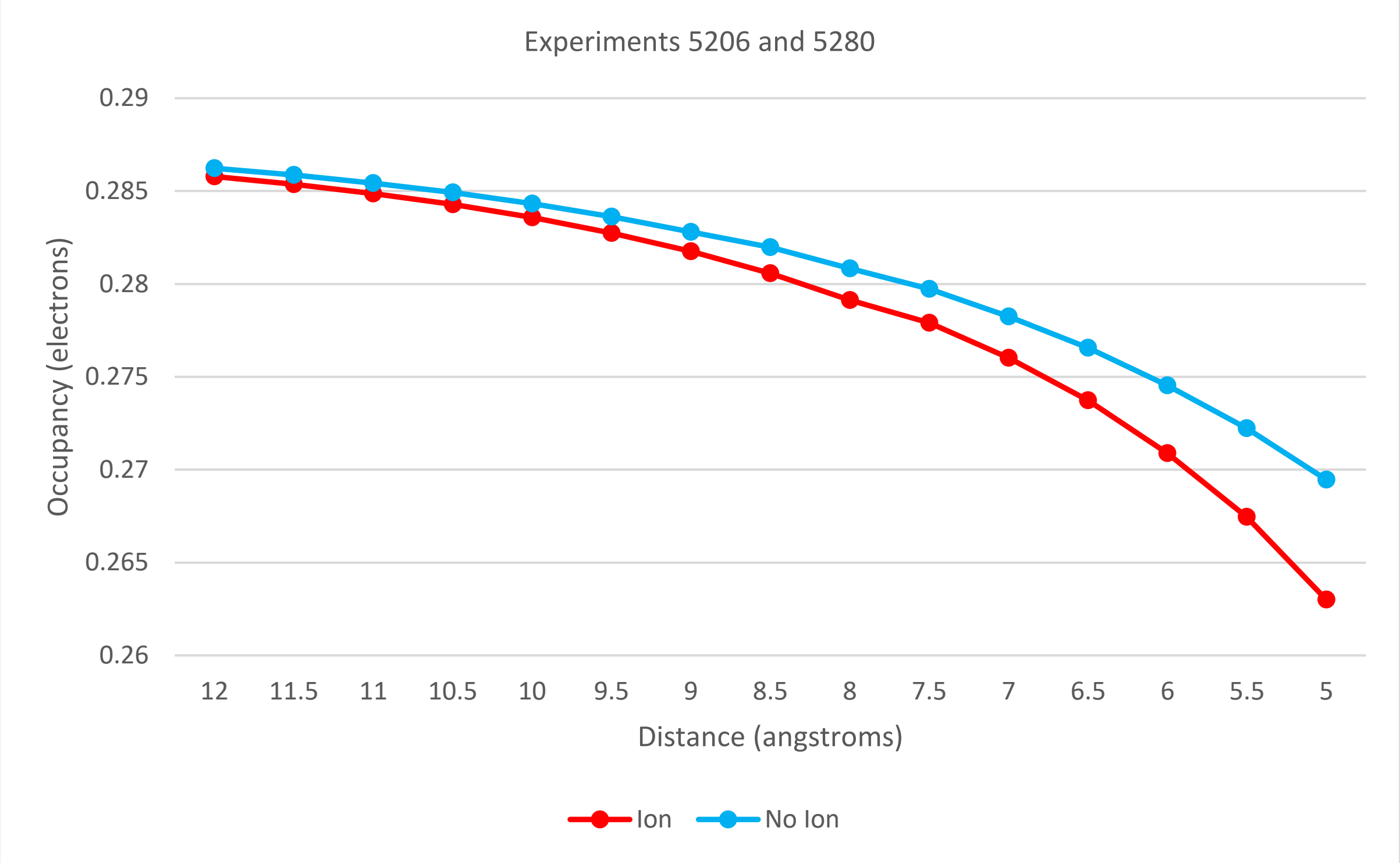

Figure 39. C-O(pi)* NBO Occupancy in N-methylformamide with O-C-N-CA and O-C-N-H Dihedrals Constrained to Result of Li+ on O-C-N Normal at Distance from N with Ion Present and Ion Removed at SCS-MP2/aug-cc-pVTZ

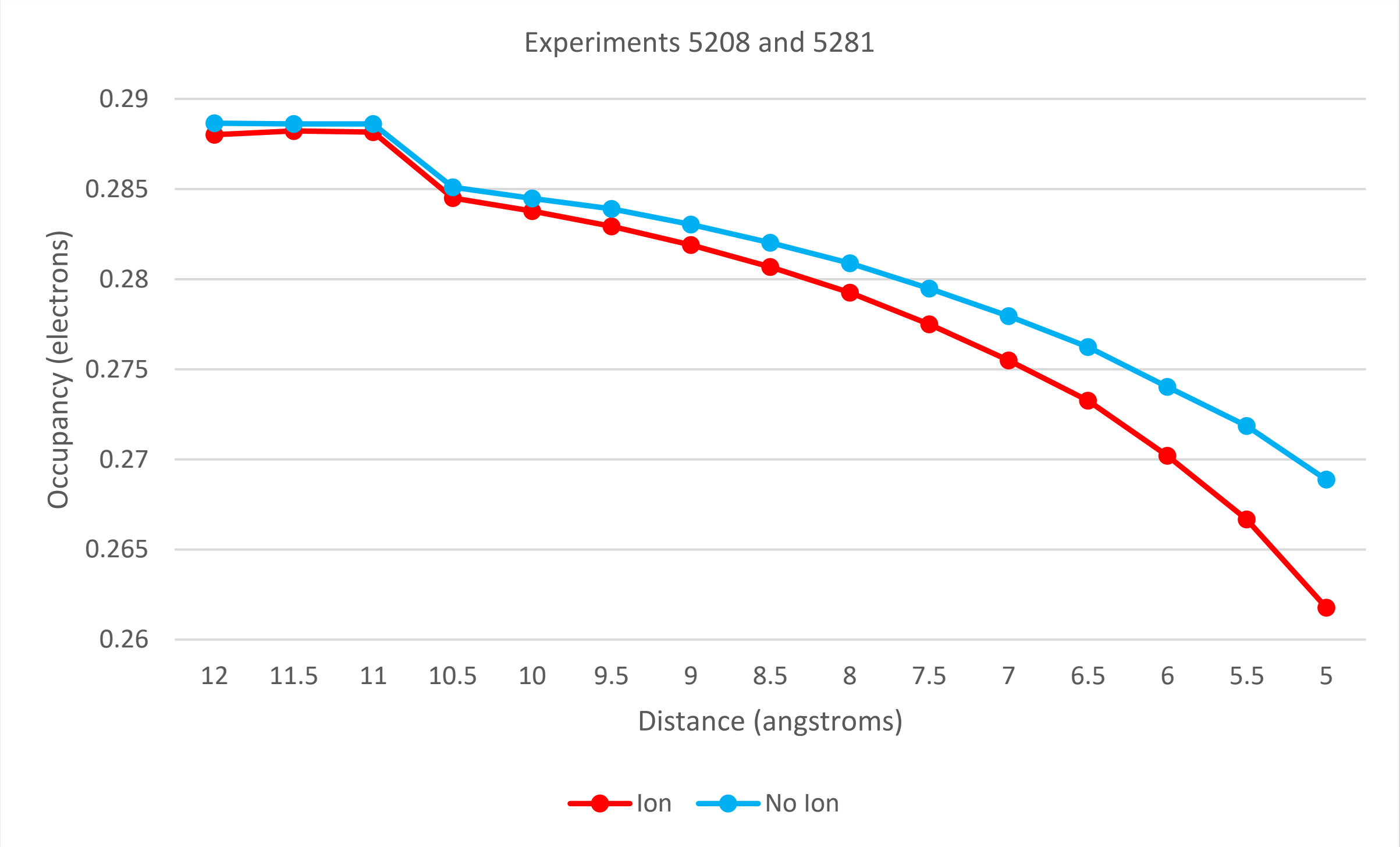

Figure 40. C-O(pi)* NBO Occupancy in N-methylformamide with O-C-N-CA and O-C-N-H Dihedrals Constrained to Result of Ammonium Ion on O-C-N Normal at Distance from N with Ion Present and Ion Removed at SCS-MP2/aug-cc-pVTZ

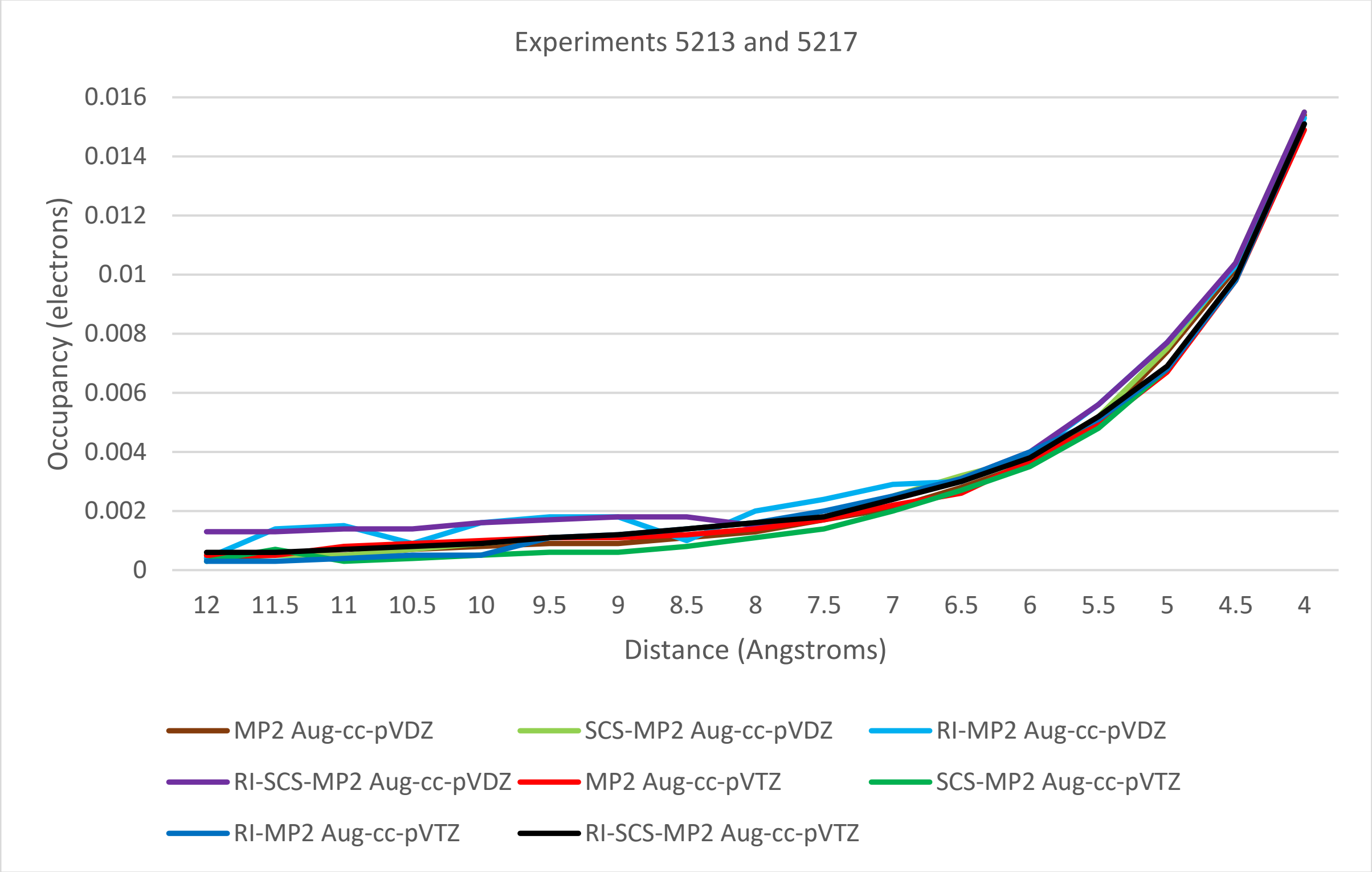

Figure 41. C-O(pi)* Occupancy in N-methylformamide with Ammonium Nitrogen at Distance from N Normal to O-C-N Plane Subtracted from that when Ammonium Not Present but O-C-N-H and O-C-N-CA Dihedrals Retained

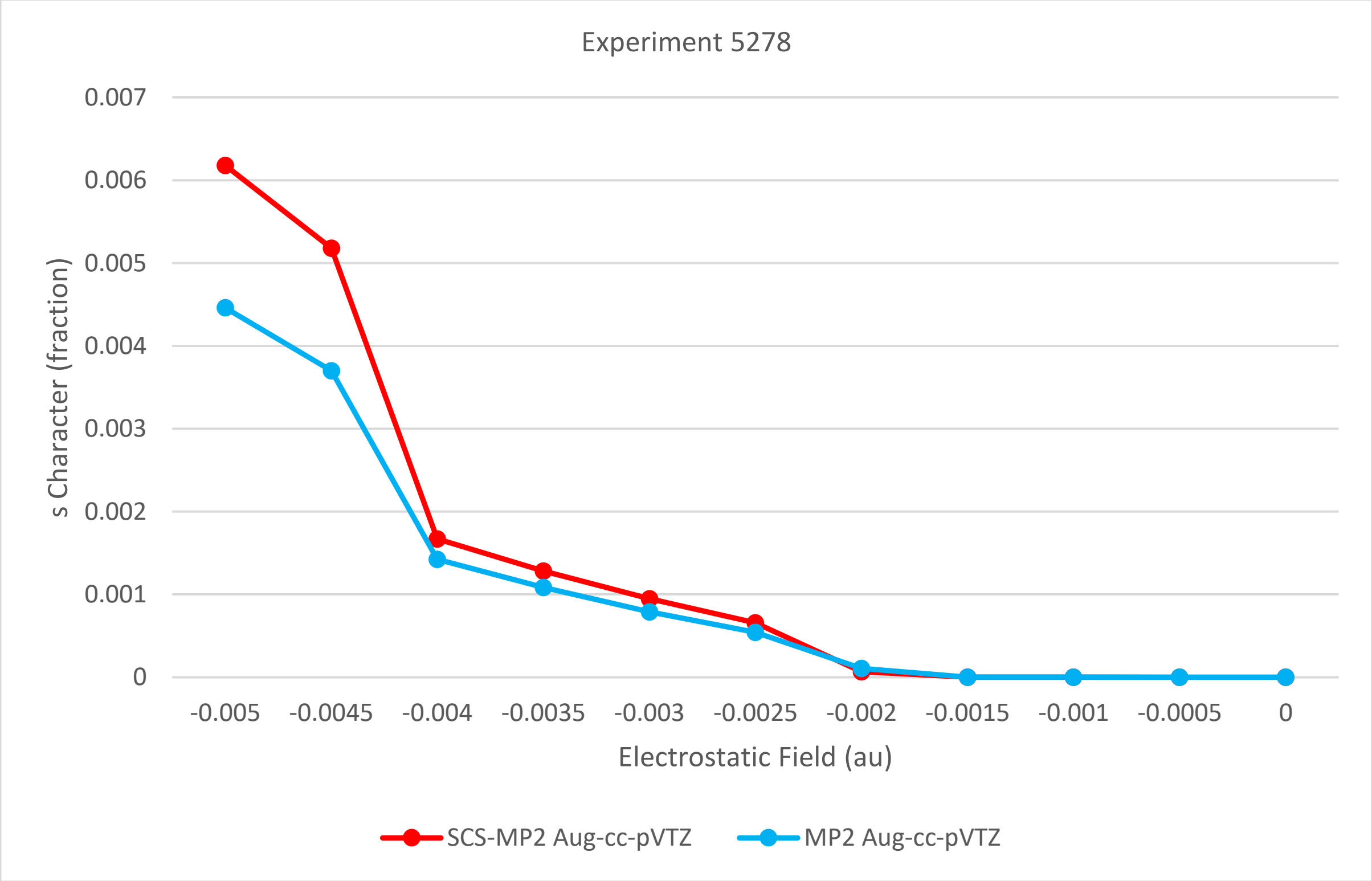

Figure 42. N(lp) NBO s Character in N-methylformamide with Unconstrained O-C-N-CA and O-C-N-H in O-C-N Normal Uniform Electrostatic Field

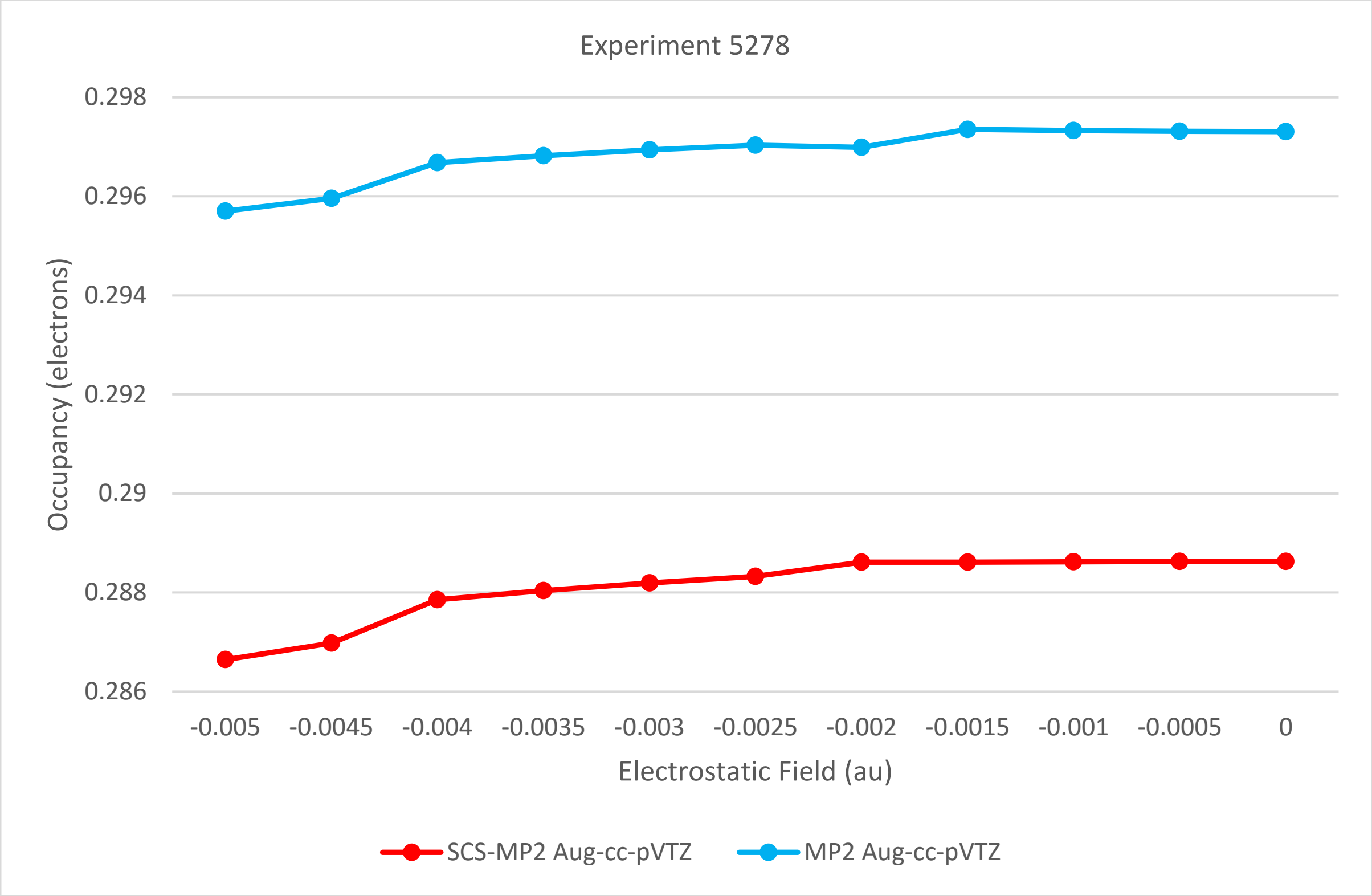

Figure 43. C-O(pi)* NBO Occupancy in N-methylformamide with Unconstrained O-C-N-CA and O-C-N-H in O-C-N Normal Uniform Electrostatic Field

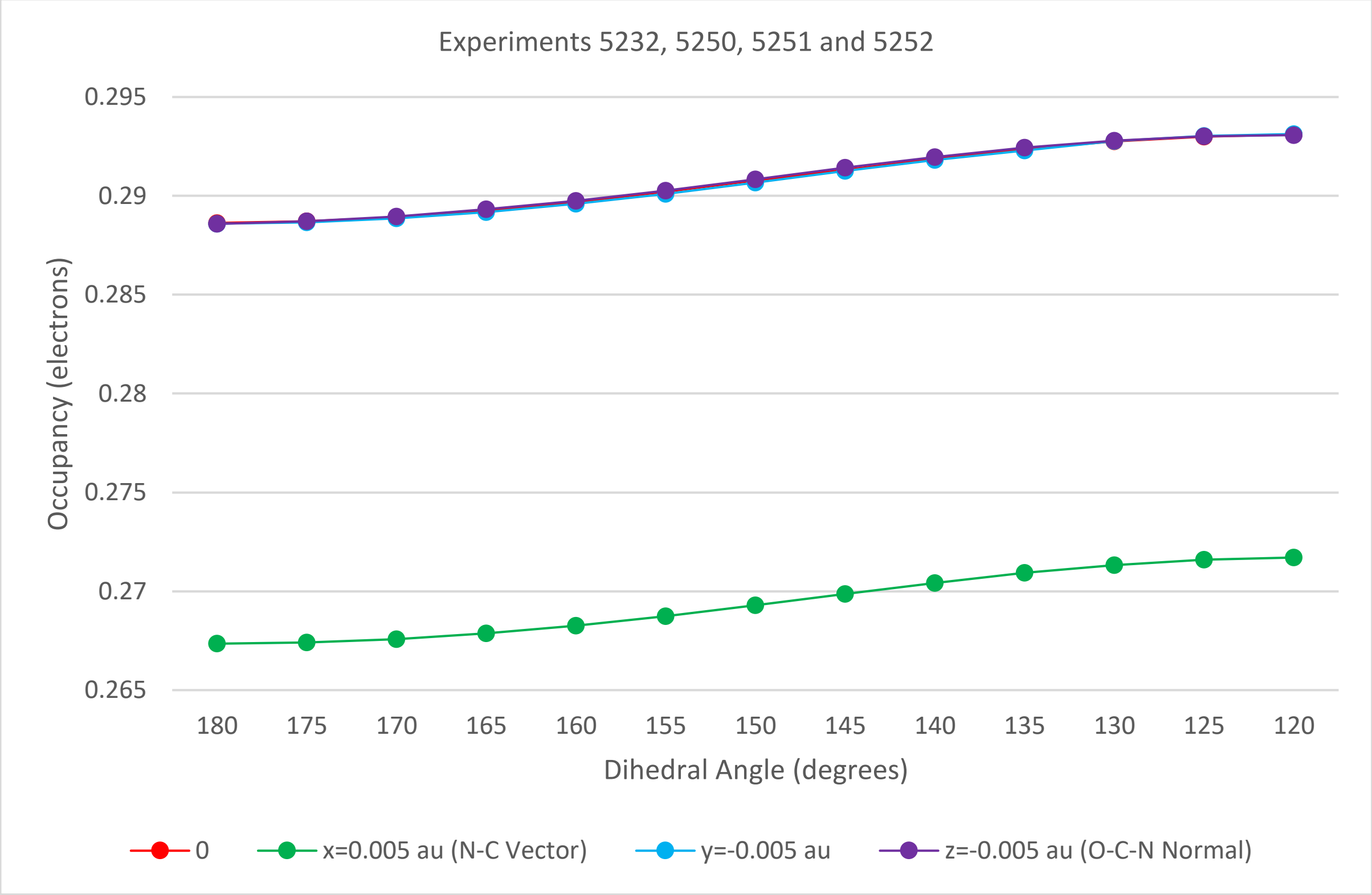

Figure 44. C-O(pi)* NBO Occupancy in N-methylformamide at C-N-CA-HA Dihedral in Uniform Electrostatic Field at SCS-MP2/aug-cc-pVTZ

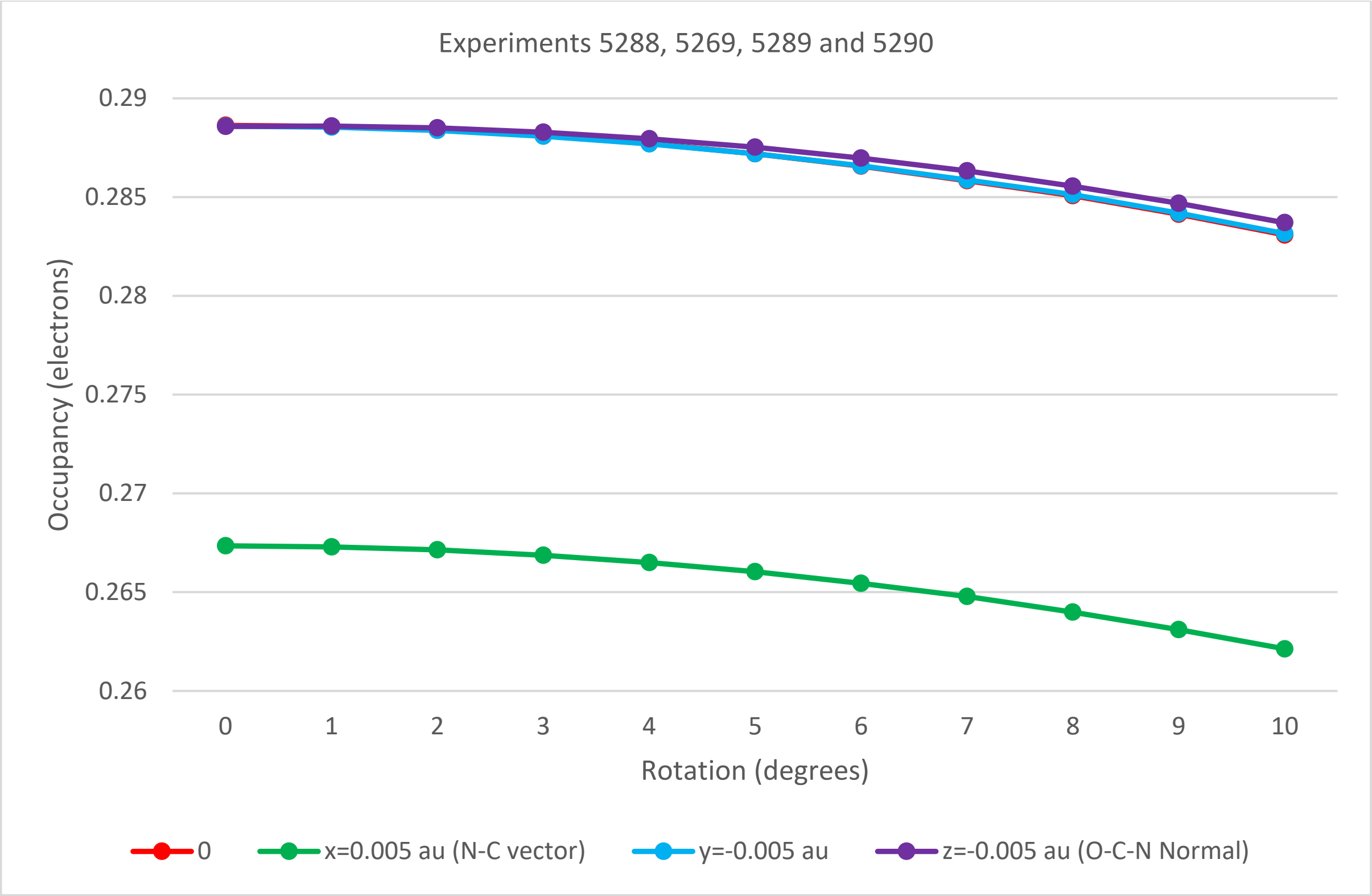

Figure 45. C-O(pi)* NBO Occupancy in N-methylformamide with Rotation About C-N Bond in Uniform Electrostatic Field at SCS-MP2/aug-cc-pVTZ

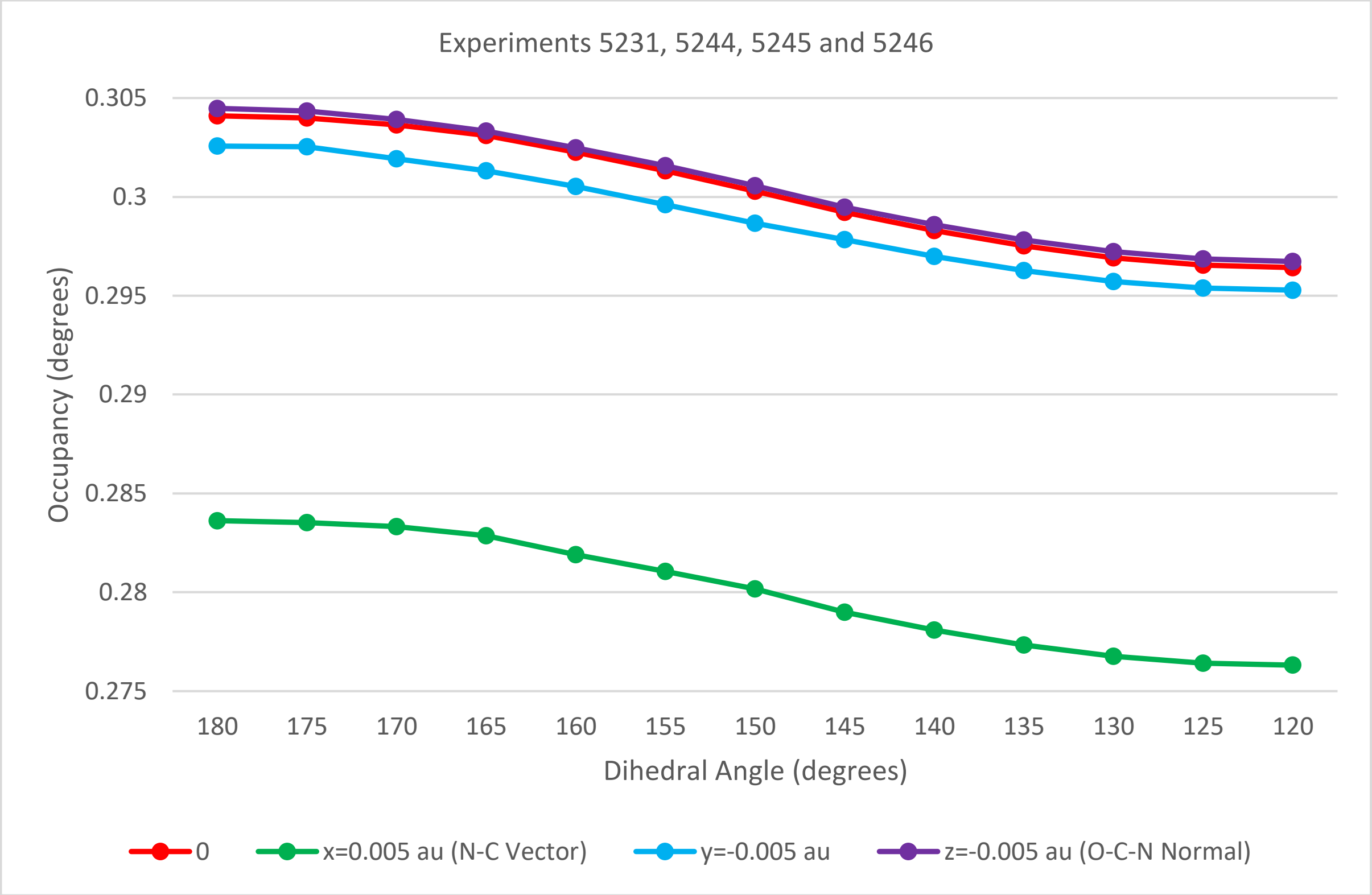

Figure 46. C-O(pi)* NBO Occupancy in N-methylethanamide with O-C-CA-H Dihedral Rotated in Uniform Electrostatic Field at SCS-MP2/aug-cc-pVTZ

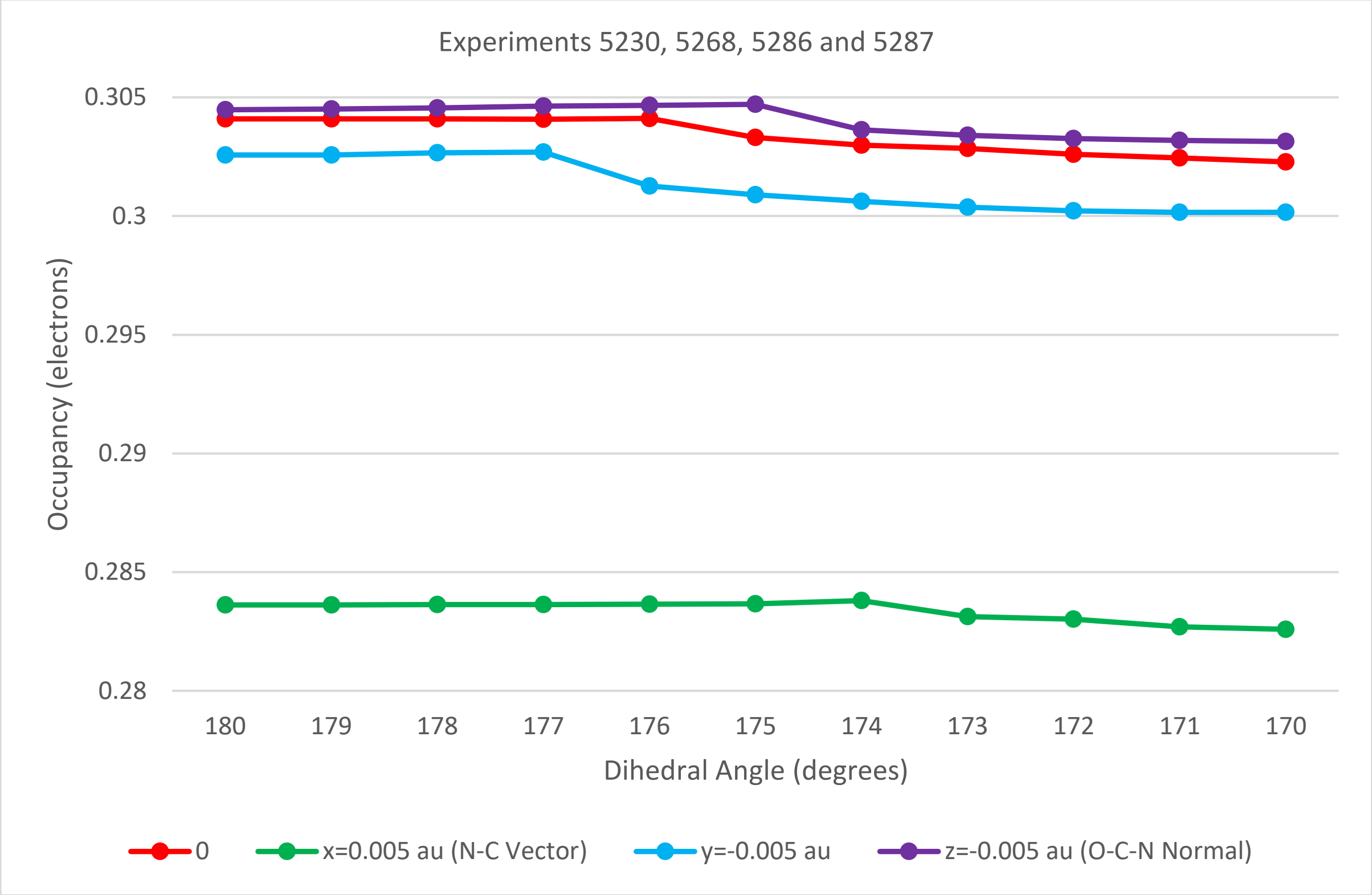

Figure 47. C-O(pi)* NBO Occupancy in N-methylethanamide Pyramidalized by Rotating CA-N-C-CA Dihedral Angle while Maintaining CA-N-C-O at Zero in Uniform Electrostatic Field at SCS-MP2/aug-cc-pVTZ

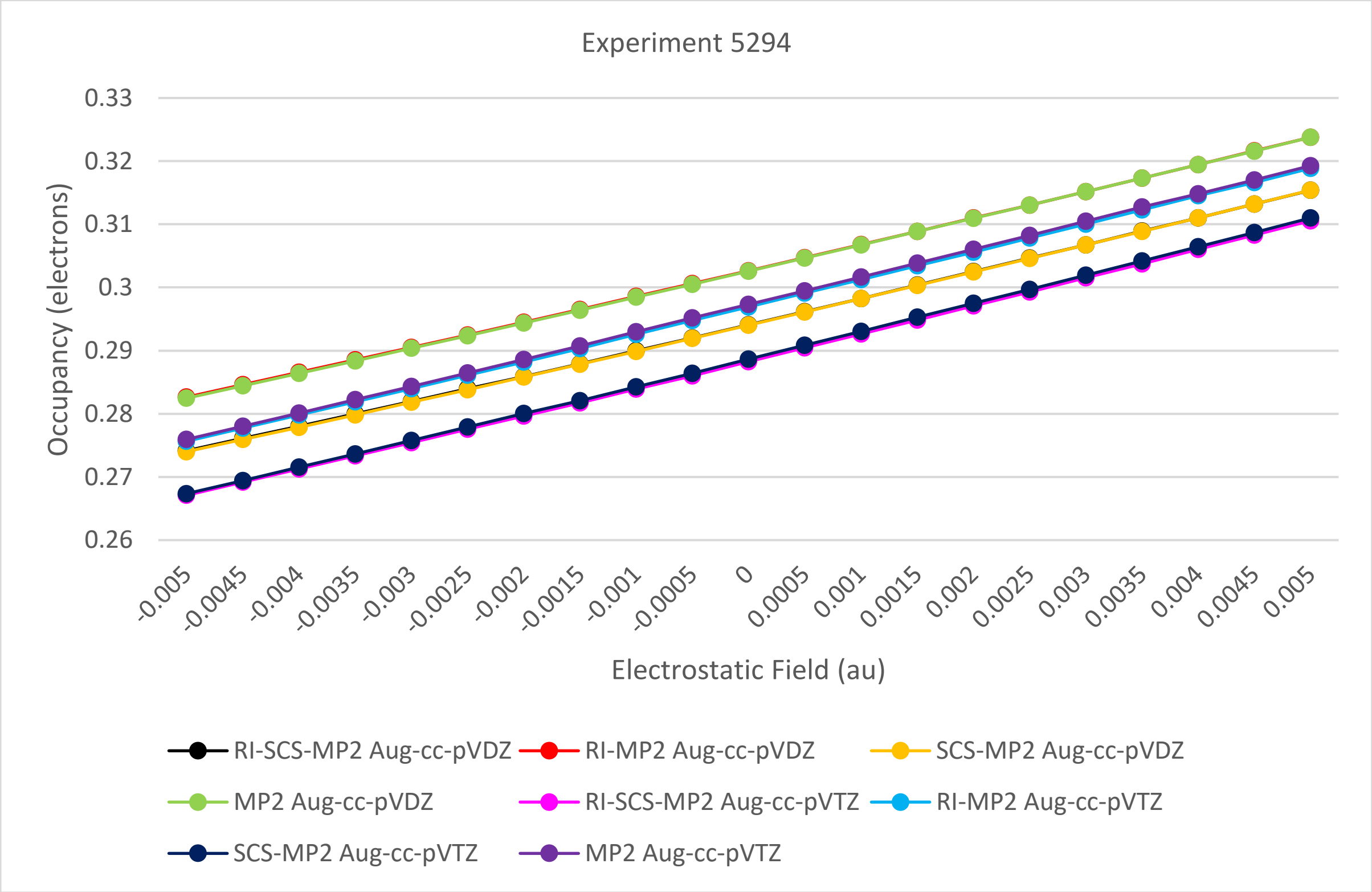

Figure 48. C-O(pi)* NBO Occupancy in N-methylformamide in Uniform Electrostatic Field with C-N Vector

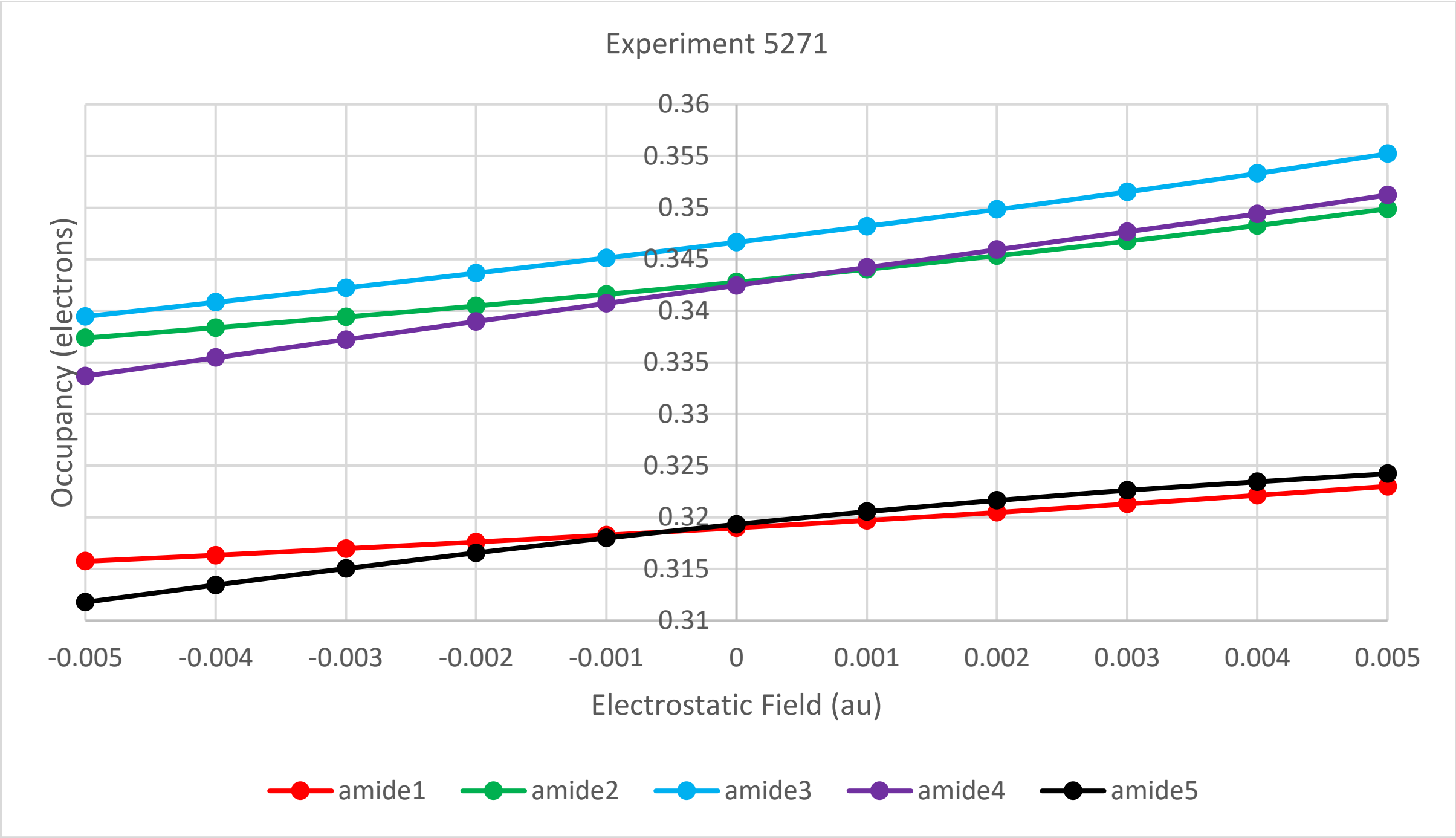

Figure 49. C-O(pi)* NBO Occupancy in Hydrogen Bonded N-methylformamides with Common N-C Vectors and O-C-N-H Planes in Uniform Electrostatic Field Orthogonal to N-C Vector in the O-C-N-H Plane at RI-SCS-MP2/aug-cc-pVDZ and Coulomb and Correlation Auxiliary Basis Sets

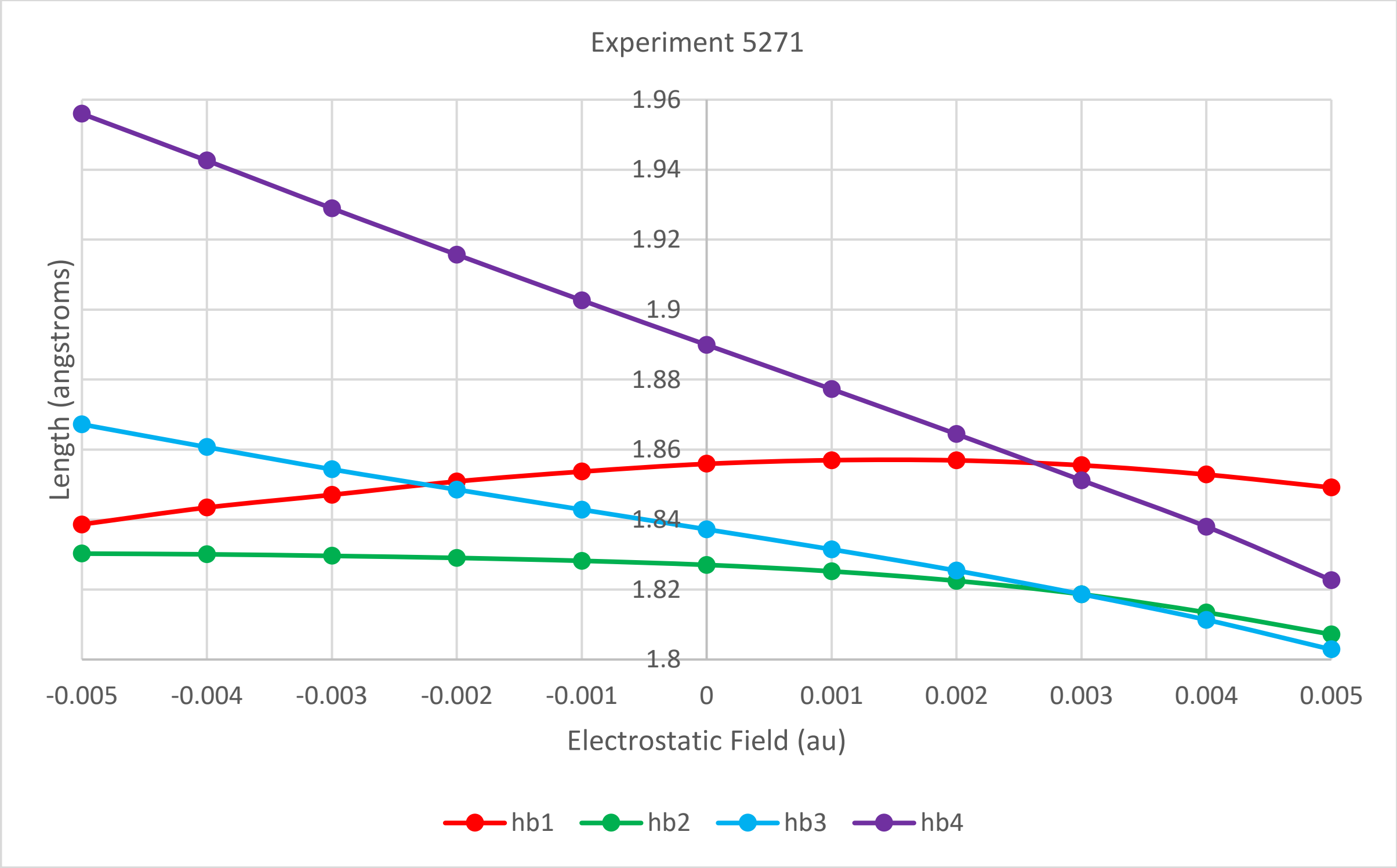

Figure 50. Hydrogen Bond Length in a Chain of N-methylformamides with Common N-C Vectors and O-C-N-H Planes in Uniform Electrostatic Field Orthogonal to N-C Vectors in O-C-N-H Plane at RI-SCS-MP2/aug-cc-pVDZ and Coulomb and Correlation Auxiliary Basis Sets

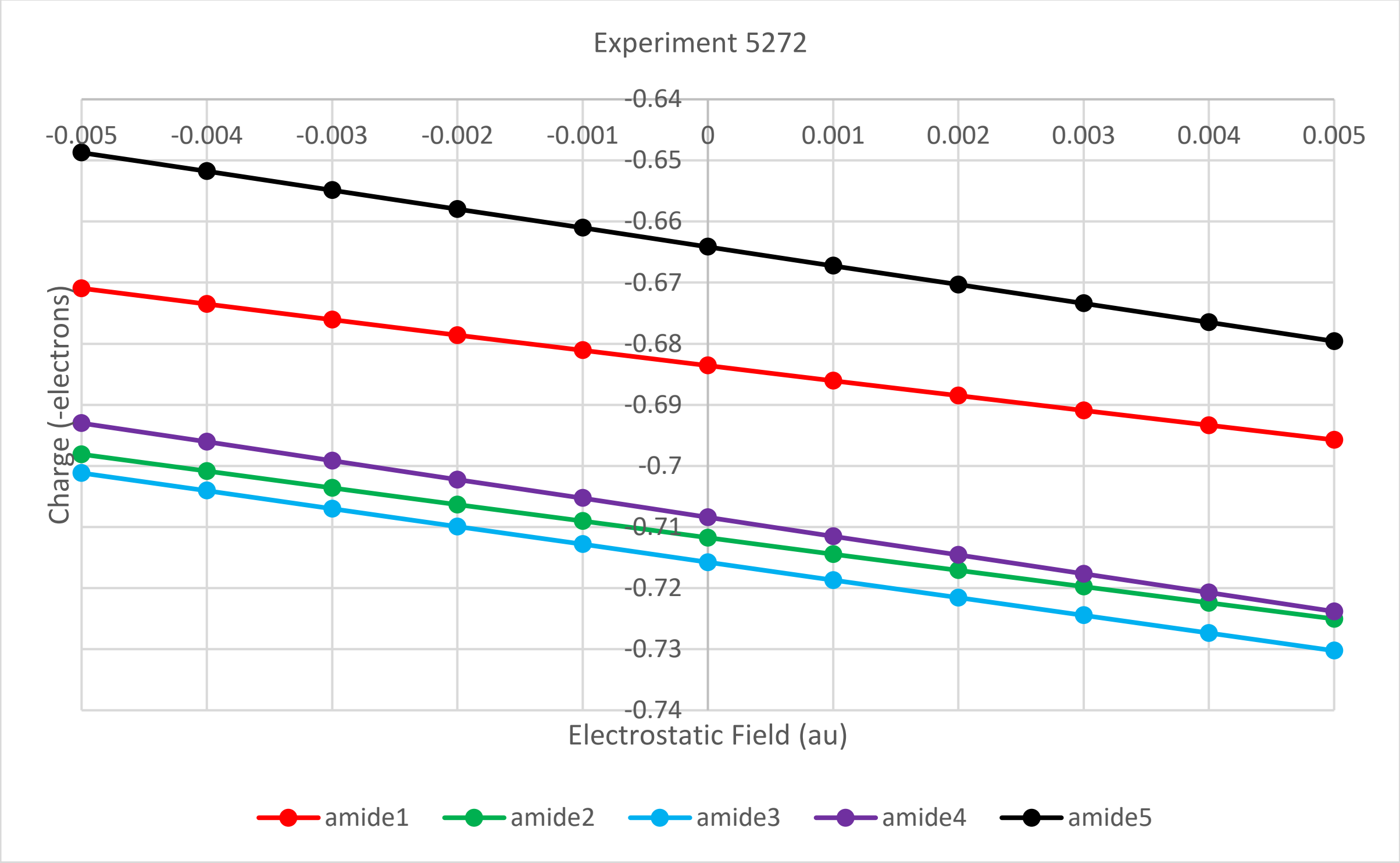

Figure 51. Oxygen Natural Atomic Charge in Hydrogen Bonded Chain of N-methylformamides with Common N-C Vectors and O-C-N-H Planes in Uniform Electrostatic Field Orthogonal to H-N Vectors in O-C-N-H Plane at RI-SCS-MP2/aug-cc-pVDZ and Coulomb and Correlation Auxiliary Basis Sets

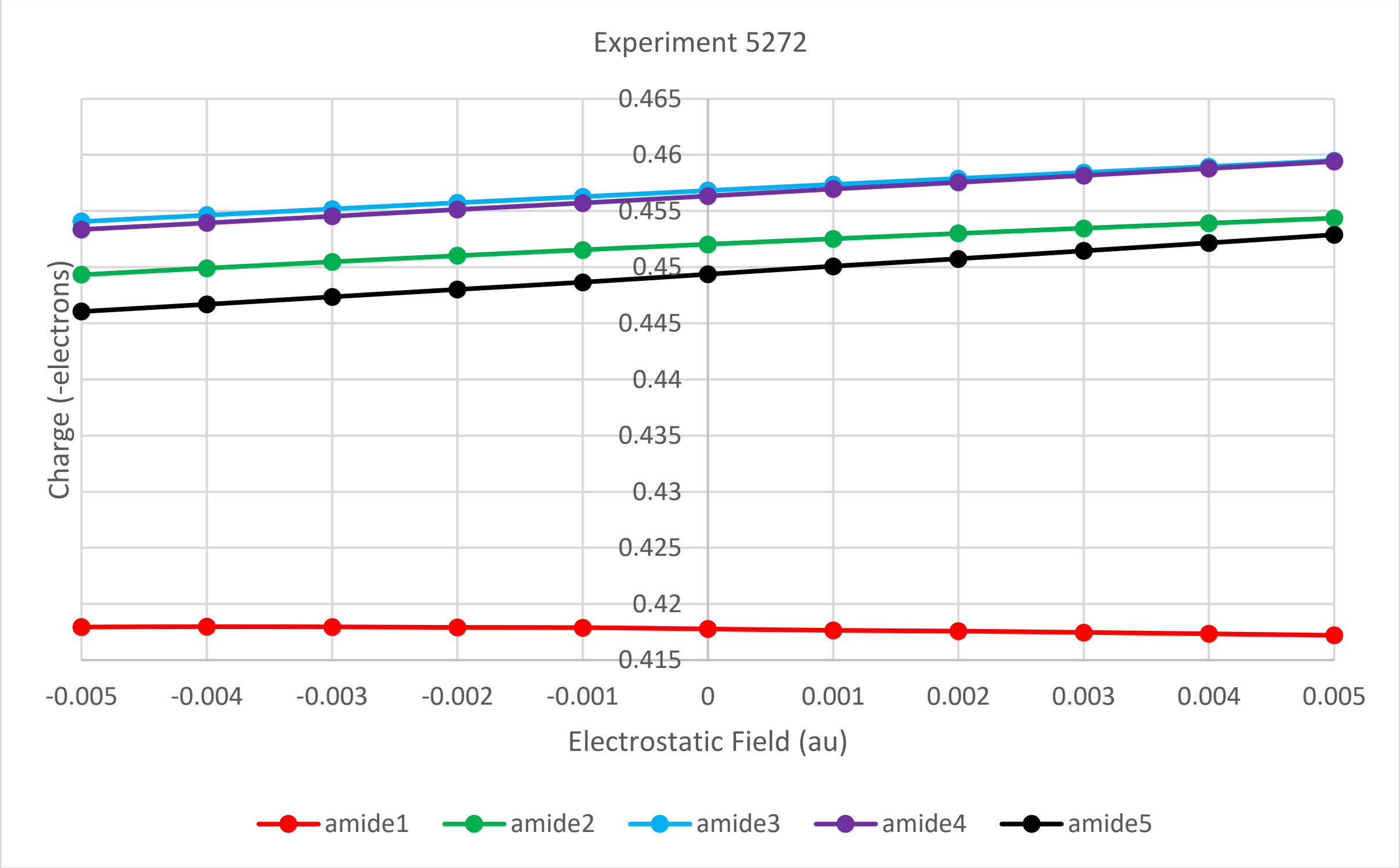

Figure 52. Amide Proton Natural Atomic Charge in Hydrogen Bonded Chain of N-methylformamides with Common N-C Vectors in Uniform Electrostatic Field Orthogonal to H-N Vectors in O-C-N-H Plane at RI-SCS-MP2/aug-cc-pVDZ and Coulomb and Correlation Auxiliary Basis Sets

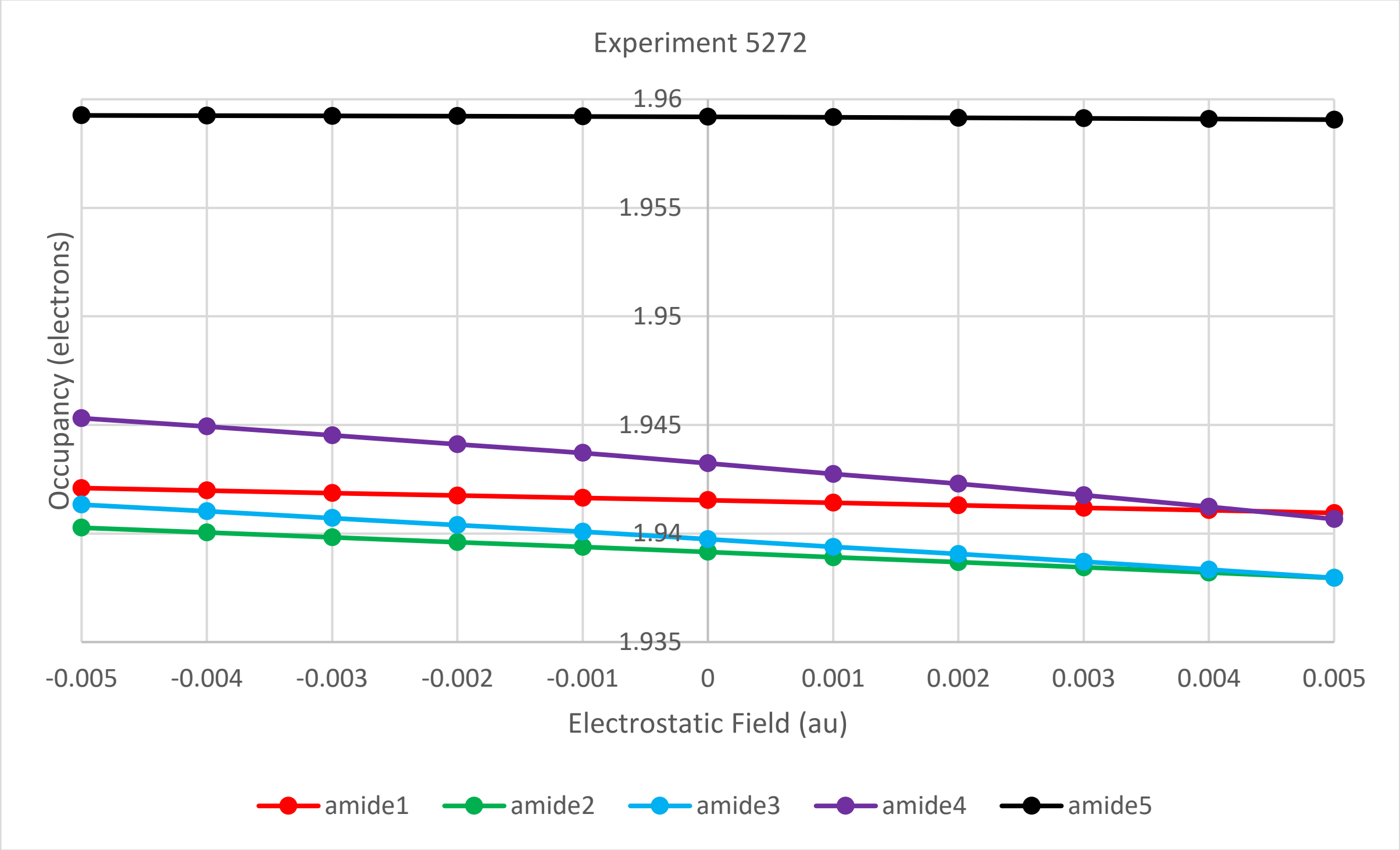

Figure 53. O(sigma-lp) NBO Occupancy in Hydrogen Bonded Chain of N-methylformamides with Common N-C Vectors and O-C-N-H Planes in Uniform Electrostatic Field Orthogonal to H-N Vectors in O-C-N-H Plane at RI-SCS-MP2/aug-cc-pVDZ and Coulomb and Correlation Auxiliary Basis Sets

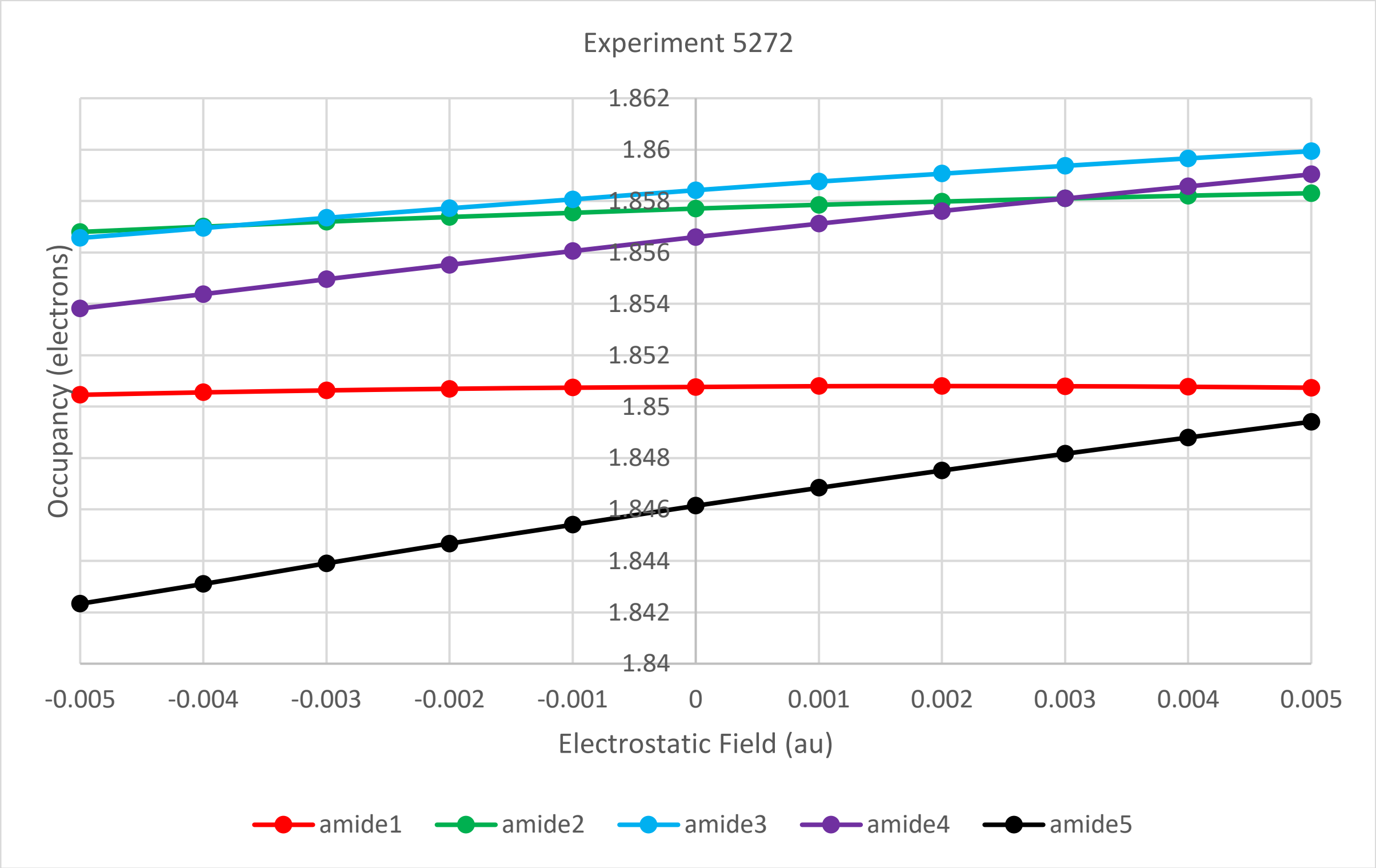

Figure 54. O(pi-lp) NBO Occupancy in Hydrogen Bonded Chain of N-methylformamides with Common N-C Vectors and O-C-N-H Planes in Uniform Electrostatic Field Orthogonal to H-N Vectors in O-C-N-H Plane at RI-SCS-MP2/aug-cc-pVDZ and Coulomb and Correlation Auxiliary Basis Sets

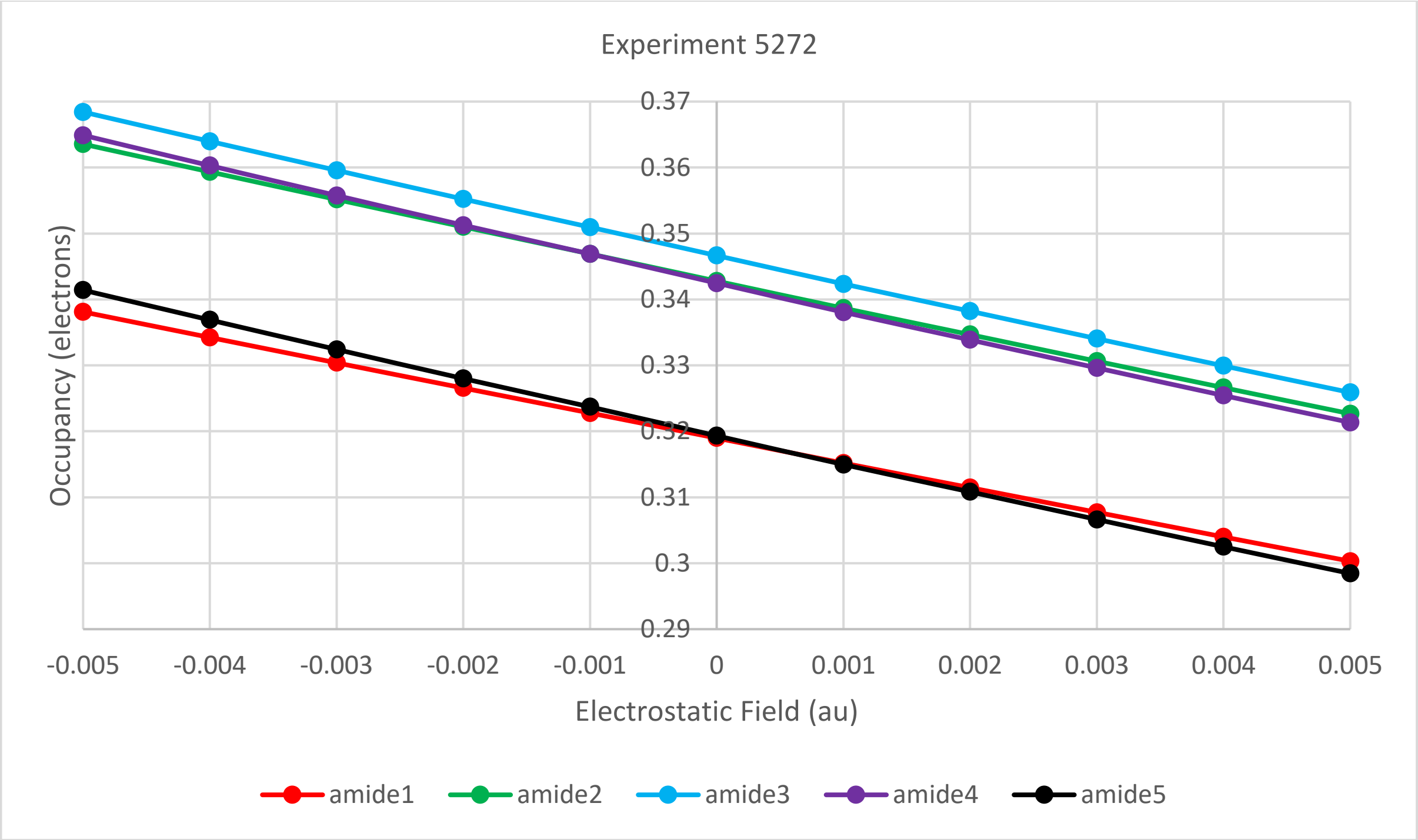

Figure 55. C-O(pi)* NBO Occupancy in Hydrogen Bonded N-methylformamides with Common N-C Vectors and O-C-N-H Planes in Uniform Electrostatic Field Orthogonal to H-N Vectors (28.861 degrees from N-C Vectors) in O-C-N-H Plane at RI-SCS-MP2/aug-cc-pVDZ and Coulomb and Correlation Auxiliary Basis Sets

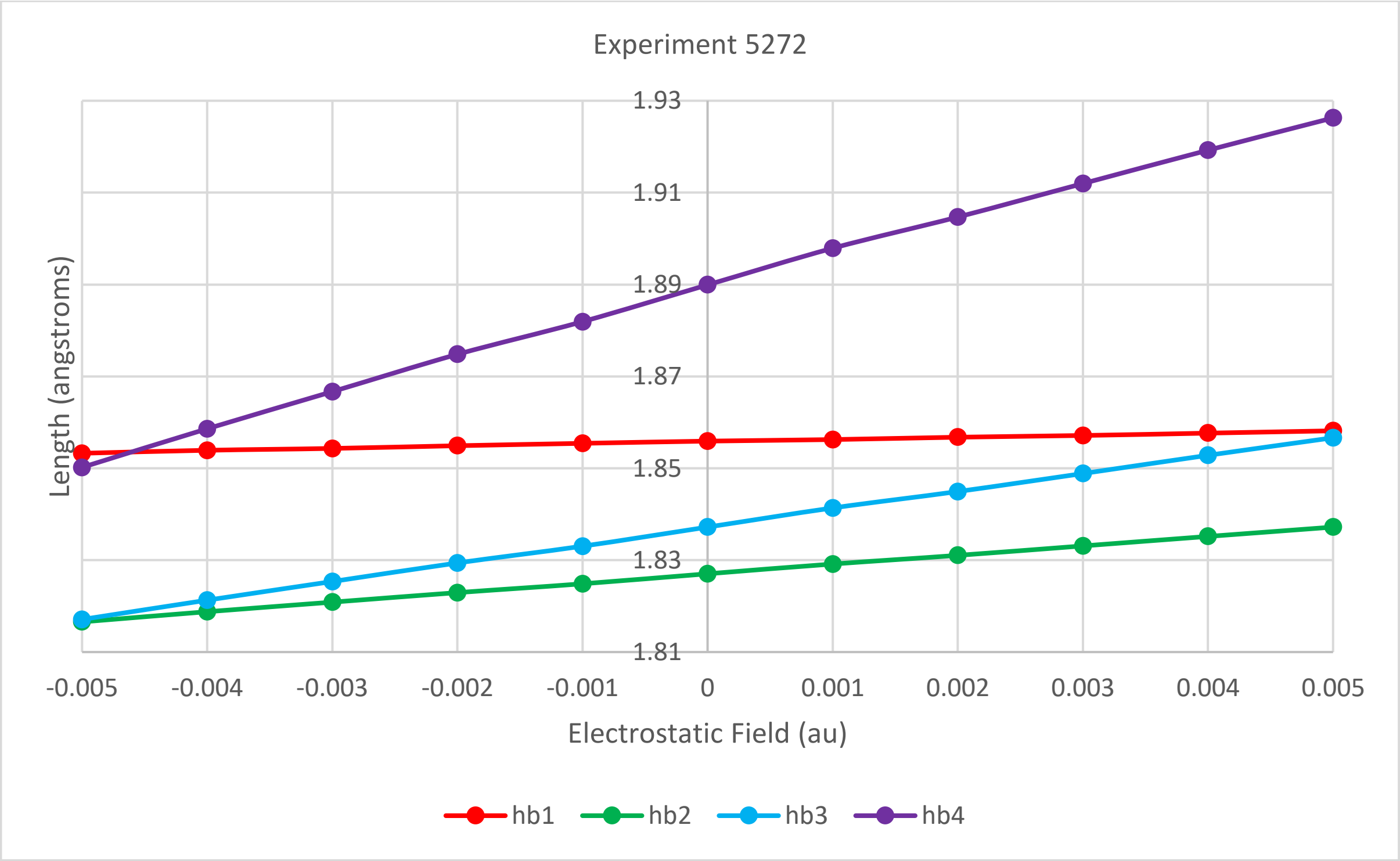

Figure 56. Hydrogen Bond Length in Chain of N-methylformamides with Common N-C Vectors and O-C-N-H Planes in Uniform Electrostatic Field Orthogonal to H-N Vectors in O-C-N-H Plane at RI-SCS-MP2/aug-cc-pVDZ and Coulomb and Correlation Auxiliary Basis Sets

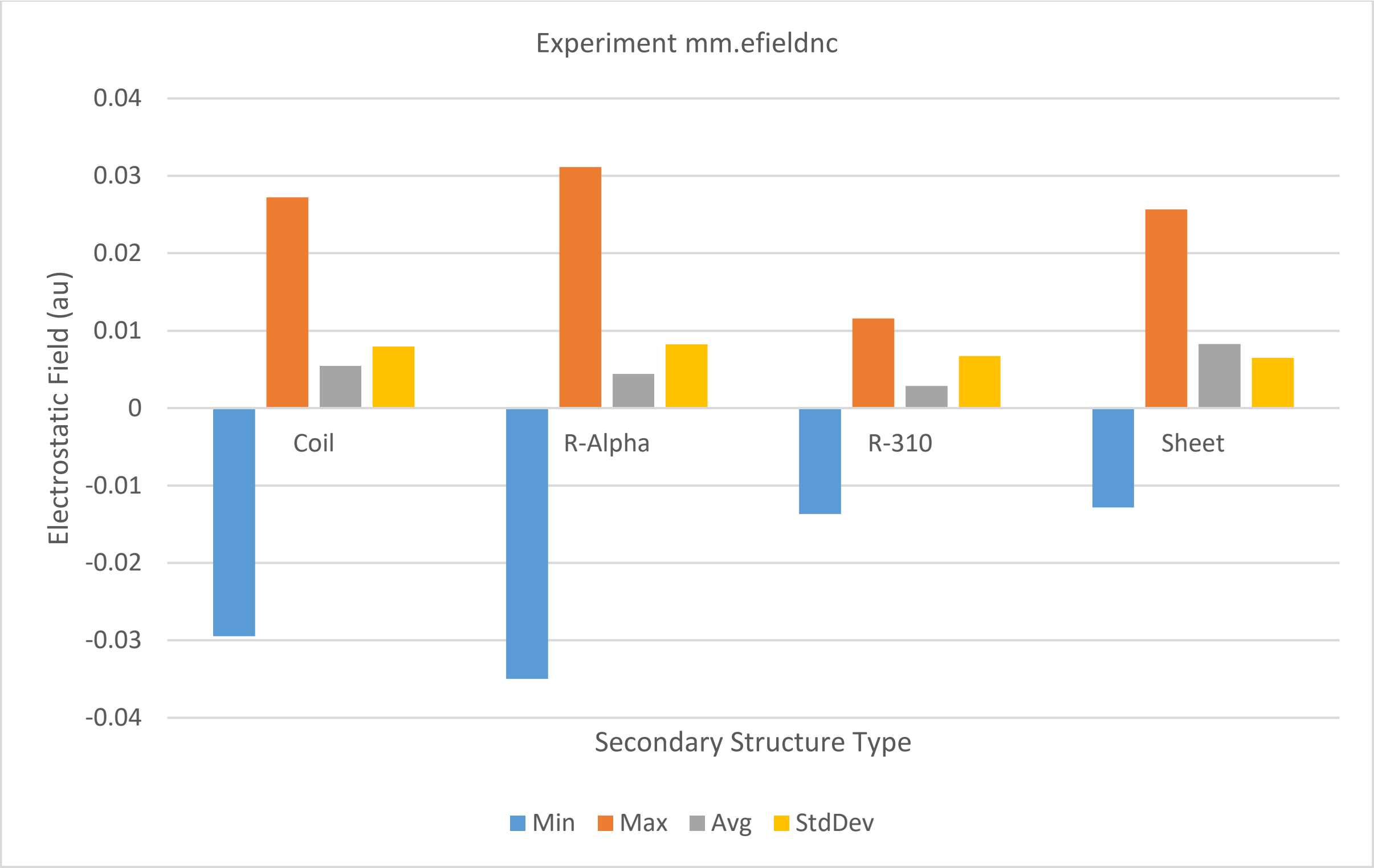

Figure 57. C-N Vector Component of Electrostatic at C in Backbone Amides of 10 Small Proteins as Calculated with AMOEBAPRO-2013 and Generalized Kirkwood Continuous Solvent

# 13 Appendix 2

Table 2. Hydrogen Bond Length (Angstroms), Dihedral Angles (Degrees) and C-O(pi)* NBO Occupancy (Electrons), with Central Atom of Molecule Constrained to O-C-N Normal from N in N-methylformamide at SCS-MP2/aug-cc-pVTZ

| Molecule | Length | O-C-N-H | O-C-N-CA | Occupancy |
|---|---|---|---|---|
| methane | 2.678 | 179.941 | 0.652 | 0.28791 |
| ammonia | 2.4 | -170.687 | -4.424 | 0.28121 |
| hydrogen sulfide | 2.357 | -170.353 | -5.043 | 0.27617 |
| water | 2.213 | -164.869 | -8.225 | 0.26296 |
| ammonium | 1.812 | -153.949 | -19.062 | 0.1961 |